%% file: pap.tex
\documentclass[number,times]{elsarticle}
\pdfoutput=1
\makeatletter
\def\ps@pprintTitle{%
     \let\@oddhead\@empty
     \let\@evenhead\@empty
     \def\@oddfoot{\footnotesize\itshape
       \parbox{\textwidth}{
         \copyright\ 2016. This accepted manuscript is made available under the CC-BY-NC-ND 4.0 licence: \url{http://creativecommons.org/licenses/by-nc-nd/4.0/}. Published in International Journal of Approximate Reasoning (2016) pp.~30--57, \url{http://dx.doi.org/10.1016/j.ijar.2016.03.001}
         }
       }%
     \let\@evenfoot\@oddfoot}
\makeatother

\usepackage{standalone}

\usepackage[british]{babel} %

\usepackage{array}
\usepackage{amsfonts,amsthm,amssymb,xfrac}
\usepackage{sty/notation}
\usepackage[table]{xcolor}
\usepackage[left]{eurosym} %
\usepackage{subfig}
\usepackage{wrapfig}

\DeclareRobustCommand{\VAN}[3]{#2 #1}
\usepackage{tikz}
\usetikzlibrary{external}
\usetikzlibrary{shapes.geometric,arrows}
\usetikzlibrary{circuits.ee.IEC}
\usetikzlibrary{calc}
\usetikzlibrary{patterns}
\usetikzlibrary{positioning}
\usepackage{pgfplots}
\pgfplotsset{compat=1.5} 

\input{declarations.tex}

\newcommand{\commentout}[1]{}

\newcommand{\TODOhide}[1]{} %

\newcommand{\entropystepsize}{1/18} %

\begin{document}
\begin{frontmatter}

\title{%
Robust Probability Updating
\tnoteref{t1}}

\author[uva]{Thijs van Ommen\corref{cor1}}
\ead{T.vanOmmen@uva.nl}

\author[cwi]{Wouter M. Koolen}
\ead{wmkoolen@cwi.nl}

\author[ul]{Thijs E. Feenstra}

\author[cwi,ul]{Peter D. Gr\"{u}nwald}
\ead{pdg@cwi.nl}

\address[uva]{Universiteit van Amsterdam, Science Park 904, 1098 XH Amsterdam, The Netherlands}
\address[cwi]{Centrum Wiskunde \& Informatica, Science Park 123, 1098 XG Amsterdam, The Netherlands}
\address[ul]{Universiteit Leiden,  Niels Bohrweg 1, 2333 CA Leiden, The Netherlands}

\tnotetext[t1]{%
This work is adapted from dissertation \cite[Chapters 6 and 7]{vanOmmen_phd}, which extends MSc.\ thesis~\cite{FeenstraThesis}.
}

\cortext[cor1]{Corresponding author}

\begin{abstract}%
  This paper discusses an alternative to conditioning that may be used
  when the probability distribution is not fully specified. It does not require
  any assumptions (such as CAR: coarsening at random) on the unknown
  distribution. The well-known Monty Hall problem is the simplest
  scenario where neither naive conditioning nor the CAR assumption
  suffice to determine an updated probability distribution. This paper
  thus addresses a generalization of that problem to arbitrary
  distributions on finite outcome spaces, arbitrary sets of
  `messages', and (almost) arbitrary loss functions, and provides
  existence and characterization theorems for robust
  probability updating strategies.
  We find that for logarithmic loss, optimality is characterized by
  an elegant condition, which we call \emph{RCAR (reverse coarsening at random)}.
  Under certain conditions, the same condition also characterizes
  optimality for a much larger class of loss functions, and we obtain an objective and general answer to how one should update probabilities in the light of new information.
\end{abstract}


\end{frontmatter}

\section{Introduction}\label{sec:feenstraintro}

There are many situations in which a decision maker receives
incomplete data and still has to reach conclusions about these data.
One type of incomplete data is \emph{coarse} data: instead of the real
outcome of a random event, the decision maker observes a subset of the
possible outcomes, and knows only that the actual outcome is an
element of this subset. An example frequently occurs in
questionnaires, where people may be asked if their date of birth lies
between 1950 and 1960 or between 1960 and 1970 et cetera. Their exact
year of birth is unknown to us, but at least we now know for sure in
which decade they are born. We introduce a simple and concrete motivating  instance of coarse data with the following example.

\begin{example}[Fair die]\label{ex:fairdie}
  Suppose I throw a fair die. I get to see the result of the throw, but
  you do not. Now I tell you that the result lies in the set $\set{1,
    2, 3, 4}$. This is an example of coarse data.
  You know that I used a fair die and that what I tell you is true.
  Now you are asked to give the probability that I rolled a 3. Likely,
  you would say that the probability of each of the remaining possible
  results is $1/4$. This is the knee-jerk reaction of someone who
  studied probability theory, since this is standard
  \emph{conditioning}. But is this always correct?

  Suppose that there is only one alternative set of results I could
  give you after rolling the die, namely the set $\set{3, 4, 5, 6}$. I
  can now follow a \emph{coarsening mechanism}: a procedure that tells
  me which subset to reveal given a particular result of the die roll.
  If the outcome is 1, 2, 5, or 6, there is nothing for me to choose.
  Suppose that if the outcome is 3 or 4, the coarsening mechanism I
  use selects set $\set{1, 2, 3, 4}$ or set $\set{3, 4, 5, 6}$ at
  random, each with probability $1/2$. If I throw the die 6000 times,
  I expect to see the outcome 3 a thousand times. Therefore I expect
  to report the set $\set{1, 2, 3, 4}$ five hundred times after I see
  the outcome 3. It is clear that I expect to report the set $\set{1,
    2, 3, 4}$ 3000 times in total. So for die rolls where I told you
  $\set{1, 2, 3, 4}$, the probability of the
  true outcome being 3 is actually $1/6$ with this coarsening
  mechanism. We see that the prediction of $1/4$ from the first paragraph was not
  correct, in the sense that the probabilities computed there do not
  correspond to the long-run relative frequencies. We conclude that
  the knee-jerk reaction is not always correct.
\end{example}

In Example~\ref{ex:fairdie} we have seen that standard conditioning
does not always give the correct answers. \citet{HeitjanRubin1991}
answer the question under what circumstances standard conditioning of
coarse data is correct. They discovered a necessary and sufficient
condition of the coarsening mechanism, called \emph{coarsening at
  random (CAR)}. A coarsening mechanism satisfies the CAR condition
if, for each subset $y$ of the outcomes, the probability of choosing
to report $y$ is the same no matter which outcome $x \in y$ is the
true outcome.
It depends on the arrangement of possible revealed subsets whether a
coarsening mechanism exists that satisfies CAR. It holds
automatically if the subsets that can be revealed partition the sample
space. As noted by \citet{GrunwaldHalpern2003} however, as soon as
events overlap, there exist distributions on the space for which CAR
does not hold.  In many such situations it even cannot hold; see
\citet{GillGrunwald2008} for a complete characterization of the ---
quite restricted --- set of situations in which CAR can hold. No
coarsening mechanism satisfies the CAR condition for
Example~\ref{ex:fairdie}.

We hasten to add that we neither question the validity of conditioning
nor do we want to replace it by something else.  The real problem lies
not with conditioning, but with conditioning within the wrong sample
space, in which the coarsening mechanism cannot be represented. If we
had a distribution $P$ on the correct, larger space, which allows for
statements like `the probability is $\alpha$ that I choose $\{1,2,3,4\}$ to reveal if the
outcome is $3$', then conditioning would give the correct
results. The problem with coarse data is though that we often do not
have enough information to identify $P$ --- e.g.\ we do not know the
value of $\alpha$ and do not want to assume that it is
$1/2$. Henceforth, we shall refer to conditioning in the overly simple
space as `naive conditioning'. In this paper we propose update rules
for situations in which naive conditioning gives the wrong answer, and
conditioning in the right space is problematic because the underlying
distribution is partially unknown. These are invariably situations in
which two or more of the potentially observed events overlap.

We illustrate this further with a famously counter-intuitive example:
the Monty Hall puzzle, posed by \citet{Selvin1975} and popularized
years later in Ask Marilyn, a weekly column in Parade Magazine by
Marilyn vos Savant \cite{vosSavant1990}.

\begin{example}[Monty Hall]\label{ex:montyhallfeenstra}
  Suppose you are on a game show and you may choose one of three
  doors. Behind one of the doors a car can be found, but the other two
  only hide a goat. Initially the car is equally likely to be behind
  each of the doors. After you have picked one of the doors, the host
  Monty Hall, who knows the location of the prize, will open one of
  the other doors, revealing a goat. Now you are asked if you would
  like to switch from the door you chose to the other unopened door.
  Is it a good idea to switch?

  At this moment we will not answer this question, but we show that
  the problem of choosing whether to switch doors is an example of the
  coarse data problem. The unknown random value we are interested in
  is the location of the car: one of the three doors. When the host
  opens a door different from the one you picked, revealing a goat,
  this is equivalent to reporting a subset. The subset he reports is
  the set of the two doors that are still closed. For example, if he
  opens door 2, this tells us that the true value, the location of the
  car, is in the subset $\set{1, 3}$. Note that if you have by chance
  picked the correct door, there are two possible doors Monty Hall can
  open, so also two subsets he can report. This implies that Monty has
  a choice in reporting a subset. How does Monty's coarsening
  mechanism influence your prediction of the true location of the car?

  The CAR condition can only be satisfied for very particular
  distributions of where the prize is: the probability that the prize
  is hidden behind the initially chosen door must be either 0 or 1,
  otherwise no CAR coarsening mechanism exists \citep[Example
  3.3]{GrunwaldHalpern2003}\footnote{This uses the
    \emph{weak} version
    of CAR in the terminology of \citet{Jaeger2005AoS}, in which
    outcomes with probability 0 are exempt from the equality
    constraint. A \emph{strong} CAR coarsening mechanism does not
    exist regardless of the probabilities with which the prize is
    hidden.}. If the prize is hidden in any other way, for example
  uniformly at random as we assume, then CAR cannot hold, and naive
  conditioning will result in an incorrect conclusion for at least one
  of the two subsets.
\end{example}

Examples~\ref{ex:fairdie} and~\ref{ex:montyhallfeenstra} are just two
instances of a more general problem: the number of outcomes may be
arbitrary; the initial distribution of the true outcome may be any
distribution; and the subsets of outcomes that may be reported to the
decision maker may be any family of sets. Our goal is
to define general procedures that tell us how to update the
probabilities of the outcomes after making a coarse observation, in
such situations where naive conditioning is not adequate.
We are aiming for modular methods that do not enforce a particular interpretation of probability. 
In Example~\ref{ex:fairdie}, we saw ``objective''
probabilities: the original distributions were known, and the updated
probabilities we found could again be interpreted as frequencies over many repetitions of
the same experiment. The original distribution of the outcomes could however
also express a subjective prior belief of how likely each outcome is.
For example, in
Example~\ref{ex:montyhallfeenstra}, the uniform distribution of
the location of the car requires an assumption on the frequentist's
part, while it may be a reasonable choice of prior for a subjective Bayesian
\citep{Gill2011}. In this case, the updated probabilities after a
coarse observation take the role of the Bayesian posterior
distribution. In any case, we will refer to the initial probability
of an outcome, regardless of observations, as the
\emph{marginal}
probability.

Without any assumptions on the quizmaster's strategy (i.e.\ the
coarsening mechanism), the conditional distributions of outcomes given
observations will be unknown, and this uncertainty cannot be fully
expressed by a single probability distribution over the outcomes. One way to deal with this is by means of imprecise probability, i.e.\ by explicitly tracking all possible quizmaster strategies and their effects \cite{deCooman200475}. We however focus on obtaining a single (precise) updated probability. To get such a single answer, we could make some assumption about how the
quizmaster chooses his strategy. Assuming that the coarsening
mechanism satisfies CAR is one such approach, but as we saw in the two
examples, there are scenarios where this assumption cannot hold. We
instead take a \emph{worst-case approach}, treating the coarsening of
the observation and the subsequent probability update as a game
between two players: the \emph{quizmaster} and the \emph{contestant}
(named for their roles in the Monty Hall scenario).
The subset of outcomes communicated by the quizmaster to the
contestant will be called the \emph{message}.

In this fictional game, the quizmaster's goal is the opposite of the
contestant's, namely to make predicting the true outcome as hard as
possible for the contestant. This type of zero-sum game, in which some information is revealed to the contestant was also considered by~\citet{GruHal2011} and~\citet{OzdePeck2008}.  Such situations are rare in practice: The
sender of a message might be motivated by interests other than
informing us (for example, a newspaper may be trying to optimize its
sales figures, %
or a company may want to present its performance in the best light),
but rarely by trying to be as uninformative as possible (though see
Section~\ref{sec:lossinvariance}, where we
consider the case that the players' goals are not diametrically
opposed).
In other situations, the `sender' might not be a rational
being at all, but just some unknown process. Yet this game is a useful
way to look at the problem of updating our probabilities even if we do
not believe that the coarsening mechanism is chosen adversarially: if
we simply do not know how `nature' chooses which message to give us
and do not want to make any assumptions about this, then choosing the
worst-case (or \emph{minimax}) optimal probability update as defined
here guarantees that we incur at most some fixed expected loss, while
any other probability update may lead to a larger expected loss
depending on the unknown coarsening mechanism. While from a Bayesian point of view, such a choice might at first seem overly pessimistic, we note that in all cases we consider, our approach is fully consistent with a Bayesian one --- our results can be interpreted as recommending a certain prior on the quizmaster's assignment of messages to outcomes, which in simple cases (such as Monty Hall) coincides with a prior that Bayesians would be tempted to adopt as well.

We will employ a loss function to measure how well the quizmaster and
the contestant are doing at this game. Our results apply to a wide
variety of loss functions. For an analysis of the Monty Hall game, 0-1
loss would be appropriate, as the contestant must choose a single
door; this is the approach used by \citet{Gill2011} and \citet{Gnedin2011_gametheory}. Other loss
functions, such as logarithmic loss
and Brier loss (see e.g.\ \citet{GrunwaldDawid2004}), also allow the
contestant to formulate their prediction of where the prize is hidden
as an arbitrary probability distribution over the outcomes. 

We model probability updating as a game as follows. An outcome $x$ is drawn with known marginal probability $\marg_x$ and shown to the quizmaster, who picks a consistent message $y \ni x$ using his coarsening mechanism $P(y \mid x)$. Seeing only $y$, the contestant makes a prediction in the form of a probability mass function $Q(\cdot \mid y)$ on outcomes. Then $x$ is revealed and the prediction quality measured using the loss function $L$. The quizmaster/contestant aim to maximize/minimize the expected loss
\begin{equation}\label{eq:intro.objective}
\sum_{x} p_x \sum_{y \ni x} P(y \mid x) L\del[\big]{x, Q(\cdot \mid y)}
.
\end{equation}

For the
Monty Hall game with logarithmic or Brier, i.e.\ squared loss, the \optimal{}
answer for the contestant is to put probability $1/3$ on his initially
chosen door and $2/3$ on the other door. (These probabilities agree
with the literature on the Monty Hall game.) Surprisingly, we will see
(in Example~\ref{ex:23} on page~\pageref{ex:23}) that for very similar
games, logarithmic and Brier loss may lead to two different answers!

We will find that for finite outcome spaces, both
players in our game have \optimal{} strategies for many loss
functions: the quizmaster has a strategy that makes the contestant's
prediction task as hard as possible, and the contestant has a strategy
that is guaranteed to give good predictions no matter how the
quizmaster coarsens. We give characterizations that allow us to recognize
such strategies, for different conditions on the loss functions.

\begin{example_cont}[\ref{ex:fairdie}]
  For logarithmic loss, the \optimal{} prediction of the die roll conditional on the revealed subset is found with the help of
  Theorem~\ref{thm:charP_local}. The \optimal{} prediction given that
  you observe the set $\set{1, 2, 3, 4}$ is: predict outcomes 1 and 2
  each with probability $1/3$, and predict 3 and 4 each with
  probability $1/6$. Symmetrically, given that you observe the set $\set{3, 4, 5,
    6}$, the \optimal{} prediction is: 3 and 4 with probability $1/6$,
  and 5 and 6 with probability $1/3$.

  These probabilities correspond with the uniform coarsening mechanism given
  earlier. However, it is a good prediction even if you do not know
  what coarsening mechanism I am using. An intuitive argument for this
  is the following: If I wanted, I could use a very extreme coarsening
  mechanism, always choosing to reveal the set $\set{1, 2, 3, 4}$ when
  the die comes up 3 or 4. But this is balanced by the possibility
  that I might be using the opposite coarsening mechanism, which
  always reveals $\set{3, 4, 5, 6}$ if the result is 3 or 4. The
  \optimal{} prediction given above hedges against both possibilities.
\end{example_cont}

\commentout{
\subsection{Caveats on the use of the word `conditioning'}

Since this paper is concerned with making \anoptimal{} prediction
conditional on a set of outcomes, we want to highlight the use of the
word \emph{conditioning}. Above, we used the word \emph{standard}
conditioning for using the conditional information in the standard
way: with random variables $\rv{X}$ the true outcome and $\rv{Y}$ the
coarse observation (a set of outcomes), computing $P(\rv{X} = x \mid
\rv{Y} = y)$ by $P(\rv{X} = x \mid \rv{X} \in y) = P(\rv{X} = x) /
P(\rv{X} \in y)$.

In the two examples, we saw that this does not always give the correct
probabilities given a coarse observation. 
For instance in Example~\ref{ex:fairdie},
$P(\rv{X} \mid \rv{X} \in y)$ is the uniform distribution on $y$, while
$P(\rv{X} \mid \rv{Y} = y)$ cannot simultaneously be uniform for both
$y = \set{1,2,3,4}$ and $y = \set{3,4,5,6}$.
In such situations, we call
the former computation \emph{naive conditioning}. The formula for
conditional probabilities does give the correct result if instead of
on the outcome space, we work in a larger space: the space of all
combinations of an outcome and a set. The problem we face in this
paper is that we do not know the probabilities of all these
combinations, as they depend on the unknown coarsening mechanism.

}

\subsection{Overview of contents}\label{sec:mh_contents}

In Section~\ref{sec:defs}, we will give a precise definition of the
`conditioning game' we described. In
Section~\ref{sec:optimal}, we find general conditions on the
loss function under which \optimal{} strategies for the quizmaster and contestant exist, and we characterize such strategies. (See Figure~\ref{fig:charP} for a
visual illustration of the concepts used in this section.) If
stronger conditions hold, \optimal{} strategies for both players may
be easier to recognize. This is explored for two classes of loss
functions in Section~\ref{sec:simplerresults}; in particular, we find that for local proper loss functions (among which logarithmic loss), \optimal{} strategies for the quizmaster
are characterized by a simple condition on their probabilities that we call the
\emph{RCAR (reverse CAR)} condition. RCAR bears a striking similarity to CAR, but with the roles of outcomes and messages interchanged.
Also, by Lemma~\ref{lem:affinetransform}, if a betting
game is played repeatedly and the
contestant is allowed to distribute investments over different
outcomes and to reinvest all capital gained so far in each round, then
the same strategy is optimal, \emph{regardless of the pay-offs!}
An overview of the theorems and the conditions under which they apply
is given in Table~\ref{tab:loss_function_summary} on
page~\pageref{tab:loss_function_summary}.

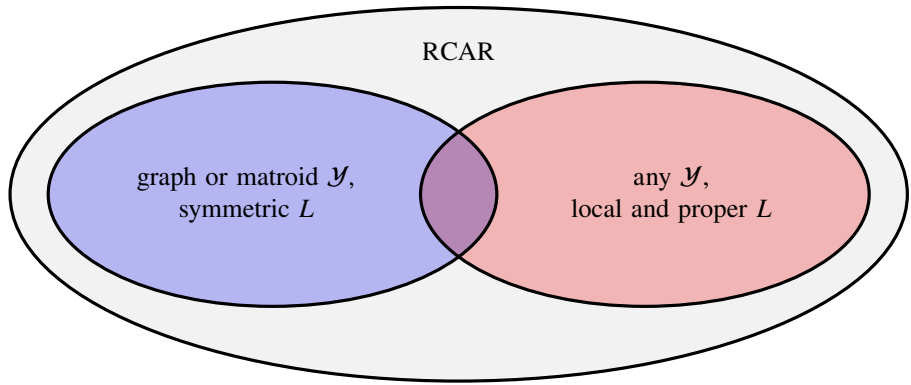
\begin{figure}
\centering
\input{cartography.tex}
\caption{Classes of games for which the RCAR condition characterizes optimality. $\Y$ is the set of messages and $L$ is the loss function. \label{fig:cartography}}
\end{figure}

Then in Section~\ref{sec:RCARgeneral} we look at the influence of the set of available messages, imposing only the minimal symmetry requirement on the loss function.  We prove that for graph and matroid games (and only these) the optimality condition is again RCAR. As RCAR is independent of the loss function, for such games probability updating can hence be meaningfully defined and performed completely agnostic of the task at hand.
Many examples are included to illustrate (the limits of) the
theoretical results. Section~\ref{sec:mh_conclusion} gives some
concluding remarks.
An overview of the results in this work is presented in Figure~\ref{fig:cartography}.  
All proofs are given in \ref{app:proofs}.

\paragraph{Highlight}
A central feature of probability distributions is that they summarize what a decision maker would do under a variety of circumstances (loss functions): for each particular loss function the decision maker minimizes expected loss (maximizes utility) using the same distribution no matter what specific loss function is used. Since we generalize conditioning by using a minimax approach, one might expect that for different loss functions one ends up with different updated probabilities. Still, we show that for a rich selection of scenarios optimality is characterized by the RCAR condition, which is independent of the loss function. As a result, our updated probabilities are \emph{application-independent}, and we may hence think of them --- if we are willing to take a cautious (minimax) approach --- as expressing what an experimenter \emph{should} believe after having received the data.

We isolate two distinct classes of scenarios where such application independence obtains. First, games with graph and matroid message sets (extending Monty Hall) and symmetric loss functions. Second, graphs with arbitrary message sets and proper local loss functions, including the symmetric logarithmic loss as well as its asymmetric generalizations appropriate for Kelly gambling with arbitrary payoffs.
In these scenarios our application-independent update rule has an objective appeal, and we feel that its importance may transcend that of being ``merely'' minimax optimal.

\bigskip
\noindent
This work is an extension of \citet{FeenstraThesis} to loss functions
other than logarithmic loss, and to the case where the \optimal{}
strategy for the quizmaster assigns probability 0 to some combinations
of outcomes $x$ and messages $y$ with $x \in y$. It can also be seen
as a concrete application of the ideas in \citet{GrunwaldDawid2004}
about minimax optimal decision making and its relation to entropy.
A more extensive discussion of worst-case optimal probability updating
can be found in \citet{vanOmmen_phd}; in particular, there the
question of efficient algorithms for determining \optimal{} strategies
is also considered.

\section{Definitions and problem formulation}\label{sec:defs}

A \emph{(probability updating) game} $\G$ is defined as a quadruple
$(\X, \Y, \marg{}, L)$, where
$\X$ is a finite set,
$\Y$ is a family of distinct subsets of $\X$
with $\bigcup_{y \in \Y} y = \X$,
$\marg{}$ is a strictly positive probability mass function on $\X$, and
$L$ is a function $L: \X \times \Delta_\X \to [0, \infty]$, where
$\Delta_\X$ is the set of all probability mass functions on $\X$.
We call $\X$ the \emph{outcome space}, $\Y$ the \emph{message
  structure}, $\marg{}$ the \emph{marginal
  distribution}, and $L$ the
\emph{loss function}. 
It is clear that outcomes with zero marginal probability $\marg{}$ 
do not contribute to the objective \eqref{intro.objective}, so we may exclude them without loss of generality. Let us illustrate these definitions by applying them to our example.

\begin{example_cont}[\ref{ex:montyhallfeenstra}]
  We assume the car is hidden uniformly at random behind one of the
  three doors. With this assumption, we can abstract away the initial
  choice of a door by the contestant: by symmetry, we can assume
  without loss of generality that he always picks door 2. Then the
  probability updating game starts with the quizmaster opening door 1
  or 3, thereby giving the message ``the car is behind door 2 or 3''
  or ``the car is behind door 1 or 2'', respectively. This can be
  expressed as follows in our formalization:
  \begin{itemize}
  \item outcome space $\X = \set{x_1, x_2, x_3}$;
  \item message space $\Y = \set{y_1, y_2}$ with $y_1 = \set{x_1,x_2}$
    and $y_2 = \set{x_2,x_3}$;
  \item marginal distribution $\marg{}$ uniform on $\X$.
  \end{itemize}
  If a loss function $L$ is also given, this fully specifies a game.
  One example is randomized 0-1 loss, which is given by $L(x, Q) = 1 -
  Q(x)$. Here $x$ is the true outcome, and $Q$ is the contestant's
  prediction of the true outcome in the form of a probability
  distribution. Thus the prediction $Q$ is awarded a smaller loss if it
  assigned a larger probability $Q(x)$ to the outcome $x$ that actually
  obtained. We will see other examples of loss functions in
  Section~\ref{sec:stdloss}.
\end{example_cont}

A function from some finite set $S$ to the reals $\R = (-\infty, \infty)$ corresponds to an $\lvert
S \rvert$-dimensional vector when we fix an order on the elements of
$S$. We write $\R^S$ for the set of such functions/vectors.
Even if no order on $S$ is specified, this allows us to apply concepts
from linear algebra to $\R^S$ without ambiguity. For example, we may
say that some set is an affine subspace of $\R^S$.
(This identification and the resulting notation are also used by
\citet{SchrijverA}.)

Using this correspondence, we identify the elements of $\Delta_\X$
with the $\lvert \X \rvert$-dimensional vectors in the unit simplex,
though we use ordinary function notation $P(x)$ for its elements.
The probability mass function $\marg{}$ that is part of a game's
definition is also a vector in $\Delta_\X$. %
Vector notation $\marg{}_x$ will be used to refer to its elements to
set $p$ apart from $P$, which will denote distributions chosen by
the quizmaster rather than fixed by the game.

For any message $y \subseteq \X$, we define $\Delta_y = \setc{P \in
  \Delta_\X}{P(x) = 0 \text{ for } x \not\in y}$. Note that these are
vectors of the same length as those in $\Delta_\X$, though contained
within a lower-dimensional affine subspace. %

A loss function $L$ is called
\emph{proper}\label{def:proper} if
$P \in \argmin_{Q \in \Delta_\X} \Exp_{\rv{X} \sim P} L(\rv{X}, Q)$ for all $P \in
\Delta_\X$, and \emph{strictly proper} if this minimizer is unique
(this is standard terminology; see for instance
\citet{GneitingRaftery2007}).  Thus if a predicting agent believes the
true distribution of an outcome to be given by some $P$, such a loss
function will encourage him to report $Q = P$ as his prediction.

\subsection{Strategies}\label{sec:strategies}

Strategies for the players are specified by conditional distributions:
a strategy $P$ for the quizmaster consists of distributions on $\Y$,
one for each possible $x \in \X$, and a strategy $Q$ for the
contestant consists of distributions on $\X$, one for each possible $y
\in \Y$. These strategies define how the two players act in any
situation: the quizmaster's strategy defines how he chooses a message
containing the true outcome (the coarsening mechanism), and the
contestant's strategy defines his prediction for each message he might
receive.

We write $P(\cdot \mid x)$ for the distribution on $\Y$ the
quizmaster plays when the true outcome is $x \in \X$. Because
$\marg{}_x > 0$, this conditional distribution can be recovered from
the joint $P(x, y) \isdef P(y \mid x)\marg{}_x$; we will use this joint
distribution to specify a strategy for the quizmaster. If $P(y) \isdef
\sum_{x \in y} P(x, y) > 0$, we may also write $P(\cdot \mid y)$ for
the vector in $\Delta_y$ given by $P(x \mid y) \isdef P(x,y)/P(y)$. No
such rewrites can be made for $Q$, as no marginal $Q(y)$ is specified
by the game or by the strategy $Q$. To shorten notation and to
emphasize that $Q$ is not a joint distribution, we write $Q_{\mid y}$
rather than $Q(\cdot \mid y)$ for the distribution that the contestant
plays in response to message $y$.

We restrict the quizmaster to conditional distributions $P$ for which
$P(y \mid x) = 0$ if $x \not\in y$; that is, he may not `lie' to the
contestant. %
We make no similar requirement on the contestant's choice of $Q$,
though for proper loss functions, and in fact all other loss functions
we will consider in our examples, the contestant can gain nothing from
using a strategy $Q$ for which $Q_{\mid y}(x) > 0$ where $x \not\in
y$.

\begin{wrapfigure}[6]{r}{.35\textwidth}
  \vspace{-.5\baselineskip}
  \begin{equation}\label{eq:montyhalltable}
  \begin{array}{r|ccc|}
    P & x_1 & x_2 & x_3 \\
    \hline
    y_1 & 1/3 & 1/6 & - \\
    y_2 & - & 1/6 & 1/3 \\
    \hline
    \marg{}_x & 1/3 & 1/3 & 1/3 \\
    \hline
  \end{array}
  \end{equation}
  \vspace{-.5\baselineskip}
\end{wrapfigure}
\noindent \par{} %
\vspace{-.5\baselineskip}

\begin{example_cont}[\ref{ex:montyhallfeenstra}]
  The table to the right specifies all aspects of a
  game except for its loss function: its outcome space (here, for the
  Monty Hall game, $\X =
  \set{x_1, x_2, x_3}$), message space ($\Y = \set{y_1, y_2}$ with
  $y_1 = \set{x_1,x_2}$ and $y_2 = \set{x_2,x_3}$) and marginal
  distribution ($\marg{}$ uniform on $\X$).
  In this table we have filled in a strategy $P$ for the quizmaster in
  the form of a joint distribution on pairs of $x$ and $y$. The cells
  in the table where $x \not\in y$ are marked with a dash to indicate
  that $P$ may not assign positive probability there. The
  probabilities in each column sum to the marginal probabilities at
  the bottom, so this joint distribution $P$ has the correct marginal
  distribution on the outcomes. For this particular strategy, if the
  true outcome is $x_2$, the quizmaster will give message $y_1$ or
  $y_2$ to the contestant with equal probability.

\end{example_cont}

More formally, write $\xiny$ as an abbreviation for the set of
pairs $\setc{(x, y)}{y \in \Y, x \in y}$. In the case of the Monty Hall
game, there are four such pairs: $\xiny = \set{(x_1, y_1), (x_2,
  y_1), (x_2, y_2), (x_3, y_2)}$.
The notation $\R_{\geq 0}^{\xiny}$ represents %
the set of all
functions from $\xiny$ to $\R_{\geq 0}$. If $P$ is an element of this
set and $(x, y) \in \xiny$, the value of $P$ at $(x, y)$ is denoted by
$P(x, y)$. For $(x, y)$ with $x \not\in y$, the notation $P(x, y)$
does not correspond to a value of the function, but is taken to be $0$.

We again identify the elements of $\R_{\geq 0}^{\xiny}$ with vectors.
Thus the mass function $P$ shown in \eqref{montyhalltable} is
identified with a four-element vector
$(\sfrac{1}{3},\allowbreak\sfrac{1}{6},\allowbreak\sfrac{1}{6},\allowbreak\sfrac{1}{3})$.
(We
could have chosen a different ordering instead.)

We define the set $\P$ of strategies for the quizmaster as $\setc{P
  \in \R_{\geq 0}^{\xiny}}{\sum_{y \ni x} P(x, y) = \marg{}_x
  \text{ for all } x}$; this is a convex set. The set of strategies
for the contestant is $\Q \isdef \Delta_\X^\Y = \setc{(Q_{\mid y})_{y \in
    \Y}}{Q_{\mid y} \in \Delta_\X \text{ for each }y \in \Y}$.

For given strategies $P$ and $Q$, the expected loss the contestant
incurs \eqref{intro.objective} is
\begin{equation}\label{eq:expectedloss}
\sum_{x \in \X} \marg{}_x \;\smash{\sum_{\mathclap{y \in \Y : x \in y}}}\; P(y \mid x) L(x, Q_{\mid y})
= \Exp_{\rv{X} \sim \marg{}} \Exp_{\rv{Y} \sim P(\cdot \mid \rv{X})} L(\rv{X}, Q_{\mid \rv{Y}})
= \Exp_{(\rv{X}, \rv{Y}) \sim P} L(\rv{X}, Q_{\mid \rv{Y}}).
\end{equation}
We allow $L$ to take the value $\infty$; if this value occurs with
positive probability, then the contestant's expected loss is infinite.
However, for terms where the probability is zero, we define $0 \cdot
\infty = 0$, as is consistent with measure-theoretic probability.

We model the probability updating problem as a zero-sum game between two players with objective \eqref{expectedloss}: the
quizmaster chooses $P \in \P$ to maximize \eqref{expectedloss}, while
simultaneously (that is, without knowing $P$) the contestant chooses
$Q \in \Q$ to minimize that quantity. The game $(\X, \Y, \marg{}, L)$
is common knowledge for the two players.

If the contestant knew the quizmaster's strategy, he would pick a
strategy $Q$ that for each $y$ minimizes the expected loss of
predicting $x$ given $y$.
When the contestant receives a message and knows the distribution $P
\in \Delta_\X$ over the outcomes given that message, this expected
loss is written as
\begin{equation}\label{eq:defHL}
H_L(P) \isdef \inf_{Q \in \Delta_\X} \sum_x P(x) L(x, Q)
  = \inf_{Q \in \Delta_\X} \Exp_{\rv{X} \sim P} L(\rv{X}, Q).
\end{equation}
This is the \emph{generalized entropy} of $P$ for loss function $L$
\citep{GrunwaldDawid2004}. (Note that in the preceding display, $P$
and $Q$ are not strategies but simply distributions over $\X$.) If the
contestant picks his strategy $Q$ this way, \eqref{expectedloss}
becomes the \emph{expected generalized entropy} of the quizmaster's
strategy $P \in \P$:
\begin{equation}\label{eq:quizmasterobjective}
\sum_{y \in \Y} P(y) H_L(P(\cdot \mid y)),
\end{equation}
where we again define terms with $P(y) = 0$ as 0. We say a strategy
$P$ is \emph{\optimal{} for the quizmaster} if it maximizes this
expected generalized entropy over all $P \in \P$. We call the version
of the game where the quizmaster has to play first the \emph{maximin}
game, where the order of the words `max' and `min' reflects the order
in which they appear in the expression for the value of this game as
well as the order in which the maximizing and minimizing players take
their turns.

Similarly, if the contestant were to play first (the \emph{minimax}
game), his goal might be to find a strategy $Q$ that minimizes his
worst-case expected loss
\begin{equation}\label{eq:contestantobjective}
\max_{P \in \P} \sum_{x \in y} P(x, y) L(x, Q_{\mid y})
= \max_{P \in \P} \Exp_{(\rv{X}, \rv{Y}) \sim P} L(\rv{X}, Q_{\mid \rv{Y}}).
\end{equation}
(In this case, the maximum is always achieved so we can write $\max$
rather than $\sup$: for each $x$, the quizmaster can choose $P$ that
puts all mass on a $y \ni x$ with the maximum loss.) We call a
strategy \emph{\optimal{} for the contestant} if it achieves the
minimum of \eqref{contestantobjective}.

It is an elementary result from game theory that if \optimal{} strategies
$P^*$ and $Q^*$ exist for the two players, their expected losses are
related by
\begin{equation}\label{eq:weakduality}
\sum_{y \in \Y} P^*(y) H_L(P^*(\cdot \mid y))
\leq \max_{P \in \P} \sum_{x \in y} P(x, y) L(x, Q^*_{\mid y})
\end{equation}
\citep[Lemma 36.1: ``maximin $\leq$ minimax'']{Rockafellar1970}. The
inequality expresses that in a sequential game where one of the
players knows the other's strategy before choosing his own, the player
to move second may have an advantage.

In the next section, we will see that in many probability updating
games, \optimal{} strategies for both players exist (but may not be
unique), and the maximum expected generalized entropy \emph{equals}
the minimum worst-case expected loss:
\begin{equation}\label{eq:strongduality}
\sum_{y \in \Y} P^*(y) H_L(P^*(\cdot \mid y))
= \max_{P \in \P} \sum_{x \in y} P(x, y) L(x, Q^*_{\mid y}).
\end{equation}
When this is the case, we say that the minimax theorem
holds \citep{Neumann1928, Ferguson1967}. We remark here that our
setting, while a zero-sum game, differs from the usual setting of
zero-sum games in some respects: We consider possibly infinite loss
and (in general) infinite sets of strategies available to the players,
but do not allow the players to randomize over these strategies.
Randomizing over $\P$ would not give the quizmaster an advantage, as
$\P$ is convex and he could just play the corresponding convex
combination directly; because \eqref{expectedloss} is linear in $P$, this results in the same expected loss.
(Another way to view this is that, essentially,
the quizmaster \emph{is} randomizing, over a finite set of
strategies.) For the contestant, $\Q$ is also convex, but in general
(depending on $L$), playing a convex combination of strategies does
not correspond to randomizing over those strategies. The two do
correspond in the case of randomized 0-1 loss, where $L$ is linear. If
$L$ is convex, then playing the convex combination is at least as good
for him as randomizing (and if $L$ is strictly convex, better), so
allowing randomization would again not give an advantage.

When \eqref{strongduality} holds, any pair of \optimal{} strategies
$(P^*, Q^*)$ forms a \emph{(pure strategy) Nash equilibrium}, a
concept introduced by \citet{NashEquilibrium}: neither player can
benefit from deviating from their \optimal{} strategy if the other
player leaves his strategy unchanged. This means that the definitions
of \optimal{}ity given above are also meaningful in the game we are
actually interested in, where the players move simultaneously in the
sense that neither knows the other's strategy when choosing his own.

\subsection{Three standard loss functions}\label{sec:stdloss}

Three commonly used loss functions are logarithmic loss, Brier loss,
and randomized 0-1 loss. These are defined as follows
\citep{GrunwaldDawid2004}:

\begin{figure} %
\centering
\subfloat[Logarithmic loss (natural base)\label{fig:logloss}]{
\includegraphics[width=2.25in]{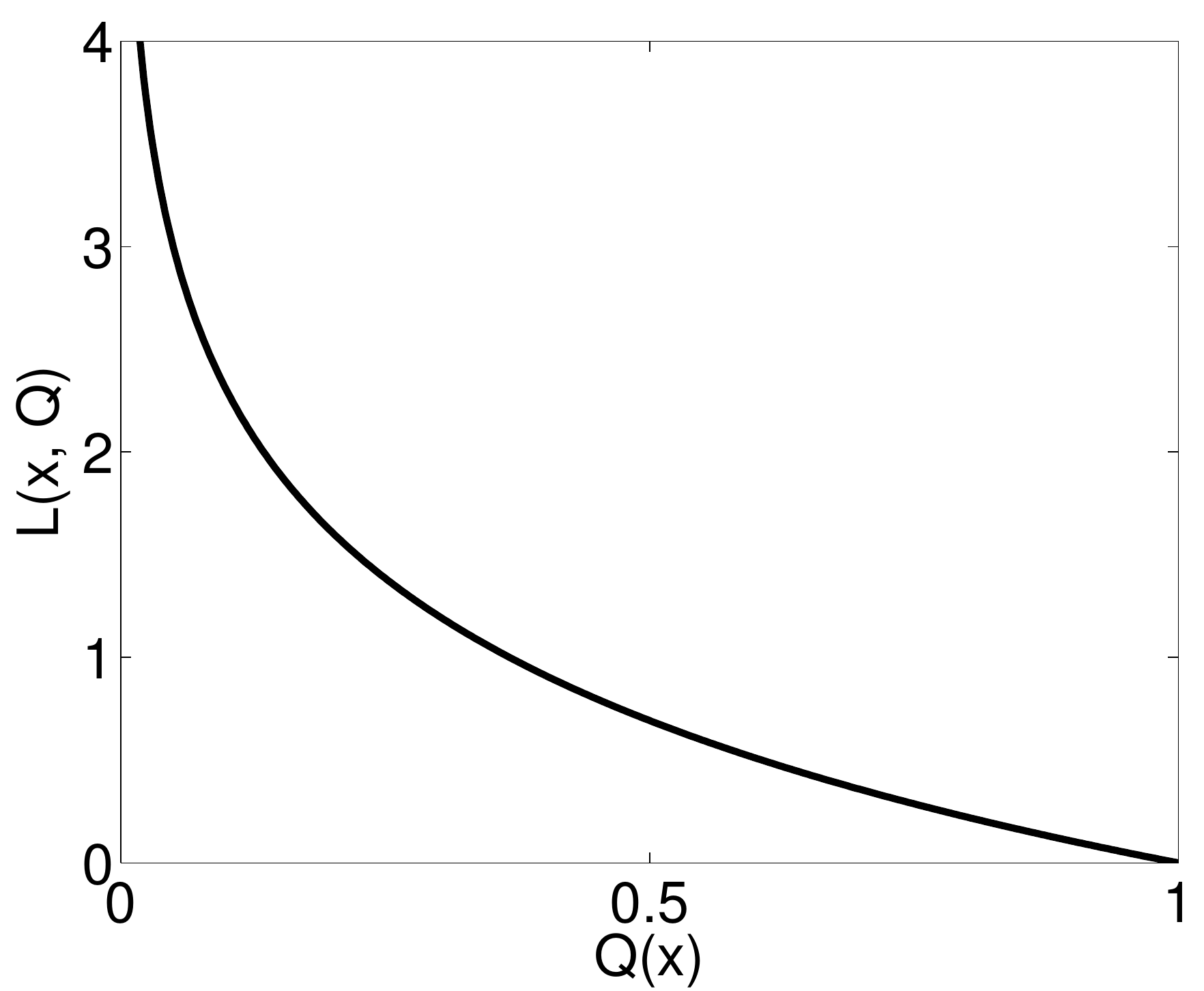} %
\quad
\includegraphics[width=2.25in]{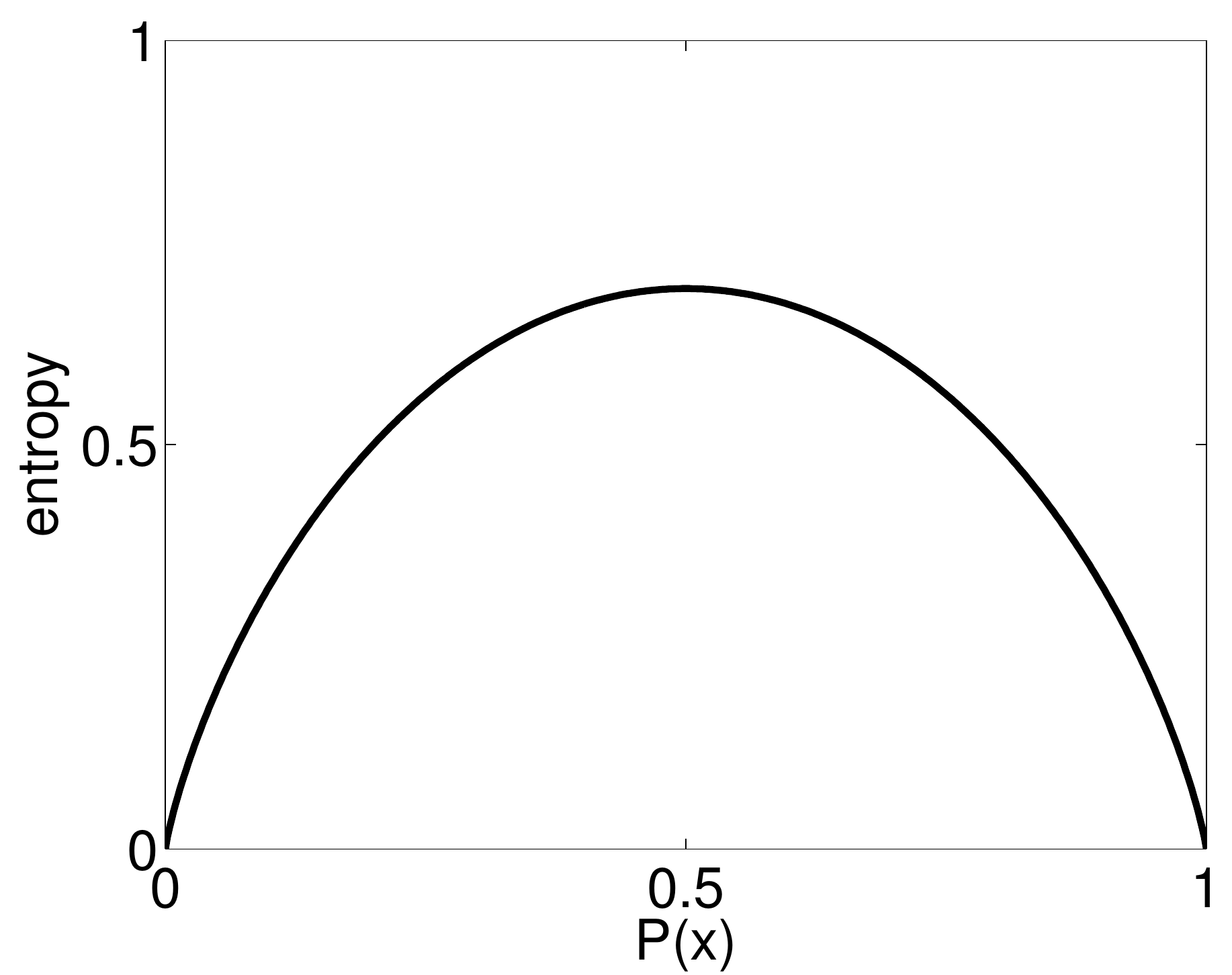}
}
\\
\subfloat[Brier loss\label{fig:brierloss}]{
\includegraphics[width=2.25in]{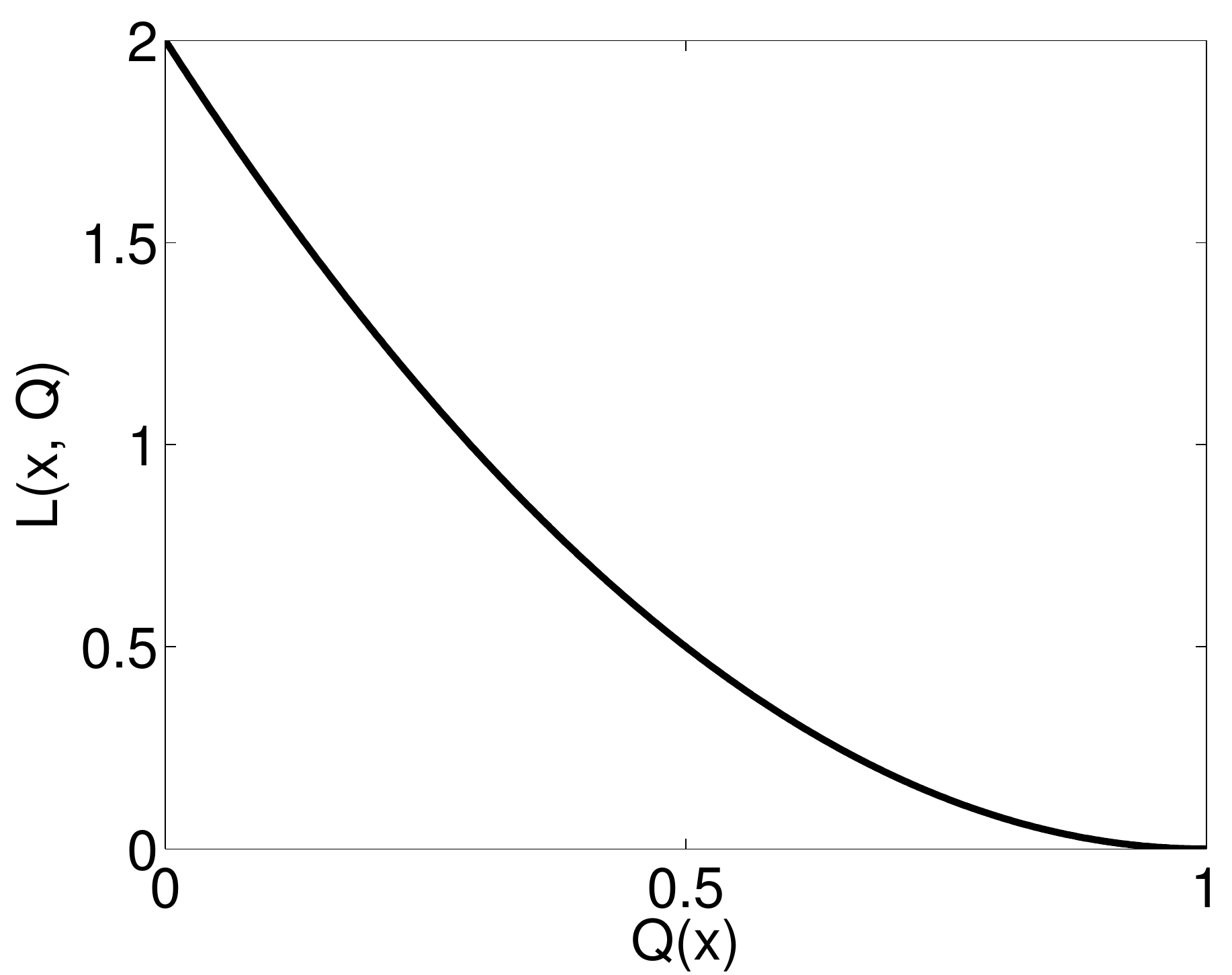} %
\quad
\includegraphics[width=2.25in]{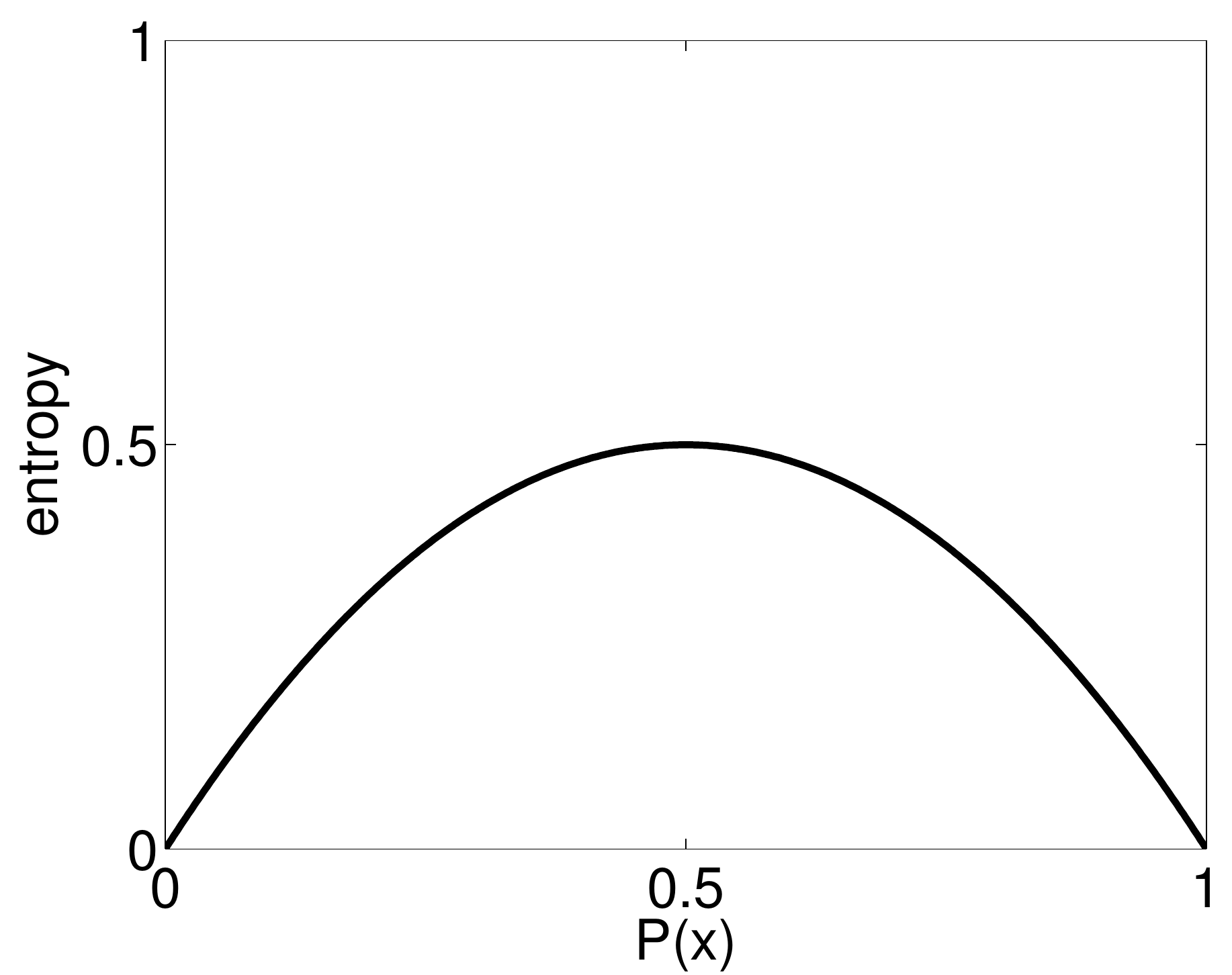}
}
\\
\subfloat[Randomized 0-1 loss\label{fig:01loss}]{
\includegraphics[width=2.25in]{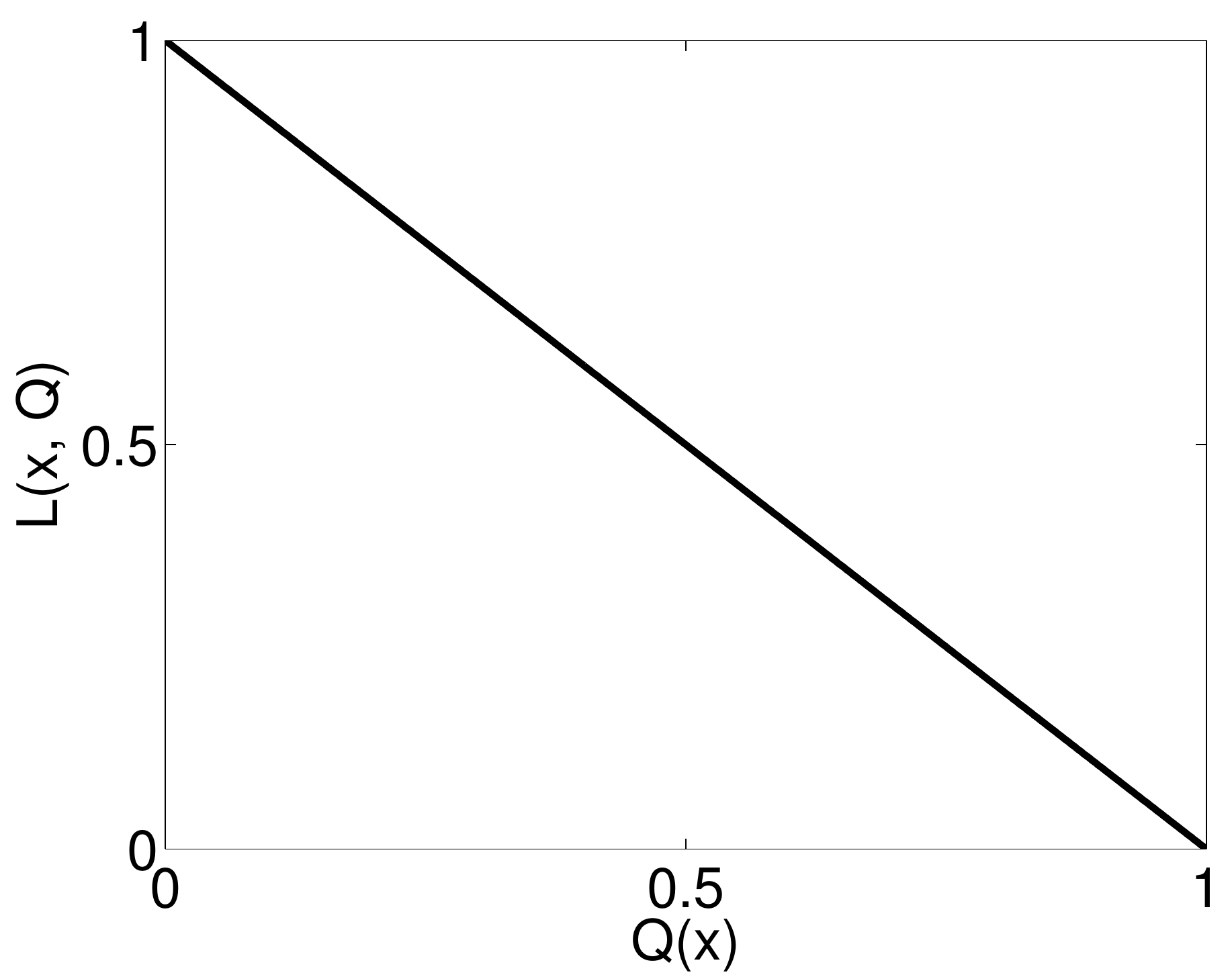} %
\quad
\includegraphics[width=2.25in]{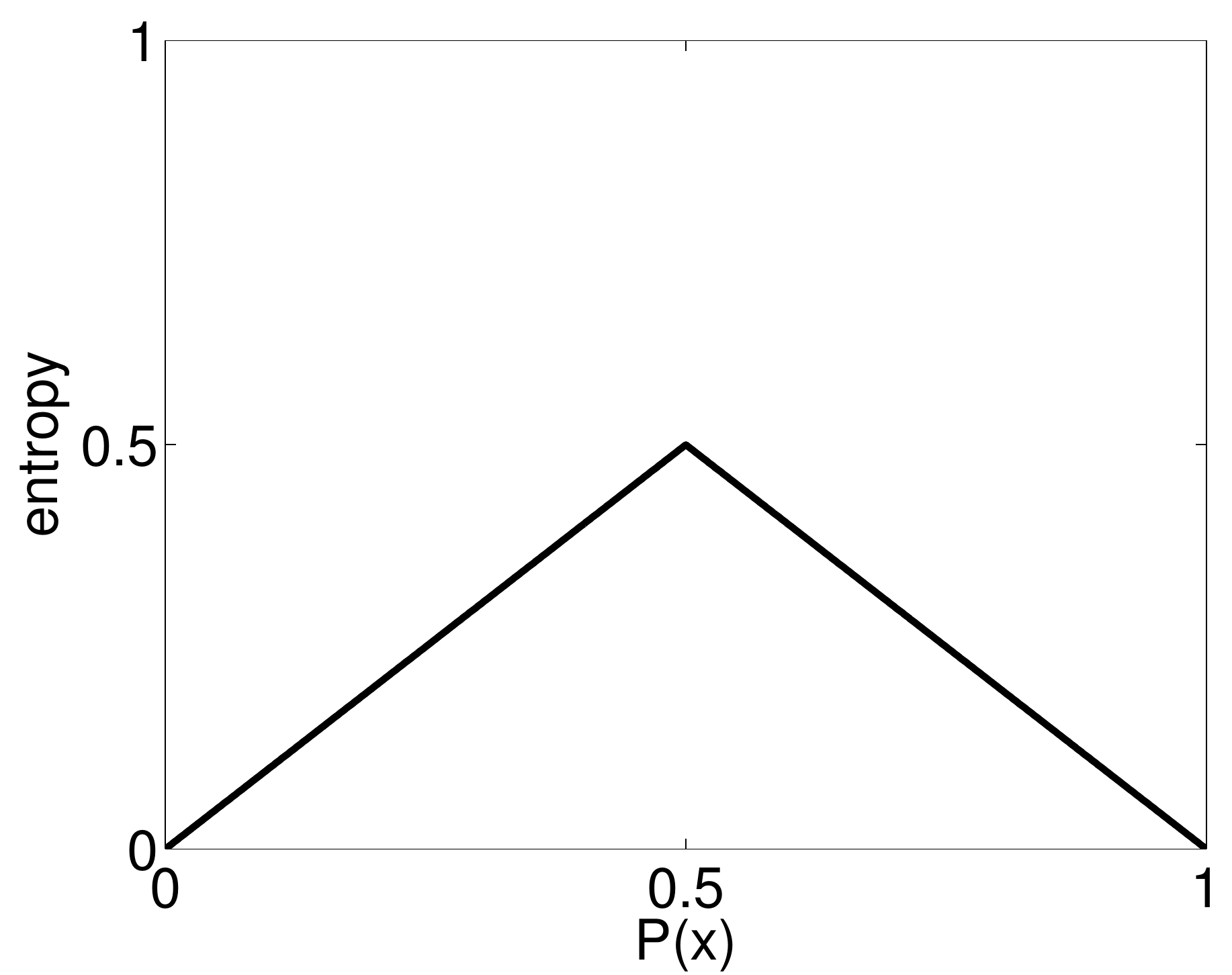}
}
\caption{Three standard loss functions on a binary prediction. The left figures show the loss $L(x,Q)$ when probability $Q(x)$ is assigned to true outcome $x \in \set{0,1}$. The right figures show the entropy $H_L(P)$.}
\end{figure}
\emph{Logarithmic loss} is a strictly proper loss function, given by
\[
L(x, Q) = -\log Q(x).
\]
Its entropy is the Shannon entropy $H_L(P) = \sum_x - P(x) \log P(x)$. The functions $L$
and $H_L$ are displayed in Figure~\ref{fig:logloss} for the case of a
binary prediction (i.e.\ a prediction between two possible outcomes).
The (three-dimensional) graph of $H_L$ for the case of three outcomes
will appear in Figure~\ref{fig:charP} on page~\pageref{fig:charP}.

\emph{Brier loss} is another strictly proper loss function,
corresponding to squared Euclidean distance:
\[
L(x, Q)
= \sum_{\mathclap{x' \in \X}} \left(\indicator_{x' = x} - Q(x')\right)^2
= (1 - Q(x))^2 + \;\sum_{\mathclap{x' \in \X, x' \neq x}}\;Q(x')^2.
\]
Its entropy function is $H_L(P) = 1 - \sum_{x \in \X}P(x)^2$; $L$ and
$H_L$ are displayed in Figure~\ref{fig:brierloss} for a binary prediction. Note that for $3$ outcomes and beyond, the Brier loss on outcome $x$ is not simply a function of $Q(x)$, it depends on the entire distribution $Q$.

The third loss function we will often refer to is \emph{randomized 0-1
  loss}, given by
\[
L(x, Q) = 1 - Q(x).
\]
It is improper: an optimal response $Q$ to some distribution $P$ puts
all mass on outcome(s) with maximum $P(x)$. Its entropy function is
$H_L(P) = 1 - \max_{x \in \X} P(x)$ (see Figure~\ref{fig:01loss}). It
is related to \emph{hard 0-1 loss}, which requires the contestant to
pick a single outcome $x'$ and gives loss 0 if $x' = x$ and 1
otherwise. Randomized 0-1 loss essentially allows the contestant to
randomize his prediction: $L(x, Q)$ equals the expected value of hard
0-1 loss when $x'$ is distributed according to $Q$. An important
difference between games with hard and randomized 0-1 loss
will be shown later in Example~\ref{ex:hard01}.

\subsection{On duplicate messages and outcomes}\label{sec:defnotes}

Our definition of a game rules out duplicate messages in $\Y$, which
would not meaningfully change the options of either player as the two
messages represent the same move for the quizmaster; this will be made
precise in Lemma~\ref{lem:messagemerging}. The definition does allow
duplicate outcomes: pairs of outcomes $x_1, x_2 \in \X$ such that $x_1
\in y$ if and only if $x_2 \in y$ for all $y \in \Y$. We will see
later (in Example~\ref{ex:23}) that games with such outcomes cannot
generally be solved in terms of games without, and thus we must
analyse them in their own right.

\section{\Optimal{} strategies}\label{sec:optimal}

In this section, we present characterization theorems that
allow \optimal{} strategies for the quizmaster and contestant to be recognized for a
large class of loss functions.
In order to be applicable to a
wide range of loss functions, this section is rather technical,
and the characterizations of \optimal{} strategies we find here are
not always easy to use (though the abstract results in these sections
are illustrated by concrete examples in Sections~\ref{sec:Pappl} and
\ref{sec:Qappl}). We will find simpler characterizations for smaller
classes of loss functions in Section~\ref{sec:simplerresults}. An
overview of these results is given in
Table~\ref{tab:loss_function_summary}.
\begin{table}[t]
\caption{Results on \optimal{} strategies for different loss functions}\label{tab:loss_function_summary}
\begin{center}
\begin{tabular}{p{\widthof{$H_L$ finite and continuous;}}@{\quad}p{\widthof{characterization of $P^*$ simplified
  further}}@{\quad}p{\widthof{hard 0-1 loss}}}%
  \hline
  Conditions on $L$ & Results & Example \\
  \hline $H_L$ finite and continuous & $P^*$ exists and is
  characterized by Theorem~\ref{thm:existP_charP}
  & hard 0-1 loss \\[0.9cm]
  $H_L$ finite and continuous; all minimal supporting \newline hyperplanes
  realizable & $Q^*$ exists and a Nash equilibrium exists by
  Theorem~\ref{thm:existQ}; $Q^*$ characterized by
  Theorem~\ref{thm:charQ}
  & randomized 0-1 loss \\[0.9cm]%
  $L$ proper and continuous; $H_L$ finite and continuous & all the
  above simplified by Theorem~\ref{thm:proper}
  & Brier loss \\[0.9cm]
  $L$ local and proper; \newline $H_L$ finite and continuous
  & characterization of $P^*$ simplified
  further by Theorem~\ref{thm:charP_local} (RCAR condition)
  & logarithmic \newline loss \\
  \hline
\end{tabular}
\end{center}
\end{table}

We will need the following properties of $H_L$ throughout our theory:
\begin{lemma}\label{lem:HL}
  For all loss functions $L$, if $H_L$ is finite,
  then it is also concave and lower semi-continuous. If $L$ is finite
  everywhere, then $H_L$ is finite, concave, and continuous.
\end{lemma}
(When we talk about (semi-)continuity, this is always with respect to the extended real line topology of losses, as in \citet[Section 7]{Rockafellar1970}.)

\subsection{\Optimal{} strategies for the quizmaster}\label{sec:optimalquizmaster}

We start by studying the probability updating game from the perspective of the quizmaster.
Using just the concavity of the quizmaster's
objective \eqref{quizmasterobjective} (which is a linear combination
of concave generalized entropies), we can prove the following
intuitive result.
\begin{lemma}[Message subsumption]\label{lem:messagemerging}
  Suppose that for $P \in \P$ there are two messages $y_1, y_2 \in \Y$
  such that any outcome $x \in y_2$ with $P(x, y_2) > 0$ is also in
  $y_1$. Then if $H_L$ is concave, the quizmaster can do at least as
  well without using $y_2$. More precisely, $P'$ given by
  \begin{equation*}
    P'(x, y) = \begin{cases}
      P(x, y_1) + P(x, y_2) & \text{for } y = y_1;\\
      0 & \text{for } y = y_2;\\
      P(x, y) & \text{otherwise.}
    \end{cases}
  \end{equation*}
  is also in $\P$ and its expected generalized entropy is at least as
  large as that of $P$. In particular, if $P$ is \optimal{}, then so is
  $P'$.
\end{lemma}
In particular, if $y_1 \supset y_2$, any strategy $P$ can be replaced
by a strategy $P'$ with $P'(y_2) = 0$ without making things worse for
the quizmaster. Thus the quizmaster, who wants to maximize the
contestant's expected loss, never needs to use a message that is
contained in another.

A \emph{dominating hyperplane} to a function $f$ from $D \subseteq
\R^\X$ to $\R$ is a hyperplane in $\R^\X \times \R$ that is nowhere
below $f$. A \emph{supporting hyperplane} to $f$ (at $P$) is a
dominating hyperplane that touches $f$ at some point $P$.\footnote{We
  deviate slightly from standard terminology here: what we call a
  supporting hyperplane to a concave function $f$ is usually called a
  supporting hyperplane to $\setc{(u, v) \in \R^\X \times \R}{v \leq
    f(u)}$, the hypograph of $f$.} A concave function has at least one
supporting hyperplane at every point \citep[Theorem
  11.6]{Rockafellar1970}, but it may be vertical. A nonvertical
hyperplane can be described by a linear function $\ell: \R^\X \to \R$:
$\ell(P) = \alpha + \sum_x P(x) \lambda_x$, where $\alpha \in \R$ and
$\lambda \in \R^\X$.

While $H_L$ is defined as a function on $\Delta_\X$, we will often
need to talk about supporting hyperplanes to the function $H_L$
restricted to $\Delta_y$ for some message $y \in \Y$.
We use the notation $H_L \restriction \Delta_y$ for the restriction of
$H_L$ to the domain $\Delta_y$. (Recall that we defined $\Delta_y$ as
a subset of $\Delta_\X$.) A supporting hyperplane to $H_L \restriction
\Delta_y$ is not a supporting hyperplane to $H_L$ itself if it goes
below $H_L$ at some $P \in \Delta_\X \setminus \Delta_y$.

A \emph{supergradient} is a generalization of the gradient: a
supergradient of a concave function at a point is the gradient of a
supporting hyperplane. If $H_L \restriction \Delta_y$ is finite and
continuous (and thus concave by Lemma~\ref{lem:HL}), then for any vector $\lambda \in
\R^\X$, a unique supporting hyperplane to $H_L \restriction \Delta_y$
can be found having that vector as its gradient, by choosing $\alpha$
appropriately in $\ell(P) = \alpha + \sum_x P(x) \lambda_x$
\citep[Theorem 27.3]{Rockafellar1970}. It will often be convenient in
our discussion to talk about supporting hyperplanes rather than
supergradients because they fix this choice of $\alpha$.

\begin{theorem}[Existence and characterization of $P^*$]\label{thm:existP_charP}
  For $H_L$ finite and upper semi-continuous (thus continuous),
  \anoptimal{} strategy for the quizmaster
(that is, a $P \in \P$ maximizing \eqref{quizmasterobjective})
  exists, and $P^*$ is such
  a strategy if and only if there exists a $\lambda^* \in \R^\X$ such
  that
  \begin{equation*}
    H_L(P') \leq \sum_{x \in y} P'(x) \lambda^*_x
    \quad\text{for all $y \in \Y$ and $P' \in \Delta_y$},
  \end{equation*}
  with equality if $P^*(y) > 0$ and $P' = P^*(\cdot \mid y)$. That is,
  for $y$ with $P^*(y) > 0$, the linear function $\sum_{x \in y} P(x)
  \lambda^*_x$ defines a supporting hyperplane to $H_L \restriction
  \Delta_y$ at $P^*(\cdot \mid y)$, and a dominating hyperplane for
  other $y$.

  A vector $\lambda^* \in \R^\X$ that satisfies the above for some
  \optimal{} $P^*$ satisfies it for all \optimal{} $P^*$ and is called a
  \emph{Kuhn-Tucker vector} (or \emph{KT-vector}).
\end{theorem}

Section~\ref{sec:Pappl} includes several examples illustrating the
application of Theorem~\ref{thm:existP_charP}; a
graphical illustration of the theorem is also included there
(Figure~\ref{fig:charP}). We will see in
Section~\ref{sec:optimalcontestant} that KT-vectors form the bridge
between \optimal{} strategies for the quizmaster and for the
contestant.

\subsubsection{Application to standard loss functions}\label{sec:Pappl}

The generalized entropy for logarithmic loss has only vertical
supporting hyperplanes at the boundary of $\Delta_y$ for any $y \in
\Y$. These hyperplanes do not correspond to any KT-vector $\lambda^*
\in \R^\X$, from which it follows that for any $y$ with $P^*(y) > 0$,
the \optimal{} strategy will not have $P^*(\cdot \mid y)$ at the
boundary of $\Delta_y$. The same is not generally true: we will see
below how for randomized 0-1 loss (in
Example~\ref{ex:montyhallfeenstra} on page~\pageref{ex:montyhall2}, and
Example~\ref{ex:23}) and Brier loss (in
Example~\ref{ex:outcomediscard}), games may have \anoptimal{} strategy
for the quizmaster that has $P^*(y) > 0$, yet $P^*(x \mid y) = 0$ for
some $y \in \Y$, $x \in y$.

Of the three loss functions we saw earlier, Brier loss and 0-1 loss
are finite, so by Lemma~\ref{lem:HL}, all conditions of
Theorem~\ref{thm:existP_charP} are satisfied for them. Logarithmic
loss is infinite when the obtained outcome was predicted to have
probability zero. The generalized entropy is still finite, because for
any true distribution $P$, there exist predictions $Q$ that give
finite expected loss (in particular, $Q = P$ does this). The entropy
is also continuous: $-P(x) \log P(x)$ is continuous as a function of
$P(x)$ with our convention that $0 \cdot \infty = 0$, and $H_L$ is the
sum of such continuous functions. Thus we can apply
Theorem~\ref{thm:existP_charP} to analyse the Monty Hall problem for
each of these three loss functions.

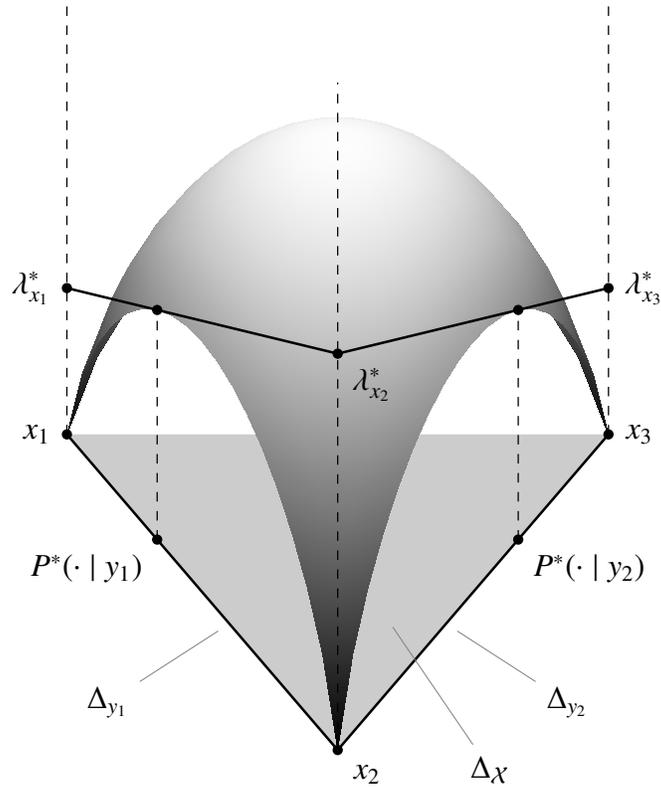
\begin{figure} %
\centering
\input{dome.tex}
\caption[Characterization of the \optimal{} strategy for the
quizmaster in the Monty Hall game with logarithmic loss]{The
  \optimal{} strategy for the quizmaster in the Monty Hall game with
  logarithmic loss, as characterized by
  Theorem~\ref{thm:existP_charP}. The triangular base is the full
  simplex $\Delta_\X$, on which the entropy function $H_L$ is defined
  (this is the grey dome); the points labelled $x_1$, $x_2$ and $x_3$
  are the elements of this simplex putting all mass on that single
  outcome; and the line segments $\Delta_{y_1}$ and $\Delta_{y_2}$ are
  the subsets of $\Delta_\X$ consisting of all distributions supported on $y_1$
  and $y_2$ respectively. Restricted to the domain $\Delta_{y_1}$, the
  vector $\lambda^*$ defines a linear function (having height
  $\lambda^*_x$ at each $x \in \X$) that is a supporting hyperplane to
  $H_L$ at $P^*(\cdot \mid y_1)$ (and similar for $y_2$). Note that
  when the linear function defined by $\lambda^*$ is extended to all
  of $\Delta_\X$, it may go below $H_L$ there, but not in
  $\Delta_{y_1}$ or $\Delta_{y_2}$.}\label{fig:charP}
\end{figure}
\begin{example_cont}[\ref{ex:montyhallfeenstra}]\label{ex:montyhall2}
  For Monty Hall, the strategy $P^*$ of choosing a message uniformly
  when the true outcome is $x_2$ is \optimal{} for the quizmaster, for
  all three loss functions. It is easy to verify that the theorem is
  satisfied by this strategy combined with the appropriate KT-vector:
  \begin{align*}
    \text{for logarithmic loss: } & \lambda^* = \left(-\log\frac{2}{3}, -\log \frac{1}{3}, -\log \frac{2}{3}\right);\\
    \text{for Brier loss: } & \lambda^* = \left(\frac{2}{9}, \frac{8}{9}, \frac{2}{9}\right);\\
    \text{for randomized 0-1 loss: } & \lambda^* = (0, 1, 0).
  \end{align*}
  The situation for logarithmic loss is illustrated in
  Figure~\ref{fig:charP}.

  We also find that for logarithmic loss and Brier loss, $P^*$ is the
  unique \optimal{} strategy, as the hyperplanes specified by
  $\lambda^*$ touch the generalized entropy functions at only one
  point each. For randomized 0-1 loss, on the other hand, all
  quizmaster strategies are \optimal{}, as the hyperplane specified by
  $\lambda^*$ touches $H_L \restriction \Delta_{y_1}$ at all $P(\cdot
  \mid y_1)$ with $P(x_1 \mid y_1) \geq 1/2$.
\end{example_cont}

\newpage

\begin{wrapfigure}{r}{0pt}
$
  \begin{array}{r|cccc|}
    P & x_1 & x_2 & x_3 & x_4 \\
    \hline
    y_1 & 1/5 & 1/5 & - & - \\
    y_2 & - & 0 & 0  & - \\
    y_3 & - & - & 1/5 & 2/5 \\
    \hline
    \marg{}_x & 1/5 & 1/5 & 1/5 & 2/5  \\
    \hline
  \end{array}
$
\end{wrapfigure}
\noindent \par{} %
\vspace{-1.5\baselineskip}

\begin{example}[The quizmaster discards a message]\label{ex:messagediscard}
  Consider a different game, with $\X = \set{x_1, x_2, x_3, x_4}$, $\Y
  = \set{\set{x_1, x_2}, \set{x_2, x_3}, \set{x_3, x_4}}$, $\marg{}$
  given by $\marg{}_{x_4} = 2/5$ and $\marg{}_x = 1/5$ elsewhere, and
  $L$ logarithmic loss. In the terminology of the Monty Hall puzzle,
  there is no initial choice by the contestant that determines what
  moves are available to the quizmaster, but the quizmaster will again
  leave two doors closed: the one hiding the car, and another adjacent
  to it. Then one strategy for the quizmaster is to never give message
  $y_2$ to the contestant; i.e.\ to pick the strategy $P \in
  \P$ with $P(y_2) = 0$ shown to the right.
  The depicted strategy $P$ is \optimal{}: When applying the theorem,
  we see that the KT-vector $\lambda^* = (\log 2, \log 2, \log 3,
  -\log(2/3))$ gives supporting hyperplanes to $H_L \restriction
  \Delta_{y_1}$ and $H_L \restriction \Delta_{y_3}$, but a
  non-supporting dominating hyperplane to $H_L \restriction
  \Delta_{y_2}$. This strategy can be seen to be intuitively
  reasonable because when the contestant receives message $y_3 =
  \set{x_3, x_4}$, he knows that the probability of the true outcome
  being $x_4$ is at least twice as large as the probability of it
  being $x_3$. By always giving message $y_3$ when the true outcome is
  $x_3$, the quizmaster can keep this difference from becoming larger.

  $P$ is also the unique \optimal{} strategy for Brier loss (as shown by
  the same analysis) and for randomized 0-1 loss (where the KT-vector
  is not unique: $(a, 1-a, 1, 0)$ for any $a \in [0, 1]$ is a
  KT-vector).
\end{example}

In the previous examples, the \optimal{} strategies $P$ coincided for
logarithmic and Brier loss. The following example shows that this is
not always the case.

\begin{wrapfigure}{r}{0pt}
$
  \begin{array}{r|cccc|}
    P & x_1 & x_2 & x_3 & x_4 \\
    \hline
    y_1 & 1/3 & 1/6 & - & - \\
    y_2 & - & 1/6 & 1/6  & 1/6\\
    \hline
    \marg{}_x & 1/3 & 1/3 & 1/6 & 1/6  \\
    \hline
  \end{array}
$
\end{wrapfigure}
\noindent \par{} %
\vspace{-.5\baselineskip}

\begin{example}[Dependence on loss function]\label{ex:23}
  Consider the family of games with $\X = \set{x_1, x_2, x_3, x_4}$,
  $\Y = \set{\set{x_1, x_2}, \set{x_2, x_3, x_4}}$, $\marg{}_{x_1} = \marg{}_{x_2}
  = 1/3$, and $\marg{}_{x_3} = \marg{}_{x_4} = 1/6$:
  This game is also similar to Monty Hall, but now one door has been
  `split in two': the quizmaster will either open door 1, or doors 3
  and 4. The strategy $P$ shown to the right is \optimal{} for logarithmic
  loss, but not for Brier
  loss: for both loss functions, there is a
  unique supporting hyperplane for both $y$ that touches $H_L
  \restriction \Delta_y$ at $P(\cdot \mid y)$ for $P$ as shown in the
  table, but for Brier loss, these two hyperplanes do not have the
  same height at the common outcome $x_2$. (Using
  Theorem~\ref{thm:proper} from page~\pageref{thm:proper}, we can find
  the \optimal{} strategy for the quizmaster under Brier loss by
  solving a quadratic equation with one unknown; this strategy has
  $P(x_2, y_1) = 11/3 - 2\sqrt{3} \approx 0.20$ and $P(x_2, y_2) =
  2\sqrt{3} - 10/3 \approx 0.13$.)

  For randomized 0-1 loss, neither the \optimal{} strategy nor the KT-vector
  are unique: the KT-vectors are $(0, 1, a, 1-a)$ for any $a \in [0,
  1]$; the \optimal{} strategies are the $P$ given above, the strategy
  that always gives message $y_1$ when the true outcome is $x_2$, and
  all convex combinations of these.
\end{example}

\newpage

\begin{wrapfigure}{r}{0pt}
  $
  \begin{array}{r|llll|} %
    P & x_1 & x_2 & x_3 & x_4 \\
    \hline
    y_1 & 0.45 & 0.05 & - & - \\
    y_2 & - & 0 & 0.25 & 0.25 \\
    \hline
    \marg{}_x & 0.45 & 0.05 & 0.25 & 0.25  \\
    \hline
  \end{array}
$
\end{wrapfigure}
\noindent \par{} %
\vspace{-1.5\baselineskip}

\begin{example}[The quizmaster discards a message-outcome
  pair]\label{ex:outcomediscard}
  Again consider the game from the previous example, but now with a
  different marginal as shown to the right.
  The strategy $P$ is \optimal{} for Brier loss, with KT-vector
  \[
  \lambda^* = (0.02, 1.62, 0.5, 0.5).
  \]
  $P$ displays another curious property (that we also saw for
  randomized 0-1 loss in the previous example): while the quizmaster
  uses message $y_2$ for some outcomes, he does not use it in
  combination with outcome $x_2$. In the theorem, the hyperplane on
  $\Delta_{y_2}$ is supporting at $P(\cdot \mid y_2)$, but is not a
  tangent plane: compared to the tangent plane, it has been `lifted
  up' at the opposite vertex of the simplex $\Delta_{y_2}$ ($x_2$) to
  the same height as the supporting hyperplane on $\Delta_{y_1}$.

  This behaviour cannot occur in games with logarithmic loss: as we
  observed at the beginning of Section~\ref{sec:Pappl}, if
  \anoptimal{} strategy $P^*$ has $P^*(y) > 0$ for some $y \in \Y$,
  then it must assign positive probability to $P^*(x \mid y)$ for all $x
  \in y$.
\end{example}

\subsection{\Optimal{} strategies for the contestant}\label{sec:optimalcontestant}

We now turn our attention to \optimal{} strategies for the contestant.
To this end, we look at the relation between the KT-vectors that
appeared in Theorem~\ref{thm:existP_charP} and the set of strategies
$\Q$ the contestant can choose from.

\subsubsection{Realizable hyperplanes}

For any $y \in \Y$, %
$\Delta_y$ is defined in Section~\ref{sec:defs} as a $(\lvert y \rvert
- 1)$-dimensional subset of $\R_{\geq 0}^\X$. Thus a linear function
$\ell: \Delta_y \to \R$ can be extended to a linear
function~$\bar{\ell}$ %
on the domain $\R_{\geq 0}^\X$ in different ways. Hence
many different vectors $\lambda \in \R^\X$ representing supporting hyperplanes
will correspond to what we can view as a single supergradient, because
the hyperplanes agree on $\Delta_y$. We can make the extension unique
by requiring $\bar{\ell}$ to be zero at the origin and at the vertices
of the simplex $\Delta_{\X \setminus y}$.
Because such a normalized function $\bar{\ell}: \R_{\geq 0}^\X \to \R$
obeys $\bar{\ell}(0) = 0$, it can be written as $\bar{\ell}(P) =
P\trans\lambda$ for some $\lambda$. These functions are thus uniquely
identified by their gradients $\lambda$, allowing us to refer to them
using `the (supporting) hyperplane $\lambda$'. Let $\Lambda_y$ be the
set of all gradients of such normalized functions that represent
dominating hyperplanes to $H_L \restriction \Delta_y$; in formula, let
\[
\Lambda_y = \setc{\lambda
  \in \R^\X}{\lambda_x = 0\text{ for }x \not\in y,\text{ and }\forall
  P \in \Delta_y : P\trans\lambda \geq H_L(P)}.
\]
For each nonvertical supporting hyperplane of $H_L \restriction
\Delta_y$, clearly the gradient is in $\Lambda_y$; that is, all finite
supergradients of this restricted function have a normalized
representative in $\Lambda_y$. The set also includes vectors $\lambda$
for which $P\trans\lambda > H_L(P)$ for all $P \in \Delta_y$, which do
not correspond to supporting hyperplanes.

Not all vectors $\lambda \in \Lambda_y$ may be available to the
contestant as responses to a play of $y \in \Y$ by the quizmaster. As
a trivial example, consider logarithmic loss and a vector $\lambda$
with $\sum_{x \in y} e^{-\lambda_x} < 1$ and $\lambda_x = 0$ for $x
\not\in y$. Then $\lambda \in \Lambda_y$ because the hyperplane
defined by $\lambda$ is dominating to $H_L \restriction \Delta_y$
(thus the expected loss from $\lambda$ is \emph{larger} than what the
contestant could achieve), but clearly there is no distribution $Q \in
\Delta_\X$ that results in these losses on $x \in y$. We say that a
vector $\lambda \in \Lambda_y$ is \emph{realizable on $y$} if there
exists a $Q \in \Delta_\X$ such that $L(x, Q) = \lambda_x$ for all $x
\in y$, and then we say that such a $Q$ \emph{realizes} $\lambda$.

A partial order on vectors $\lambda, \lambda' \in \R^\X$ is given by:
$\lambda \leq \lambda'$ if and only if $\lambda_x \leq \lambda'_x$ for
all $x \in \X$. We write $\lambda < \lambda'$ when $\lambda \leq
\lambda'$ and $\lambda \neq \lambda'$. For all $y \in \Y$,
this partial order has the following property: For $\lambda, \lambda'
\in \Lambda_y$, we have $\lambda \leq \lambda'$ if and only if for all
$P \in \Delta_y$, $P\trans\lambda \leq P\trans\lambda'$ (since any linear function is maximized over the simplex at a vertex). Therefore if
$Q, Q' \in \Delta_\X$ realize $\lambda, \lambda' \in \Lambda_y$
respectively and $\lambda \leq \lambda'$, the contestant is never hurt
by using $Q$ instead of $Q'$ as a prediction given the message $y$.

Any minimal element with respect to this partial order defines a
supporting hyperplane to $H_L \restriction \Delta_y$. For $P$ in the
relative interior of $\Delta_y$, the converse also holds: all
supporting hyperplanes at $P$ are minimal elements. This is not the
case for $P$ at the relative boundary of $\Delta_y$, where some
supporting hyperplanes (the ones that `tip over' the boundary) are not
minimal.

\begin{lemma}\label{lem:lambdareduction}
  If $H_L$ is finite and continuous on $\Delta_y$, then the following
  hold: %
  \begin{enumerate}
  \item If $\lambda \in \Lambda_y$ is not a supporting hyperplane to
    $H_L \restriction \Delta_y$, then there exists a supporting
    hyperplane $\lambda' \in \Lambda_y$ with $\lambda' < \lambda$;
  \item If $\lambda \in \Lambda_y$ is a supporting hyperplane to $H_L
    \restriction \Delta_y$ at $P$ but is not minimal in $\Lambda_y$,
    then there exists a minimal $\lambda' < \lambda$ in $\Lambda_y$;
  \item If $\lambda \in \Lambda_y$ is a supporting hyperplane to $H_L
    \restriction \Delta_y$ at $P$, then any $\lambda' \leq \lambda$ in
    $\Lambda_y$ is a supporting hyperplane at $P$ and obeys
    $\lambda'_x = \lambda_x$ for all $x \in y$ with $P(x) > 0$.
  \end{enumerate}
\end{lemma}
Thus the contestant never needs to play a $Q_{\mid y}$ realizing a
non-minimal element of $\Lambda_y$.

\subsubsection{Existence}

With the help of Lemma~\ref{lem:lambdareduction}, we can formulate
sufficient conditions for the existence of \anoptimal{} strategy $Q^*$
for the contestant that, together with $P^*$ for the quizmaster, forms
a Nash equilibrium.

\begin{theorem}[Existence of $Q^*$]\label{thm:existQ}
  Suppose that $H_L$ is finite and continuous and that for all $y \in \Y$, all minimal supporting hyperplanes
  $\lambda \in \Lambda_y$ to $H_L \restriction \Delta_y$ are
  realizable on $y$. Then there exists \anoptimal{} strategy $Q^* \in
  \Q$ for the contestant that achieves the same expected loss in the
  minimax game as $P^*$
  achieves in the maximin
  game: $(P^*, Q^*)$ is a Nash
  equilibrium.
\end{theorem}

We will see in Theorem~\ref{thm:proper} that a (or rather, at least
one) Nash equilibrium exists for logarithmic loss and Brier loss.
The existence of a Nash equilibrium in games with randomized 0-1 loss
is shown by the following consequence of Theorem~\ref{thm:existQ}.
\begin{proposition}\label{prop:existQ_01}
  In games with randomized 0-1 loss, a Nash equilibrium exists.
\end{proposition}

The following example shows what may go wrong if some supporting
hyperplanes are not realizable.

\newpage
\begin{wrapfigure}{r}{0pt}
$
  \begin{array}{r|ccc|}
    P^* & x_1 & x_2 & x_3 \\
    \hline
    y_1 & 1/6 & 1/6 & - \\
    y_2 & - & 1/6 & 1/6 \\
    y_3 & 1/6 & - & 1/6 \\
    \hline
    \marg{}_x & 1/3 & 1/3 & 1/3 \\
    \hline
  \end{array}
$
\end{wrapfigure}
\noindent \par{} %
\vspace{-1.5\baselineskip}

\begin{example}[Hard 0-1 loss]\label{ex:hard01}
  Consider the game with $\X$, $\Y$ and $\marg$ as shown in the table,
  and with hard 0-1 loss (so that the contestant is not allowed to
  randomize):
  \[
    L(x,Q) = \begin{cases}
      0 & \text{if } Q(x) = 1;\\
      1 & \text{otherwise.}
    \end{cases}
  \]
  This loss function has the same entropy function as randomized 0-1
  loss, so the two loss functions are the same from the quizmaster's
  perspective. The table shows the unique \optimal{} strategy for the
  quizmaster, with KT-vector $\lambda^* = (\sfrac{1}{2}, \sfrac{1}{2}, \sfrac{1}{2})$ and
  expected loss $1/2$. For randomized 0-1 loss, the (as we will see
  below: unique) \optimal{} strategy for the contestant would be to
  respond to any message $y$ with the uniform distribution on $y$.
  However, for all $y \in \Y$, $\lambda$ given by $\lambda_x =
  \indicator_{x \in y} \lambda^*_x$ is not realizable on $y$ under hard 0-1
  loss, so Theorem~\ref{thm:existQ} does not apply. In fact, for any
  strategy $Q$ the contestant might use, there exists a strategy $P$
  for the quizmaster that gives expected loss $2/3$ or larger (because
  for at least two outcomes $x$, there must be a $y \ni x$ such that
  $L(x, Q_{\mid y}) = 1$). Thus the inequality \eqref{weakduality} is
  strict: there is no Nash equilibrium, and \anoptimal{} strategy for
  either player is optimal only in the minimax/maximin sense.

  This example also shows that the condition on existence of
  supporting hyperplanes in Theorem~\ref{thm:existQ} cannot be
  replaced by the weaker condition that the infimum appearing in the
  definition \eqref{defHL} of $H_L$ is always attained.
\end{example}

\paragraph{Games without Nash equilibria} We will now briefly go into
the situation seen in the preceding example, where
Theorem~\ref{thm:existQ} does not apply.

While for some games with $L$ hard 0-1 loss, no Nash equilibrium may
exist, \optimal{} strategies for the contestant do exist, and can be
characterized using stable sets of a graph. A \emph{stable
  set} is a
set of nodes no two of which are adjacent \citep[Chapter
  64]{SchrijverB}. Consider the graph with node set $\X$ and with an
edge between two nodes if and only if they occur together in some
message. A set $S \subseteq \X$ is stable in this graph if and only if
there exists a strategy $Q \in \Q$ for the contestant such that for
all $x \in S$, $\max_{y \in \Y : x \in y} L(x, Q_{\mid y}) = 0$, and
equal to $1$ otherwise. The worst-case loss obtained by this strategy
is $1 - \sum_{x \in S} \marg{}_x$.  Thus finding the \optimal{}
strategy $Q$ for the contestant is equivalent to finding a stable set
$S$ with maximum weight.  Algorithmically, this is an NP-hard problem
in general, though polynomial-time algorithms exist for certain
classes of graphs, including perfect (this includes bipartite) graphs
and claw-free graphs \citep{SchrijverB}.

With the exception of two examples in Section~\ref{sec:propcont}
illustrating the limits of our theory, we will not look at games
without Nash equilibria any more from now on.

\subsubsection{Characterization and nonuniqueness}\label{sec:Qappl}

The concept of a KT-vector, which helped characterize \optimal{}
strategies for the quizmaster in Theorem~\ref{thm:existP_charP}, now
returns for a similar role in the characterization of \optimal{}
strategies for the contestant.
\begin{theorem}[Characterization of $Q^*$]\label{thm:charQ}
  Under the conditions of Theorem~\ref{thm:existQ} ($H_L$ finite and continuous, all minimal supporting hyperplanes realizable), a strategy $Q^* \in \Q$ is \optimal{} for the
  contestant if and only if the vector given by $\lambda_x \isdef \max_{y
    \ni x} L(x, Q^*_{\mid y})$ is a KT-vector.
\end{theorem}

If the loss $L(x, Q^*_{\mid y})$ equals $\lambda_x$ for all $x \in y$, then
the \optimal{} strategy $Q^*$ is an equalizer
strategy
\citep{Ferguson1967}: the expected loss of $Q^*$ does not depend on
the quizmaster's strategy. Not all games have an equalizer strategy as
\optimal{} strategy, as Example~\ref{ex:triangle_discard} below shows.

The following examples demonstrate that \anoptimal{} strategy for the
contestant is in general not unique.

\begin{wrapfigure}{r}{0pt}
$
  \begin{array}{r|cccc|}
     & x_1 & x_2 & x_3 & x_4 \\
    \hline
    y_1 & \qblack{} & \qblack{} & - & - \\
    y_2 & - & \qblack{} & \qblack{} & - \\
    y_3 & - & - & \qblack{} & \qblack{} \\
    y_4 & \qblack{} & - & - & \qblack{} \\
    \hline
    \marg{}_x & 1/4 & 1/4 & 1/4 & 1/4 \\
    \hline
  \end{array}
$
\end{wrapfigure}
\noindent \par{} %
\vspace{-.5\baselineskip}

\begin{example}[$\lambda^*$ not unique]\label{ex:KTnonunique}
  Consider the game with $\X$, $\Y$ and $\marg$ as in the table to the right,
  and with randomized 0-1 loss. For the quizmaster, any $P^*$ that is
  uniform given each $y$ is \optimal{}, and any $\lambda_a = (a, 1-a,
  a, 1-a)$ with $a \in [0,1]$ is a KT-vector. To each $\lambda_a$
  corresponds a unique \optimal{} $Q^*$, namely the strategy that puts
  conditional probability $1-a$ on outcome $x_1$ or $x_3$ (whichever
  is in the given message), and probability $a$ on $x_2$ or $x_4$.

  Note that if we replace randomized 0-1 loss by a strictly proper
  loss function such as logarithmic or Brier loss, the KT-vector and
  the \optimal{} strategy for the contestant become unique, while the
  same set of strategies as before continues to be \optimal{} for the
  quizmaster. This shows that the freedom for the contestant we see
  here for randomized 0-1 loss is due to the nonuniqueness of the
  KT-vector, not due to the nonuniqueness of $P^*$.
\end{example}

\begin{wrapfigure}{r}{0pt}
$
  \begin{array}{r|ccc|}
    P^* & x_1 & x_2 & x_3 \\
    \hline
    y_1 & 1/5 & 3/10 & - \\
    y_2 & - & 3/10 & 1/5 \\
    y_3 & 0 & - & 0 \\
    \hline
    \marg{}_x & 1/5 & 3/5 & 1/5 \\
    \hline
  \end{array}
$
\end{wrapfigure}
\noindent \par{} %
\vspace{-.5\baselineskip}

\begin{example}[Minimal $\lambda$ not
  unique]\label{ex:triangle_discard}
  Consider the game as shown in the table with logarithmic loss; the
  strategy $P^*$ shown in this table is the unique \optimal{} strategy
  for the quizmaster. Because logarithmic loss is proper, we know that
  $Q^*_{\mid y_1} = P^*(\cdot \mid y_1)$ and $Q^*_{\mid y_2} = P^*(\cdot
  \mid y_2)$ are optimal responses for the contestant, but this does
  not tell us what $Q^*_{\mid y_3}$ should be in \anoptimal{} strategy
  for the contestant.

  We see that $P^*$ assigns probability zero to message $y_3$, and the
  KT-vector
  \begin{equation*}
    \lambda^* = (-\log\frac{2}{5}, -\log\frac{3}{5}, -\log\frac{2}{5})
  \end{equation*}
  specifies a hyperplane that does not support $H_L$ in
  $\Delta_{y_3}$. Hence the construction of $Q^*_{\mid y_3}$ in the
  proof of Theorem~\ref{thm:existQ} allows freedom in the choice of a
  minimal element $\lambda \in \Lambda_{y_3}$ less than $(-\log 2/5,
  0, -\log 2/5)$: the valid choices are $(-\log q, 0, -\log (1-q))$
  for any $q \in [2/5, 3/5]$; each of these is realized on $y_3$ by
  $Q_{\mid y_3} = (q, 0, 1-q)$. Using Theorem~\ref{thm:charQ}, we can
  verify that these choices of $Q^*_{\mid y_3}$ define \optimal{}
  strategies: the vector $\lambda$ defined there equals the KT-vector
  $\lambda^*$ given above.

  This also shows that \optimal{} strategies for the contestant cannot
  be characterized simply as `optimal responses to $P^*$': in this
  example, $P^*(\cdot \mid y_3)$ is undefined, yet there is a
  nontrivial constraint on $Q_{\mid y_3}$ in the \optimal{} strategy
  $Q$ for the contestant.
\end{example}

\section{Results for well-behaved loss functions}\label{sec:simplerresults}

In the preceding sections, we have established characterization
results for the \optimal{} strategies of both players. While these
results are applicable to many loss functions, they have the
disadvantage of being complicated, involving supporting hyperplanes.
For some common loss functions, simpler characterizations can be
given.

\subsection{Proper continuous loss functions}\label{sec:propcont}

Recall from page~\pageref{def:proper} that for a \emph{proper} loss
function, the contestant's expected loss for a given message is
minimized if his predicted probabilities equal the true probabilities.
Such loss functions are natural to consider in our probability
updating game, as our goal will often be to find these true
probabilities. However, simplifying our theorems requires further
restrictions on the class of loss functions. In this subsection, we
consider loss functions that are both proper and continuous.

\begin{lemma}\label{lem:diffb}
  If the loss function $L(x, Q)$ is proper and continuous as a
  function of $Q$ for all $x$ and $H_L$ is finite, then $H_L$ is
  differentiable in the following sense: for all $y \in \Y$
  and all $P \in \Delta_y$, there is at most one element of
  $\Lambda_y$ that is a minimal supporting hyperplane to $H_L
  \restriction \Delta_y$ at $P$; if $P$ is in the relative interior of
  $\Delta_y$, there is exactly one. If it exists, the minimal
  supporting hyperplane at $P$ is realized by $Q_{\mid y} = P$.
\end{lemma}
The uniqueness of minimal supporting hyperplanes in $\Lambda_y$ is
equivalent to there being exactly one equivalence class of
supergradients, where supergradients are taken to be equivalent if
their corresponding supporting hyperplanes coincide on $\Delta_y$. The
property shown in the above lemma is then related to differentiability
by \citet[Theorem 25.1]{Rockafellar1970}, which says that for a
finite, concave function such as $H_L$, uniqueness of the
supergradient at $P$ is equivalent to differentiability at $P$.

\begin{theorem}\label{thm:proper}
  For $L$ proper and continuous and $H_L$ finite and continuous,
  \begin{enumerate}
  \item \optimal{} strategies for both players exist and form a Nash
    equilibrium;
  \item there is a unique KT-vector;
  \item a strategy $P^* \in \P$ for the quizmaster is \optimal{} if and
    only if there exists $\lambda^* \in \R^\X$ such that
    \begin{align*}
      L(x, P^*(\cdot \mid y)) &~=~ \lambda^*_x
      && \text{for all $x \in y$ with $P^*(x,y) > 0$,} \\
      L(x, P^*(\cdot \mid y)) &~\le~ \lambda^*_x
      && \text{for all $x \in y$ with $P^*(x,y) = 0$, $P^*(y) > 0$, and} \\
      \exists Q^*_{\mid y}: L(x, Q^*_{\mid y}) &~\le~ \lambda^*_x
      && \text{for all $x \in y$ with $P^*(y) = 0$;}
    \end{align*}
  \item a strategy $Q^*$ for the contestant is \optimal{} if and only
    if there exists \anoptimal{} $P^*$ such that for all $x$,
    \begin{equation}\label{eq:properQ}
      \max_{y \ni x} L(x, Q^*_{\mid y})
      ~=~ \max_{\mathclap{\substack{y \ni x,\\P^*(y) > 0}}}\, L(x, P^*(\cdot \mid y)),
    \end{equation}
    which holds if and only if \eqref{properQ} holds for all
    \optimal{} $P^*$.
  \end{enumerate}
\end{theorem}

Using this theorem, many observations made about logarithmic loss and
Brier loss in the examples we have seen so far can now be more easily
verified. For instance, in the \optimal{} strategy we saw in
Example~\ref{ex:outcomediscard} on page~\pageref{ex:outcomediscard},
we verify that $L(x_2, P^*(\cdot \mid y_2)) = 1.5 \leq 1.62 =
\lambda^*_{x_2} = L(x_2, P^*(\cdot \mid y_1))$.

Theorem~\ref{thm:proper} requires that $L$ is both proper and continuous. If either condition is removed then the conclusions of the theorem may fail to hold. Counterexamples for either case where no Nash equilibrium exists are given by \citet[Examples~6.J and 6.K]{vanOmmen_phd}.

While uniqueness of $\lambda^*$ was established by the theorem, we do
not have uniqueness of $P^*$ or $Q^*$. Multiple \optimal{} strategies
$Q^*$ for the contestant may exist as soon as a message is unused, as
in Example~\ref{ex:triangle_discard}. Multiple \optimal{} strategies
$P^*$ for the quizmaster are also possible, even for strictly proper
$L$: see Example~\ref{ex:KTnonunique}. The worst-case optimality criterion does not provide any guidance for selecting particular strategies in such cases. One might make use of symmetry and/or various kinds of limits to select a specific recommendation, or search for an analogue of subgame perfect equilibria. We will leave such interesting extensions to future research.

\subsection{Local loss functions}\label{sec:local}

Logarithmic loss is an example of a \emph{local} loss function: a loss
function where the loss $L(x, Q)$ depends on the probability assigned
by the prediction $Q$ to the obtained outcome $x$, but not on the
probabilities assigned to outcomes that did not occur. The following
theorem shows how for such loss functions, \optimal{}ity of the
quizmaster's strategy can be characterized purely in terms of
probabilities, without converting them to losses.

\begin{theorem}[Characterization of $P^*$ for local $L$]\label{thm:charP_local}
  For $L$ local and proper and $H_L$ finite and continuous, $P^* \in \P$
  is \optimal{} if there exists a vector $q \in [0,1]^\X$ such that
  \begin{equation}\label{eq:RCARcondition}
  \begin{aligned}
    q_x &= P^*(x \mid y) \text{ for all $y \in \Y$, $x \in y$ with $P^*(y > 0)$, and}\\
    \sum_{x \in y} q_x &\leq 1 \text{ for all $y \in \Y$}.
  \end{aligned}
  \end{equation}
  If additionally $H_L \restriction \Delta_y$ is strictly concave for
  all $y \in \Y$, only such $P^*$ are \optimal{} for $L$.
\end{theorem}
Among loss functions that are `smooth' for all $x$, logarithmic
loss
is, up to some transformations, the only proper local loss function
\citep{Bernardo1979}. We do not know what non-smooth local proper loss
functions may exist. In particular, it is conceivable (yet unlikely)
that a discontinuous $L$ exists satisfying the conditions of
Theorem~\ref{thm:charP_local}, but not those of
Theorem~\ref{thm:proper}.

If  $L$ is also continuous, then Theorem~\ref{thm:proper}
applies, and it follows that $Q^* \in \Q$ is \anoptimal{} strategy for
the contestant if $Q^*_{\mid y}(x) \geq q_x$ for all $y \in \Y$, $x \in y$.
For strictly proper loss functions such as logarithmic loss, this
fully characterizes the \optimal{} strategies for the contestant.
%
%
%

\begin{wrapfigure}{r}{0pt}
$
  \begin{array}{r|ccc|}
    P^* & x_1 & x_2 & x_3 \\
    \hline
    y_1 & 1/5 & 3/10 & - \\
    y_2 & - & 3/10 & 1/5 \\
    y_3 & 0 & - & 0 \\
    \hline
    \marg{}_x & 1/5 & 3/5 & 1/5 \\
    \hline
  \end{array}
$
\end{wrapfigure}
\noindent \par{} %
\vspace{-.5\baselineskip}

\begin{example_cont}[\ref{ex:triangle_discard}]%
  Consider again the game shown to the right with logarithmic loss. The conditionals $P^*(x \mid y)$ agree with
  the vector $q = (\sfrac{2}{5}, \sfrac{3}{5}, \sfrac{2}{5})$. For all $y \in \Y$ with $P^*(y) >
  0$, this implies that $\sum_{x \in y} q_x = 1$; for $y_3$, we see
  that this sum equals $4/5 \leq 1$. Thus $P^*$ is verified to be
  \optimal{}.
\end{example_cont}

The equality of conditionals $P^*(x \mid y)$ with the same $x$ in the
statement of Theorem~\ref{thm:charP_local} is oddly similar to the CAR
condition we saw in Section~\ref{sec:feenstraintro}, but reversing the
roles of outcomes and messages. We may say that a strategy $P^*$
satisfying \eqref{RCARcondition} is \emph{RCAR} (sometimes \emph{with
  vector $q$}), for `reverse CAR'. Note that whether a strategy is
RCAR does not depend on the loss function.

A vector $q$ is called an \emph{RCAR vector} if a strategy $P^* \in
\P$ exists such that $P^*$ and $q$ satisfy \eqref{RCARcondition}. This
definition is also independent of the loss function. If $q$ is an RCAR
vector, then $q_x > 0$ for all $x \in \X$; otherwise we would get
$P^*(x) = 0 < \marg{}_x$.
Like the KT-vector $\lambda^*$ in Theorem~\ref{thm:proper}, the RCAR
vector is unique:

\begin{lemma}\label{lem:RCARvector}
  Given $\X, \Y, \marg{}$, there exists a unique RCAR vector $q \in [0,1]^\X$.
\end{lemma}

If each message in $\Y$ contains an outcome $x$ not contained in any
other message, then any strategy $P^*$ must have $P^*(y) > 0$ for all $y
\in \Y$. Then the first line of \eqref{RCARcondition} implies that
$\sum_{x \in y} q_x = 1$ for all $y$. Thus the second line is now
satisfied automatically, allowing the theorem to be simplified for
this case:
\begin{corollary}\label{cor:strongRCAR}
  A strategy $P^* \in \P$ with $P^*(y) > 0$ for all $y \in \Y$ that
  satisfies
  \begin{equation}\label{eq:strongRCAR}
    P^*(x \mid y) = P^*(x \mid y') \text{ for all $y, y' \ni x$}
  \end{equation}
  is \optimal{} for the loss functions covered by
  Theorem~\ref{thm:charP_local}
\end{corollary}
In this case, $P^*$ is an equalizer strategy \citep{Ferguson1967}.

The symmetry between versions of CAR and RCAR is clearest in
Corollary~\ref{cor:strongRCAR}: the condition \eqref{strongRCAR} is
the mirror image of the definition of \emph{strong CAR} in
\citet{Jaeger2005AoS}. Thus we may call it \emph{strong RCAR}.
Ordinary RCAR \eqref{RCARcondition} imposes an inequality on $q$ for
messages with probability $0$, which has no analogue in the CAR
literature that we know of: the definition of \emph{weak CAR} in
\cite{Jaeger2005AoS} puts no requirement at all on outcomes with
probability $0$.

Strict concavity of $H_L$ occurred as a new condition in
Theorem~\ref{thm:charP_local}. The main loss function of interest here
is logarithmic loss, and its entropy is strictly concave. For other
loss functions, the following lemma relates strict concavity of $H_L$
to conditions we have seen before.
\begin{lemma}\label{lem:strictlyconcave}
  If $L$ is strictly proper and all minimal supporting hyperplanes
  $\lambda \in \Lambda_y$ to $H_L \restriction \Delta_y$ are
  realizable on $y$ for all $y \in \Y$, then $H_L$ is strictly concave
  on $H_L \restriction \Delta_y$ for all $y \in \Y$.
\end{lemma}

\paragraph{Affine transformations of the loss function} Previously, we
mentioned that logarithmic loss is the only local proper loss function
up to some transformations. The transformations considered in
\citet{Bernardo1979} are affine transformations, of the form
\begin{equation}\label{eq:affinetransform}
L'(x, Q) = a L(x, Q) + b_x
\end{equation}
for $a \in \R_{>0}$ and $b \in \R^\X$. (This transformation can result
in a function $L'$ that can take negative values, so that it does not
satisfy our definition of a loss function. However, our results can
easily be extended to loss functions bounded from below by an
arbitrary real number, so we allow such transformations here.)

The following lemma shows that, for logarithmic loss as well as for
other loss functions,
the transformation \eqref{affinetransform} does not
change how the players of the probability updating game should act.
\begin{lemma}\label{lem:affinetransform}
  Let $L$ be a loss function for which $H_L$ is finite and continuous,
  and let $L'$ be an affine transformation of $L$ as in
  \eqref{affinetransform}. Then a strategy $P^*$ is \optimal{} for the
  quizmaster in the game $\G' \isdef (\X, \Y, \marg{}, L')$ if and
  only if $P^*$ is \optimal{} in $\G \isdef (\X, \Y, \marg{}, L)$. If
  $\G$ also satisfies the conditions of Theorem~\ref{thm:existQ}, then
  the same equivalence holds for \optimal{} strategies $Q^*$ for the
  contestant.
\end{lemma}

Lemma~\ref{lem:affinetransform} has highly important implications when
applied to the logarithmic loss. While multiplying logarithmic loss by
a constant $a \neq 1$ merely corresponds to changing the base of the
logarithm, \emph{adding} constants $b_x$ allows the logarithmic loss
to become the appropriate loss function for a very wide class of
games. This means that the RCAR characterization of \optimal{}
strategies for logarithmic loss is also valid for all these games. We
are referring to so-called \emph{Kelly gambling} games, also known as
\emph{horse race games} \citep{CoverT91} in the literature. In such
games (with terminology adapted to our setting), for any outcome $x$
the contestant can buy a ticket which costs \EUR{1} and which pays off
a positive amount \EUR{$c_x$} if $x$ actually obtains; if some $x'
\neq x$ is realized, nothing is paid so the \EUR{1} is lost. The
contestant is allowed to distribute his capital over several tickets
(outcomes), and he is also allowed to buy a fractional nonnegative
number of tickets. For example, if $\X = \set{1,2}$ and $c_1 = c_2 =
2$, then the contestant is guaranteed
to neither win nor lose any %
money if he splits his capital fifty-fifty over both outcomes.

Now consider a contestant with some initial capital (say, \EUR{1}),
who faces an i.i.d.\ sequence $(\rv{X}_1,\rv{Y}_1), (\rv{X}_2,
\rv{Y}_2), \ldots \sim P$ of outcomes in $\X \times \Y$. At each point
in time $i$ he observes `side information' $\rv{Y}_i = y_i$ and he
distributes his capital gained so far over all $x \in \X$, putting
some fraction $Q_{\mid y_i}(x)$ of his capital on outcome $x$. Then he
is paid out according to the $x_i$ that was actually realized.
Here each $Q_{\mid y}$ is a probability distribution over $\X$,
i.e.\ for all $y \in \Y$,
all $x \in \X$, $Q_{\mid y}(x) \geq 0$ and $\sum_{x \in \X} Q_{\mid
  y}(x) = 1$. So if his capital was $U_i$ before the $i$-th round, it
will be $U_i \cdot Q_{\mid y_i}(x_i) c_{x_i}$ after the $i$-th round.
By the law of large numbers, his capital will grow (or shrink,
depending on the odds on offer) almost surely exponentially fast,
with exponent $\Exp_{\rv{X},\rv{Y} \sim P} [ \log
Q_{\mid \rv{Y}}(\rv{X}) c_{\rv{X}}] = \Exp_{\rv{X},\rv{Y} \sim P} [ \log
Q_{\mid \rv{Y}}(\rv{X}) - b_{\rv{X}}]$, where $b_x = -\log c_x$
\citep[Chapter 6]{CoverT91}.
Thus, %
the contestant's capital will grow fastest, among all constant
strategies and against an adversarial distribution $P \in \P$, if he
plays a worst-case optimal strategy for gains $\log Q(x) - b_x$, i.e.\
for loss function $L'(x,Q) = - \log Q(x) + b_x$. By
Lemma~\ref{lem:affinetransform} above, this worst-case optimal
strategy $Q^*$ is just the $Q^*$ that is also worst-case optimal for
logarithmic loss --- \emph{it does not depend on the pay-offs} (`odds'
in the horse race interpretation) $c_x$. Clearly, if data are i.i.d.\
then this continues to hold even if the pay-offs are allowed to change
over time, and even if the contestant is allowed to use different
strategies at different time points: the worst-case optimal capital
growth rate is always achieved by choosing $Q^*$ at all time points.

The upshot is that whenever (a)~the probability updating game is played
repeatedly, and (b)~the contestant is allowed to reinvest and
redistribute his capital over all outcomes at each point in time, then
his worst-case optimal strategy is equal to the worst-case optimal
$Q^*$ for logarithmic loss \emph{irrespective of the pay-offs}. This
makes the logarithmic loss, and hence the RCAR characterization,
appropriate for a very wide class of settings.

\section{The RCAR characterization for general loss functions}\label{sec:RCARgeneral}

Our results so far have focused on properties of the loss function
$L$. This section will focus on understanding a game's
message structure $\Y$.
For the purpose of summarizing our main theorems in simplified form, we restrict attention to irreducible games. Those are games which are connected (a game is disconnected if the outcome space can be partitioned such that each message is contained within one of the parts) and contain no dominated messages (a message is dominated if it is a strict subset of another message). We say that a game is a \markdef{graph game} (Section~\ref{sec:RCARgraph}) if all its messages are binary, and we call it a \markdef{matroid game} (Section~\ref{sec:RCARmatroid}) if its messages satisfy the \emph{basis exchange} property \citep[Corollary 1.2.5]{Oxley_matroid} detailed below. The abstract matroid property holds for a rich class of games, including e.g.\ \emph{negation games}, where each message excludes a single outcome. Monty Hall (Example~\ref{ex:montyhallfeenstra}) is both a graph and a matroid game. We also need a mild condition on loss functions. We call a loss function \markdef{regular} if it is invariant under permutations of the outcomes, and has finite and continuous associated entropy function $H_L$.

To solve a quizmaster-contestant game, we are typically looking for a Nash equilibrium (saddle point strategy pair). In general, not only the contestant's optimal response but also the optimal quizmaster strategy will depend on the loss function. This dependence renders probability updating an application-dependent endeavour. Our main result  in this section is that for an important class of games (and only these) the optimal quizmaster strategy is independent of the loss function. For these cases we derive a probability update rule that is application independent, like standard probability conditioning, but that applies much more broadly. The following is the summary result of this section (the precise statements can be found in Theorems~\ref{thm:binary}, \ref{thm:matroid} and~\ref{thm:nonmatroid_nongraph}).

\tikzexternaldisable
\begin{theorem}
Fix an irreducible game with set of messages $\Y$. Then
\begin{center}
\tikz[baseline] \node [rectangle, draw, text width={width("Y is a graphM")}] {
$\Y$ is a graph or a matroid
};
\quad
\tikz[baseline] \node {iff};
\quad
\tikz[baseline] \node[rectangle, text width={width("for any marginal on outcomes, the worst-caseMM")}, draw] {
for any marginal on outcomes, the worst-case optimal strategy for the quizmaster is identical for all regular loss functions
};
\end{center}
\end{theorem}
\tikzexternalenable

\bigskip\noindent
We first show a simple method of simplifying message structures in
Section~\ref{sec:decomposition}; there we will also see that if $\Y$
is a partition of $\X$, naive conditioning is \optimal{}. In
Section~\ref{sec:outcomesymmetry}, we consider symmetry properties
that \optimal{} strategies must have, provided that the loss function
also obeys a form of symmetry. Then in Sections~\ref{sec:RCARgraph} and~\ref{sec:RCARmatroid},
we show two classes of message structures for which the \optimal{}
strategy for the quizmaster can be characterized by the RCAR condition
\eqref{RCARcondition}. This is the condition that also characterizes
\optimal{} strategies for local loss functions and for Kelly gambling
with arbitrary payoffs (by Theorem~\ref{thm:charP_local} and
Lemma~\ref{lem:affinetransform}); the results in this section show
that the same characterization sometimes holds for a much more general
class of loss functions. This leads to an interesting
property of those (and only those) message structures, discussed in
Section~\ref{sec:lossinvariance}: the same strategy $P^*$ will be
optimal for the quizmaster regardless of the loss function.

We will look at the game from the perspective of the quizmaster, and
consider \optimal{} strategies $P^*$ for him. 
In games for which a
Nash equilibrium exists, the contestant's \optimal{}
strategies can be
found easily once we know $P^*$ and a KT-vector certifying its
optimality as in Theorem~\ref{thm:existP_charP}: given a KT-vector,
$Q^*$ can be constructed message-by-message to satisfy the condition
in Theorem~\ref{thm:charQ}. This is even easier in the case of proper
loss functions, where for each $y$ with $P^*(y) > 0$, an optimal
response is simply $Q^*_{\mid y} = P^*(\cdot \mid y)$. Another
advantage of looking at the game from the quizmaster's side is that
our Theorem~\ref{thm:existP_charP} characterizing \optimal{} $P^*$
requires weaker conditions than Theorem~\ref{thm:charQ} characterizing
\optimal{} $Q^*$.

\subsection{Decomposition of games}\label{sec:decomposition}

For some message structures, regardless of the marginal and loss
function, the problem of finding \anoptimal{} strategy for the
quizmaster can be solved by considering a smaller message structure
instead. It will be useful to look at such simplifications first, so
that in the coming sections we will only need to deal with
message structures that have already been simplified.

We have already seen one example of the type of result we are looking
for earlier, in Lemma~\ref{lem:messagemerging} on
page~\pageref{lem:messagemerging}, where we saw that if a message is
\emph{dominated} by another (meaning that it is a subset of the
other), then the quizmaster always has \anoptimal{} strategy that
assigns probability 0 to the dominated message.
In this subsection, we introduce a second simplification, by means of decomposition into connected components.

Connectivity is a fundamental concept from graph theory. However, in
general, our message structures are not graphs, but
\emph{hypergraphs}. Like an ordinary graph, a hypergraph is defined by
a set of nodes and a set of edges, but the edges are allowed to be
arbitrary subsets of the nodes; in a graph, all edges must contain
exactly two nodes. Thus for a probability updating game, we can talk
about the hypergraph $(\X, \Y)$, having the outcomes as its nodes and
the messages as its edges.

The terminology of connectivity can be generalized from graphs to
hypergraphs \citep{SchrijverA}. We will say that a game is connected
if its underlying hypergraph is connected. This leads to the following
definitions.

If for some game $\G = (\X, \Y, \marg{}, L)$, there is a set
$\myempty \subsetneq S \subsetneq \X$ such that for each message
$y$, either $y \subseteq S$ or $y \subseteq \X \setminus S$, then the
game can be \emph{decomposed} into two games $\G_1 = (\X_1, \Y_1,
\marg{}^{(1)}, L)$ and $\G_2 = (\X_2, \Y_2, \marg{}^{(2)}, L)$ with
$\X_1 = S$, $\X_2 = \X \setminus S$, $\Y_i = \setc{y \in \Y}{y
  \subseteq \X_i}$, and $\marg{}^{(i)}(x) = \marg{}(x) / \sum_{x' \in
  \X_i} \marg{}(x')$. If no such set $S$ exists, we say the game $\G$
is \emph{connected}.
\begin{lemma}[Decomposition]\label{lem:decomposition}
  If a game $\G$ can be decomposed into $\G_1$ and $\G_2$
  as described above, and  %
  its loss function $L$ is such that
  $H_L$ is finite and continuous
  and $H_L(P) = \inf_{Q \in \Delta_y} \sum_{x \in y} P(x)L(x,Q)$ for
  each $y \in \Y$ and each $P \in \Delta_y$,
  then a strategy $P^*$ is \optimal{} for
  the quizmaster in $\G$ if and only if there exist \optimal{}
  strategies for the quizmaster $P^*_1$ and $P^*_2$ in $\G_1$ and
  $\G_2$ respectively such that
  \begin{equation*}
    P^*(x, y) = \begin{cases}
      P^*_1(x, y) \cdot \sum_{x' \in \X_1} \marg{}(x') & \text{for } x \in \X_1;\\
      P^*_2(x, y) \cdot \sum_{x' \in \X_2} \marg{}(x') & \text{for } x \in \X_2.
    \end{cases}
  \end{equation*}
\end{lemma}
(The extra condition on $L$ is necessary to exclude some `very
improper' loss functions: those that reward the contestant for
predicting outcomes known to have probability 0.) In particular, if
the messages of $\G$ form a partition of $\X$, then $\G$ can be
decomposed into games that each contain only one message. In a game
$\G$ of this form, the quizmaster has only one strategy to choose
from. If the loss function is proper, naive conditioning is an optimal
response to this strategy, and thus worst-case optimal.

Together with Lemma~\ref{lem:messagemerging}, this lemma allows us to
reduce any game in which we want to find \anoptimal{}{} strategy for
the quizmaster to a set of connected games containing no dominated
messages. These reduced games will not contain any messages of size
one, unless one of the games consists of only that message: a message
of size one is either dominated, or it forms a trivial component
containing no other messages.

\subsection{Outcome symmetry}\label{sec:outcomesymmetry}

Sometimes, the problem of finding \anoptimal{} strategy is simplified
because certain `symmetry' properties of the message structure and
loss function allow us to conclude that \optimal{} strategies
satisfying an additional condition must have the same symmetries.

\subsubsection{Symmetry of loss functions}\label{sec:losssymmetry}

We now briefly return to the topic of loss functions to define a
property we will need next.
For a probability distribution $Q \in \Delta_\X$ and $x_1, x_2$
distinct elements of $\X$, define $Q^{x_1 \leftrightarrow x_2}$ as
\begin{equation*}
  Q^{x_1 \leftrightarrow x_2}(x) = \begin{cases}
    Q(x_2) & \text{for } x = x_1;\\
    Q(x_1) & \text{for } x = x_2;\\
    Q(x) & \text{otherwise,}
  \end{cases}
\end{equation*}
and similarly for a contestant's strategy $Q \in \Q$ by applying this
transformation to the conditional for each $y$. We say $L$ is
\emph{symmetric between $x_1$ and $x_2$} if for all $Q \in \Delta_\X$,
we have $L(x_1, Q) = L(x_2, Q^{x_1 \leftrightarrow x_2})$ and $L(x, Q)
= L(x, Q^{x_1 \leftrightarrow x_2})$ for all $x \in \X \setminus
\set{x_1, x_2}$. If $L$ is symmetric between $x_1$ and $x_2$ and
between $x_2$ and $x_3$, then it is also symmetric between $x_1$ and
$x_3$, because $((Q^{x_1 \leftrightarrow x_2})^{x_2 \leftrightarrow
  x_3})^{x_1 \leftrightarrow x_2} = Q^{x_1 \leftrightarrow x_3}$. In
words: we can apply the first symmetry, then the second, then the
first again to find that we have exchanged $x_1$ and $x_3$. We also
consider any loss function to be symmetric between $x$ and $x$ for any
$x$. So this symmetry of $L$ is an equivalence relation on $\X$, and
we are justified in talking about $L$ being symmetric on sets $S
\subseteq \X$, meaning that all pairs of elements of that set can be
exchanged.
If $L$ is symmetric on $\X$, we say it is \emph{fully
  symmetric}.

The loss functions we have seen so far were fully symmetric.
The affine transformations of loss
functions discussed at the end of Section~\ref{sec:local} may change
the symmetries of a loss function, while they do not change which
strategies are \optimal{} for the two players. This means that
sometimes, an asymmetric loss function can be transformed into an
essentially equivalent loss function with better symmetry properties. Yet not all loss functions can be transformed this way. The following two examples are about loss functions exhibiting this kind of \emph{inherent} asymmetry.

\begin{example}[Matrix loss]\label{ex:matrixloss}
  Given a $[0, \infty)$-valued $\X \times \X$ matrix of losses $A$,
  define \emph{hard matrix loss} by
  \[
    L(x,Q) = \begin{cases}
      A_{x, x'} & \text{if } Q(x') = 1 \text{ for some $x'$};\\
      \infty & \text{otherwise.}
    \end{cases}
  \]
  This generalizes hard 0-1 loss, which is obtained for the matrix $A$
  with zeroes on the diagonal and ones elsewhere (except that the
  definition above may give infinite loss for some $Q$, but a rational
  contestant would never use such $Q$). It is symmetric between $x_1$
  and $x_2$ if and only if swapping row $x_1$ with $x_2$ and column
  $x_1$ with $x_2$ results in matrix $A$ again; that is, if and only
  if $A_{x_1, x_1} = A_{x_2, x_2}$, $A_{x_1, x_2} = A_{x_2, x_1}$,
  $A_{x', x_1} = A_{x', x_2}$, and $A_{x_1, x'} = A_{x_2, x'}$, for
  all $x' \in \X \setminus \set{x_1, x_2}$.

  We can also define randomized matrix loss as an analogous
  generalization of randomized 0-1 loss, by taking an expectation over
  $Q$ in hard matrix loss:
  \[
  L(x,Q) = \sum_{x' \in \X} Q(x') A_{x, x'}.
  \]
  It has the same symmetry properties as hard matrix loss. The proof
  of Proposition~\ref{prop:existQ_01} also applies to randomized
  matrix loss without modification, showing that a Nash
  equilibrium
  exists in games using this loss function.
\end{example}

\begin{example}[Skewed logarithmic loss]
  Fix a vector $c \in \R_{\geq 0}^\X$, and define the function $F:
  \Delta_\X \to \R_{\geq 0}$ by
  \[
  F(P) \isdef -\sum_{x \in \X} c_x P(x) \log P(x).
  \]
  This is a sum of differentiable concave functions, and therefore
  differentiable and concave; if $c \in \R_{>0}^\X$, it is strictly
  concave (in fact, it is also strictly concave if $c$ contains a
  single 0). We use the construction of \emph{Bregman
    scores} in
  \citet[Section 3.5.4]{GrunwaldDawid2004} to construct a proper loss
  function $L$ having $F$ as its generalized entropy, and find
  \[
  L(x, Q)
    = F(Q) + (e_x - Q) \cdot \nabla F(Q)
    = -c_x(1 + \log Q(x)) + \sum_{x' \in \X} c_{x'} Q(x'),
  \]
  where $e_x$ is the distribution that puts all mass on $x$. This loss
  function is strictly proper if $H_L$ is strictly concave. Unlike
  logarithmic loss and its affine transformations, it is not local for
  $\lvert \X \rvert > 2$. Also, it is not generally fully symmetric,
  but is symmetric between pairs of outcomes $x_1, x_2 \in \X$ with
  $c_{x_1} = c_{x_2}$.
\end{example}

\subsubsection{Symmetry of KT-vectors}

Using the definition of symmetry of loss functions introduced in the
previous section, we can now state the following lemma.
\begin{lemma}[Loss exchange]\label{lem:lossexchange}
  Consider a game with $y_1, y_2 \in \Y$, $y_1 \setminus y_2 =
  \set{x_1}$, $y_2 \setminus y_1 = \set{x_2}$, $H_L$ finite and
  continuous and $L$ symmetric between $x_1$ and $x_2$. If
  \anoptimal{} strategy $P^*$ for the quizmaster exists with $P^*(x_1,
  y_1) > 0$, then all KT-vectors $\lambda^*$ satisfy $\lambda^*_{x_1}
  \leq \lambda^*_{x_2}$ (so if in addition $P^*(x_2,
  y_2) > 0$ then $\lambda^*_{x_1}
  = \lambda^*_{x_2}$).
\end{lemma}
When two messages $y_1, y_2 \in \Y$ satisfy $y_1 \setminus y_2 =
\set{x_1}$ and $y_2 \setminus y_1 = \set{x_2}$, we say that they
differ by the \emph{exchange} of one outcome.

It will be useful to have a term for the symmetry conditions on $L$
that allow us to apply Lemma~\ref{lem:lossexchange} to any pair of
messages in some set $\Y' \subseteq \Y$ satisfying the statement of
the lemma. We say $L$ is \emph{symmetric with respect to exchanges in
$\Y'$} if $L$ is
symmetric between any pair of outcomes $x_1, x_2$ such that messages
$y_1, y_2 \in \Y'$ exist with $y_1 \setminus y_2 = \set{x_1}$ and
$y_2 \setminus y_1 = \set{x_2}$.

\bigskip\noindent
We saw in Theorem~\ref{thm:charP_local} that for logarithmic loss,
\optimal{} strategies for the quizmaster can be characterized in terms
of a simple condition, the \emph{RCAR condition}
\eqref{RCARcondition}. We also saw that sometimes (in
Examples~\ref{ex:montyhallfeenstra} and~\ref{ex:messagediscard} on
pages~\pageref{ex:montyhall2} and~\pageref{ex:messagediscard}, %
but not in Example~\ref{ex:23}), those
same strategies were also \optimal{} for %
other loss
functions. This suggests that even for some types of games
where Theorem~\ref{thm:charP_local} does not apply,
it is possible to recognize \optimal{} strategies
using the easily verifiable
RCAR condition. We show that there are
two classes of message structures in which this is possible regardless
of the marginal, and explore the consequences %
in Section~\ref{sec:lossinvariance}.

\subsection{Graph games}\label{sec:RCARgraph}

The first of these classes consist of all message structures $\Y$ for
which each message contains at most two outcomes. After removing
singleton messages (which are either dominated or are decomposable
from the rest of the game), we have $\lvert y \rvert = 2$ for all $y
\in \Y$. This corresponds to a simple undirected graph
(that is, a graph containing no loops or multiple edges)
with a node for each outcome in $\X$ and an edge for each
message in $\Y$. For this reason, a game where each message in $\Y$
contains at most two outcomes is called a \emph{graph
  game}.
Many games we saw in previous examples were graph games. Their
underlying graphs are shown in Figure~\ref{fig:gamegraphs}.
\tikzset{xnode/.style={circle,draw,fill,inner sep=0pt,minimum
    size=3pt}}
\begin{figure}[t]
  \centering
  \subfloat[Example~\ref{ex:montyhallfeenstra} (Monty
    Hall)\label{fig:gamegraphs_3path}]
  {\begin{minipage}{.4\textwidth}
      \centering
    \begin{tikzpicture}[scale=1.8]
      \node [xnode,label=below:$x_1$] (x1) at (0,.5) {};
      \node [xnode,label=below:$x_2$] (x2) at (1,.5) {};
      \node [xnode,label=below:$x_3$] (x3) at (2,.5) {};
      \draw (x1) -- (x2) node[pos=0.5,above]{$y_1$};
      \draw (x2) -- (x3) node[pos=0.5,above]{$y_2$};
    \end{tikzpicture}
  \end{minipage}
  }
  \subfloat[Example~\ref{ex:messagediscard}\label{fig:gamegraphs_4path}]
  {
    \begin{minipage}{.4\textwidth}
      \centering
    \begin{tikzpicture}[scale=1.5]
      \node [xnode,label=below:$x_1$] (x1) at (0,.5) {};
      \node [xnode,label=below:$x_2$] (x2) at (1,.5) {};
      \node [xnode,label=below:$x_3$] (x3) at (2,.5) {};
      \node [xnode,label=below:$x_4$] (x4) at (3,.5) {};
      \draw (x1) -- (x2) node[pos=0.5,above]{$y_1$};
      \draw (x2) -- (x3) node[pos=0.5,above]{$y_2$};
      \draw (x3) -- (x4) node[pos=0.5,above]{$y_3$};
    \end{tikzpicture}
  \end{minipage}
  }

  \bigskip

  \subfloat[Examples~\ref{ex:hard01} and~\ref{ex:triangle_discard}\label{fig:gamegraphs_3cycle}]
  {
    \begin{minipage}{.4\textwidth}
      \centering
    \begin{tikzpicture}[scale=1.8]
      \node [xnode,label=above:$x_1$] (x1) at (0.5,0.866) {};
      \node [xnode,label=below:$x_2$] (x2) at (1,0) {};
      \node [xnode,label=below:$x_3$] (x3) at (0,0) {};
      \draw (x1) -- (x2) node[pos=0.5,right]{$y_1$};
      \draw (x2) -- (x3) node[pos=0.5,above]{$y_2$};
      \draw (x3) -- (x1) node[pos=0.5,left]{$y_3$};
    \end{tikzpicture}
  \end{minipage}
  }
  \subfloat[Example~\ref{ex:KTnonunique}\label{fig:gamegraphs_4cycle}]
  {
    \begin{minipage}{.4\textwidth}
      \centering
    \begin{tikzpicture}[scale=1.8]
      \node [xnode,label=above:$x_1$] (x1) at (0,1) {};
      \node [xnode,label=above:$x_2$] (x2) at (1,1) {};
      \node [xnode,label=below:$x_3$] (x3) at (1,0) {};
      \node [xnode,label=below:$x_4$] (x4) at (0,0) {};
      \draw (x1) -- (x2) node[pos=0.5,above]{$y_1$};
      \draw (x2) -- (x3) node[pos=0.5,right]{$y_2$};
      \draw (x3) -- (x4) node[pos=0.5,above]{$y_3$};
      \draw (x4) -- (x1) node[pos=0.5,left]{$y_4$};
    \end{tikzpicture}
  \end{minipage}
  }
  \caption{Underlying graphs of the graph games seen in
    the examples}\label{fig:gamegraphs}
\end{figure}
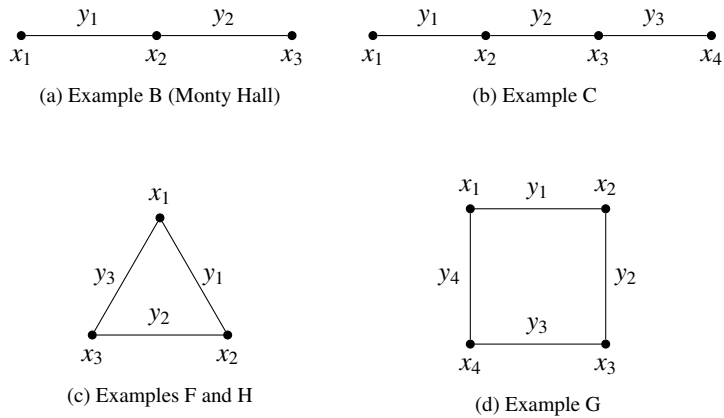

\begin{theorem}[RCAR for graph games]\label{thm:binary}
  If each message in $\Y$ contains at most two outcomes and $P^* \in \P$
  is an RCAR strategy, then $P^*$ is \optimal{} for all $L$ symmetric
  with respect to exchanges in $\setc{y \in \Y}{\lvert y \rvert = 2}$
  with $H_L$ finite and continuous. If additionally $H_L$ is strictly
  concave, only such $P^*$ are \optimal{} for $L$.
\end{theorem}

The statement of the theorem is very similar to that of
Theorem~\ref{thm:charP_local} in Section~\ref{sec:local}, and the
restrictions on $L$ in the present theorem (except for symmetry) were
also seen in the previous theorem. Sufficient conditions for these
restrictions to hold were given by Lemma~\ref{lem:HL} in
Section~\ref{sec:optimalquizmaster} ($H_L$ finite and continuous) and
Lemma~\ref{lem:strictlyconcave} in Section~\ref{sec:local} (strict
concavity of $H_L$).

The intuition behind the proof is that for binary predictions $Q$, the
probability assigned by $Q$ to one outcome determines the probability
$Q$ assigns to the other outcome. Thus all loss functions are
essentially local when used to assess such predictions, and their
behaviour is similar to logarithmic loss.

\subsection{Matroid games}\label{sec:RCARmatroid}

The other class is that of \emph{matroid games}. A
\emph{matroid} over
a finite \emph{ground set} $\X$ can be defined by a nonempty family
$\Y$ of subsets of $\X$ (the \emph{bases} of the matroid) satisfying
the \emph{basis exchange} property \citep[Corollary
1.2.5]{Oxley_matroid}: for all $y_1, y_2 \in \Y$ and $x_1 \in y_1
\setminus y_2$,
\begin{equation}\label{eq:basisexchange}
  (y_1 \setminus \set{x_1}) \cup \set{x_2} \in \Y
  \text{ for some } x_2 \in y_2 \setminus y_1.
\end{equation}
In words, for any pair of messages, if an outcome that is not in the
second message is removed from the first message, it must be possible
to replace it by an outcome from the second message that is not in the
first message, in such a way that the resulting set of outcomes is
again a message.

A matroid game is a game in which $\Y$ is the set of bases of a
matroid. The Monty Hall game (Example~\ref{ex:montyhallfeenstra}) is a
matroid game: taking one of the two messages and replacing the outcome
unique to it by the only other outcome will result in the other
message. By our definition of a game, it is required in addition to
\eqref{basisexchange} that each element of the ground set $\X$ of the
matroid occurs in some basis.

Many alternative characterizations of matroids exist. For example, a
matroid with ground set $\X$ and bases $\Y$ can also be represented by
its family of \emph{independent sets} ${\mathcal I} = \setc{I
  \subseteq \X}{I \subseteq y \text{ for some } y \in \Y}$, and a
different set of axioms analogous to \eqref{basisexchange}
characterizes whether a given set $\mathcal I$ is the family of
independent sets of some matroid.

The concept of a matroid was introduced by \citet{Whitney1935} to
study the abstract properties of the notion of dependence, as seen for
example in linear algebra and graph theory (explained below).
Different characterizations of the concept, applied to different
examples, were given independently by other authors, but then turned
out to be equivalent to matroids. One field where matroids play an
important role is combinatorial optimization. We refer to
\citet[Section 39.10b]{SchrijverB} for extensive historical notes.

We give two example classes of matroids, taken from \citet[Section
39.4]{SchrijverB}:
\begin{itemize}
\item Given an $m \times n$ matrix $A$ over some vector space, let $\X
  = \set{1, 2, \ldots, n}$ and ${\mathcal I}$ the family of all
  subsets $I$ of $\X$ such that the set of column vectors with index
  in $I$ is linearly independent. Then ${\mathcal I}$ is the family of
  independent sets of a matroid. A subset that spans the column space
  of $A$ is a basis of this matroid.
\item Given a simple undirected graph $G$, let $\X$ be its set of
  edges and ${\mathcal I}$ consist of all acyclic subsets of $\X$.
  Then ${\mathcal I}$ is the family of independent sets of a matroid.
  This matroid is called the \emph{cycle
    matroid}\label{text:cyclematroid} of $G$.
  The bases are the maximal independent sets; if $G$ is connected,
  these are its spanning trees.
\end{itemize}

One interesting class of games for which $\Y$ are the bases of a
matroid is the class of \emph{negation games}. In such a game, each
element of $\Y$ is of the form $\X \setminus \set{x}$ for some $x$.
(Not all sets of this form need to be in $\Y$.) Thus the quizmaster
will tell the contestant, ``The true outcome is \emph{not} $x$,'' as
in the original Monty Hall problem where one of the three doors is
opened to reveal a goat. A family $\Y$ of this form satisfies
\eqref{basisexchange} trivially: for $y_1, y_2$ distinct elements of
$\Y$, there is only one choice for each of $x_1$ and $x_2$, and with
these choices we get $(y_1 \setminus \set{x_1}) \cup \set{x_2} = y_2
\in \Y$.

Another class of matroids is formed by the \emph{uniform matroids}, in
which \emph{every} set of some fixed size $k$ is a basis. These also
have a natural interpretation when they occur as the message structure
of a game: the quizmaster is allowed to leave any set of $k$ doors
shut.

As the following theorem shows, matroid games share with graph games the property %
that RCAR strategies are \optimal{} for a wide variety of
loss functions.

\begin{theorem}[RCAR for matroid games]\label{thm:matroid}
  If $\Y$ are the bases of a matroid and $P^* \in \P$ is an RCAR
  strategy, then $P^*$ is \optimal{} for all $L$ symmetric with
  respect to exchanges in $\Y$ with $H_L$ finite and continuous. If
  additionally $H_L$ is strictly concave, only such $P^*$ are
  \optimal{} for $L$.
\end{theorem}

\subsection{Loss invariance}\label{sec:lossinvariance}

We saw in the preceding sections that in graph and matroid games,
\optimal{} strategies for the quizmaster are characterized by the RCAR
property. This property does not depend on what loss function is used
in the game (though the theorems do put some conditions on the loss
function, such as some symmetry requirements). Consequently, in such
games, strategies exist that are \optimal{} regardless of what loss
function is used (at least, for a large class of loss functions). We
call this phenomenon \emph{loss invariance}.

For such message structures, we can really think of the \optimal{}
strategies as `conditioning' (as a purely probability-based operation)
rather than as \optimal{} strategies for some game. This conditioning
operation can be seen as the generalization of naive conditioning to
message structures other than partitions (where naive conditioning
gives the right answer). Unlike naive conditioning, which requires
just the distribution $\marg{}$ and the message $y$ to compute $P(x \mid
y)$, we also need the message structure $\Y$ to compute that
conditional probability. But like naive conditioning, we do not need
to fix a loss function in order to talk about the worst-case optimal
prediction of $x$ given a message $y$.

A subtlety appears when improper loss functions are considered. Our
theorems show that the \optimal{} strategies for the quizmaster are
characterized independently of the loss function; however, the
\optimal{} strategies for the contestant will not necessarily coincide
with these if the loss function is not proper. In this case, loss
invariance tells us that the loss function does not affect \emph{what
  the contestant should believe} about the true outcome, but it may
affect how the contestant translates this belief into a prediction.

In the cases of graph and matroid games, our analysis of \optimal{}
strategies becomes more widely applicable in situations where the
probability updating game is really played by two players (as opposed
to being a theoretical tool for defining safe updating strategies):
\begin{itemize}
\item the same strategies continue to be \optimal{} if the two players
  use different loss functions (so that the game is no longer
  zero-sum);
\item both players will be able to play optimally without knowing the
  loss function(s) in use.
\end{itemize}
This is true for the Monty Hall game
(Example~\ref{ex:montyhallfeenstra}), which lies in the intersection
of graph and matroid games. This provides some justification for the
prevailing intuition that the Monty Hall problem should be analysed
using probability theory, without mention of loss functions.

Theorems~\ref{thm:binary} and~\ref{thm:matroid} apply only to loss
functions that are sufficiently symmetric and for which $H_L$ is
continuous and finite. We make no claim about the question whether
RCAR strategies are also \optimal{} for loss functions that do not
satisfy these properties. However, note that by
Lemma~\ref{lem:affinetransform}, sometimes affine transformations can
be used to convert an asymmetric loss function into a symmetric one
without affecting the players' strategies.

Lemma~\ref{lem:affinetransform} also shows that a limited form of loss
invariance holds regardless of the message structure. If the players
are using different affine transformations of the same loss function
(for example, of logarithmic loss; this corresponds to Kelly gambling
where the pay-offs for the contestant are different from those for the
quizmaster), both players can play optimally without knowing the
transformations in use.

An obvious question that remains is: are there any other classes of
message structure for which we have loss invariance? This is answered
in the negative by the following theorem.

\begin{theorem}\label{thm:nonmatroid_nongraph}
  If a connected game containing no dominated messages is neither a
  matroid game nor a graph game,
  then there exists a marginal such that no strategy $P$ for the
  quizmaster is \optimal{} for both logarithmic loss and Brier loss.
\end{theorem}

\section{Conclusion}\label{sec:mh_conclusion}

Conditioning is the method of choice for updating probabilities to incorporate new information. We started by reviewing why naive conditioning gives incorrect answers when the set of messages is not a partition of the set of outcomes. For in this case the association of messages to outcomes (the coarsening mechanism, or quizmaster strategy) needs to be taken into account. In general however, this mechanism is unknown. Previous work showed that conditioning is correct if the mechanism satisfies (variants of) the CAR (coarsening at random) condition. In this article we take a different route and investigate minimax probability updating strategies that are robust to the worst-case coarsening mechanism. To this end we modelled probability updating as a two-player zero sum game between a quizmaster and a contestant. In Section~\ref{sec:optimal} we discussed the optimal strategies for both players, characterizing both the hardest coarsening mechanism (quizmaster strategy) and the worst-case optimal way to incorporate new information (contestant strategy). In Section~\ref{sec:simplerresults} we specialized the results to  different classes of loss functions. A summary of
these theorems was given in Table~\ref{tab:loss_function_summary} on
page~\pageref{tab:loss_function_summary}. Then in Section~\ref{sec:RCARgeneral} we investigated graph and matroid games, and show that for these games the \optimal{} strategy does not depend on which loss function is used by both players, so that the updated probabilities are independent of the intended application.

\subsection{Future work}

There are many scenarios in
which our results currently do not apply, but to which they might be
extended. For example, the quizmaster's hard constraint $y \ni x$
could be replaced by some soft constraint, so that each message $y$
still carries information about the true outcome, but no longer in the
form of a subset of $\X$. One way to achieve this might be by affine
transformations of the loss function as discussed at the end of
Section~\ref{sec:local}, but allowing the constants to depend on both
$x$ and $y$. This could give a worst-case analogue to \emph{Jeffrey
  conditioning} or \emph{minimum relative
  entropy updating}
\citep{GrunwaldHalpern2003}.

Possible %
extensions would be to infinite outcome and message
spaces. We are also interested in probability updating in sequential (online) interaction protocols \citep{CesaBianchiL06}.

Other questions concern the comparison between different alternative
approaches the contestant might use to update his probabilities. For
example, can we bound the difference in expected loss between
\optimal{} and naive conditioning? What about ignoring the message and
always predicting with the marginal, or ignoring the constraints
imposed on the quizmaster by the marginal and predicting with the
maximum entropy distribution on $y$? (Both these strategies are overly
pessimistic.) Conversely, we might wonder how much the contestant
loses by playing \anoptimal{} strategy when the quizmaster is not
adversarial, but for instance chooses from the available messages
uniformly at random. In this context it is also of interest how to resolve the problem of non-unique worst-case optimal strategies in a principled way.

\paragraph{Acknowledgments}
Van Ommen and Gr\"unwald gratefully acknowledge support from NWO Vici grant 639.073.04. We would like to thank Teddy Seidenfeld and Erik Quaeghebeur for insightful discussions. Koolen was supported by a Queensland University of Technology Vice-Chancellor's Research Fellowship and by NWO Veni grant 639.021.439.

\DeclareRobustCommand{\VAN}[3]{#3 #1} %
\bibliography{bib/probupd}

\appendix
\renewcommand\appendixname{Appendix} %
\section{Proofs}
\label{app:proofs}

\input{pap_proofs.tex}

\end{document}

%% file: declarations.tex
\makeatletter
\renewenvironment{proof}[1][\proofname]{\par
  \pushQED{\qed}%
  \normalfont \topsep6\p@\@plus6\p@\relax
  \trivlist
  \item[\hskip\labelsep
        \bfseries
    #1\@addpunct{.}]\ignorespaces
}{%
  \popQED\endtrivlist\@endpefalse
}
\makeatother

\renewcommand{\eqref}[1]{(\ref{eq:#1})}

\newtheorem{theorem}{Theorem} 
\newtheorem{corollary}[theorem]{Corollary}
\newtheorem{lemma}[theorem]{Lemma}
\newtheorem{proposition}[theorem]{Proposition}

\theoremstyle{definition} 
\newtheorem{example}{Example} 

\newtheoremstyle{contstyle}
  {\topsep}   
  {\topsep}   
  {\normalfont} 
  {0pt}       
  {\bfseries} 
  {.}         
  {5pt plus 1pt minus 1pt} 
  {\thmname{#1}\thmnote{ #3} (continued)} 
\theoremstyle{contstyle}
\newtheorem*{example_cont}{Example}

\newcommand{\myempty}{\emptyset} 
\newcommand{\isdef}{\df} 
\newcommand{\ispdef}{\phantom{\vcentcolon}\protect\nolinebreak\mkern-1.2mu=}
\newcommand{\R}{\mathbf{R}}

\DeclareMathOperator{\Exp}{{\mathbf E}} 

\DeclareMathOperator*{\argmin}{arg\,min}

\newcommand{\trans}{^{\mkern-2mu\top\mkern-2mu}}
\newcommand{\rv}[1]{{#1}} 
\newcommand{\indicator}{\mathbf{1}}

\newcommand{\X}{{\mathcal X}} 
\newcommand{\Y}{{\mathcal Y}} 
\newcommand{\G}{{\mathcal G}} 
\renewcommand{\P}{{\mathcal P}} 
\newcommand{\Q}{{\mathcal Q}} 
\newcommand{\marg}{p}
\newcommand{\xiny}{{{\mathcal R}(\X,\Y)}} 
\newcommand{\optimal}{worst-case optimal}
\newcommand{\Optimal}{Worst-case optimal}
\newcommand{\optimalart}{a}
\newcommand{\Optimalart}{A}
\newcommand{\anoptimal}{\optimalart{} \optimal{}}
\newcommand{\Anoptimal}{\Optimalart{} \optimal{}}
\newcommand{\qblack}{*}



%% file: cartography.tex
\begin{tikzpicture}[scale=0.9,very thick, fill opacity=0.5, text opacity=1,blend mode=multiply]
\node (GM) [fill=blue!50,ellipse,draw,inner xsep=6em, inner ysep=3em] {};

\node [left=1em of GM.center,anchor=center, align=center,text width=width("$Y$ is graph or matroid")] {
graph or matroid $\mathcal Y$,\\
symmetric $L$
};

\node (LP) at (5.5,0) [fill=red!50,ellipse,draw, inner xsep=6em, inner ysep=3em] {};

\node [right=1em of LP.center, anchor=center, align=center,text width=width("$L$ is local and proper"), inner sep=2em] {
any $\mathcal Y$,\\
local and proper $L$ 
};

\node (big) at (2.75,0) [ellipse, draw, inner xsep=12em, inner ysep=5em,fill=black!10] {};
\node [below=1em of big.north] {RCAR};

\end{tikzpicture} 

%% file: dome.tex
\begin{tikzpicture}[scale=1.2,x={(-3cm,2cm)},y={(3cm,2cm)},z={(0cm,8cm)},
  declare function={
    logloss(\p)=\p?(\p*-ln(max(\p,.01))):0;
    entropy(\p,\q)=logloss(\p)+logloss(\q)+logloss(1-\p-\q);
  }]
  \begin{axis}[hide axis,clip=false,colormap/blackwhite,z buffer=none,
      x={(-3cm,3.5cm)},y={(3cm,3.5cm)},z={(0cm,4cm)}]
  \fill [fill=black!20]
    (axis cs: 0,0,0) -- (axis cs: 1,0,0) -- (axis cs: 0,1,0) -- cycle;
  \pgfmathsetmacro\mystep{\entropystepsize}
  \pgfmathsetmacro\mysecond{1-\mystep}
  \foreach \p in {1.00,\mysecond,...,-.0001}
  \foreach \q in {1.00,\mysecond,...,-.0001} {
    \pgfmathsetmacro\r{1-\p-\q}
    \ifdim \r pt > .0001pt
      \pgfmathsetmacro\pb{\p+\mystep}
      \pgfmathsetmacro\qc{\q+\mystep}
      \pgfmathsetmacro\rd{1-\pb-\qc}
      \pgfmathsetmacro\entra{entropy(\p,\q)}
      \pgfmathsetmacro\entrb{entropy(\pb,\q)}
      \pgfmathsetmacro\entrc{entropy(\p,\qc)}
      \ifdim \rd pt > -.01pt
        \pgfmathsetmacro\entrd{entropy(\pb,\qc)}
        \addplot3[patch,shader=interp]
        coordinates {
          (\pb, \qc, \entrd) (\pb, \q, \entrb) (\p, \qc, \entrc)
        };
      \fi
      \addplot3[patch,shader=interp]
      coordinates {
        (\p, \q, \entra) (\pb, \q, \entrb) (\p, \qc, \entrc)
      };
    \fi
  } 

  \draw[dashed] (axis cs: 1,0,0) -- (axis cs: 1,0,1.2);
  \draw[dashed] (axis cs: 0,0,0) -- (axis cs: 0,0,1.85);
  \draw[dashed] (axis cs: 0,1,0) -- (axis cs: 0,1,1.2);

  \pgfmathsetmacro\hiloss{-ln(1/3)}
  \pgfmathsetmacro\loloss{-ln(2/3)}
  \draw[thick]
     (axis cs: 1,0,\loloss)
  -- (axis cs: 0,0,\hiloss)
  -- (axis cs: 0,1,\loloss);
  \path (axis cs: 1,0,\loloss) node[draw,circle,inner sep=1pt,fill=black,
    label=left:$\lambda^*_{x_1}$] {};
  \path (axis cs: 0,0,\hiloss) node[draw,circle,inner sep=1pt,fill=black,
    label={below right:$\lambda^*_{x_2}$}] {};
  \path (axis cs: 0,1,\loloss) node[draw,circle,inner sep=1pt,fill=black,
    label=right:$\lambda^*_{x_3}$] {};

  \pgfmathsetmacro\entrP{entropy(1/3,0)}
  \draw[dashed] (axis cs: 2/3,0,0) -- (axis cs: 2/3,0,\entrP);
  \path (axis cs: 2/3,0,0) node[draw,circle,inner sep=1pt,fill=black,
    label={below left:$P^*(\cdot \mid y_1)$}] {};
  \path (axis cs: 2/3,0,\entrP) node[draw,circle,inner sep=1pt,fill=black] {};
  \draw[dashed] (axis cs: 0,2/3,0) -- (axis cs: 0,2/3,\entrP);
  \path (axis cs: 0,2/3,0) node[draw,circle,inner sep=1pt,fill=black,
    label={below right:$P^*(\cdot \mid y_2)$}] {};
  \path (axis cs: 0,2/3,\entrP) node[draw,circle,inner sep=1pt,fill=black] {};

  \path (axis cs: 1,0,0) node[draw,circle,inner sep=1pt,fill=black,
    label=left:$x_1$] {};
  \path (axis cs: 0,0,0) node[draw,circle,inner sep=1pt,fill=black,
    label=below right:$x_2$] {};
  \path (axis cs: 0,1,0) node[draw,circle,inner sep=1pt,fill=black,
    label=right:$x_3$] {};

  \draw[thick] (axis cs: 1,0,0)
  -- (axis cs: 0,0,0)
  -- (axis cs: 0,1,0);

  \path (axis cs: .4,0,0) node[pin=
    {[pin distance=1cm]210:$\Delta_{y_1}$}] {};
  \path (axis cs: 0,0.4,0) node[pin=
    {[pin distance=1cm]330:$\Delta_{y_2}$}] {};

  \path (axis cs: .125,0.3,0) node[pin=
    {[pin distance=1.5cm]300:$\Delta_\X$}] {};
  \end{axis}
\end{tikzpicture}

%% file: pap_proofs.tex
\begin{proof}[Proof of Lemma~\ref{lem:HL}]
  For finite $H_L$, concavity of $H_L$ %
  is shown by \citet[Proposition
  3.2]{GrunwaldDawid2004}, and lower semi-continuity by \citet[Theorem
  10.2]{Rockafellar1970} (using that the domain of $H_L$ is a
  simplex). If $L$ is finite, then picking any $Q \in \Delta_\X$ gives
  an upper bound to $H_L$, so that $H_L$ is in particular finite.
  Concavity now follows by the first claim, and continuity by
  \citet[Corollary 3.3; an important condition is
  in Corollary 3.2]{GrunwaldDawid2004}.
\end{proof}

\begin{proof}[Proof of Lemma~\ref{lem:messagemerging}]
  If $P(y_2) = 0$ then $P' = P$ and the result is trivial; if $P(y_2)
  > 0$ but $P(y_1) = 0$, then $P(\cdot \mid y_2) = P'(\cdot \mid y_1)$
  so $P$ and $P'$ have the same expected generalized entropy.
  Otherwise $P(\cdot \mid y_1)$ and $P(\cdot \mid y_2)$ are
  well-defined, and $P'(\cdot \mid y_1) = (P(y_1)P(\cdot \mid y_1) +
  P(y_2)P(\cdot \mid y_2))/(P(y_1) + P(y_2))$ is a convex combination
  of them. By concavity of $H_L$, $\sum_{y} P'(y)H_L(P'(\cdot \mid
  y)) \geq \sum_{y} P(y)H_L(P(\cdot \mid y))$.
\end{proof}

\begin{proof}[Proof of Theorem~\ref{thm:existP_charP}]
  \citet[Theorem 27.3]{Rockafellar1970} gives %
  conditions under which a convex minimization problem has a solution
  attaining the minimum. These are satisfied by $\P$ and $-H_L$: $\P$
  is nonempty, closed, convex, and bounded (thus has no direction of
  recession), and $-H_L$ is convex (Lemma~\ref{lem:HL}), %
  finite for
  all $P \in \P$ (thus proper), and lower semi-continuous (thus
  closed).

  By \citet[Corollary 28.2.2]{Rockafellar1970}, a KT-vector
  $\lambda^*$ exists, so that for the remaining claims of the theorem,
  it suffices to show that $P^*$ is \optimal{} and $\lambda^*$ is a
  KT-vector if and only if the given conditions on $(P^*, \lambda^*)$
  hold.
  To prove this, we rewrite the maximin problem to a convex
  optimization problem where $P$ may range over $P \in \R_{\geq
    0}^{\xiny}$, a strict superset of $\P$. In this larger set, $P(x
  \mid y) \isdef P(x,y)/P(y)$ still defines a conditional probability
  distribution for $y$ with $P(y) > 0$, because any scale factor
  cancels out.

  The following function extends the quizmaster's objective function
  \eqref{quizmasterobjective} (the expected generalized entropy of $P
  \in \P$) to the domain $\R_{\geq 0}^{\xiny}$:
  \begin{align*}%
    f_0(P) &\isdef \inf_{Q\in\Q} \; \sum_{y\in\Y, x \in y}\!\!\!\! P(x, y) L(x, Q_{\mid y})\\
    &\ispdef \sum_{\substack{y\in\Y:\\P(y)>0}} \inf_{Q_{\mid y} \in \Delta_\X}
    \sum_{x \in y} P(x,y) L(x, Q_{\mid y})
    ~=~ \sum_{\mathclap{\substack{y\in\Y:\\P(y)>0}}} P(y) H_L(P(\cdot \mid y)).
  \end{align*}
  Using this concave function (infimum of linear functions), the
  convex optimization problem is given by
  \begin{equation*}
    \begin{aligned} %
      & \text{maximize}
      & & f_0(P) \\
      & \text{subject to}
      & & \sum_{y\ni x} P(x,y) = \marg{}_x & & \text{ for all } x \in \X,
    \end{aligned}
  \end{equation*}
  with $P \in \R_{\geq 0}^{\xiny}$. By
  \citet[Theorem 28.3]{Rockafellar1970}, $P^* \in \R_{\geq 0}^{\xiny}$
  maximizes this and $\lambda^* \in \R^\X$ is a KT-vector if and only
  if $P^* \in \P$ and at $P^*$, the zero vector is a supergradient to
  \begin{equation}\label{eq:charP_Rocka}
    f_0(P^*)
    - \sum_{x \in \X} \lambda^*_x \left(\sum_{y \ni x} P^*(x,y) - \marg{}_x\right).
  \end{equation}
  The term being subtracted is linear, with gradient
  $\bar{\lambda} \in \R^{\xiny}$ given by
  \begin{equation}\label{eq:charP_barlambda}
    \bar{\lambda}_{x, y}
    \isdef \frac{\partial}{\partial P^*(x, y)}
    \sum_{x \in \X} \lambda^*_x \left(\sum_{y \ni x} P^*(x,y) - \marg{}_x\right)
    = \lambda^*_x.
  \end{equation}
  By \citet[Theorem 23.8]{Rockafellar1970}, $0$ is a
  supergradient to \eqref{charP_Rocka} if and only if $\bar{\lambda}$ is a
  supergradient to $f_0$ at $P^*$.

  For any $P^*$ that is not everywhere zero, we have for all $c \geq
  0$ that $f_0(cP^*) = cf_0(P^*)$, so that a supporting hyperplane to
  $f_0$ at any $P^* \in \P$ must go through the origin. Then the
  supporting hyperplane with gradient $\bar{\lambda}$ has as defining
  equation the linear expression $\sum_{x,y} P(x,y)
  \bar{\lambda}_{x,y}$.

  If $\sum_{x,y} P(x,y) \bar{\lambda}_{x,y}$ defines a supporting
  hyperplane to $f_0$ at $P^*$, then
  \begin{enumerate}
  \item at every $y \in \Y$ with $P^*(y) > 0$, it is a supporting
    hyperplane to $H_L \restriction \Delta_y$ at $P^*(\cdot \mid
    y)$, and
  \item for every $y$ with $P^*(y) = 0$, $H_L(P') \leq \sum_x P'(x)
    \bar{\lambda}_{x,y}$ for all $P' \in \Delta_y$.
  \end{enumerate}
  The converse also holds: we have for all $y \in \Y$ and $P' \in
  \Delta_y$ that $H_L(P') \leq \sum_{x \in y} P' \bar{\lambda}_{x,y}$,
  with equality if $P^*(y) > 0$ and $P' = P^*(\cdot \mid y)$; taking
  the convex combination with coefficients $P^*(y)$ shows that the
  hyperplane defined by $\sum_{x,y} P(x,y) \bar{\lambda}_{x,y}$ is
  nowhere below $f_0$ and touches it at $P = P^*$.

  For $\bar{\lambda}$ of the required form \eqref{charP_barlambda},
  this is in turn equivalent to the characterization given in the
  statement of the theorem.
\end{proof}

\begin{proof}[Proof of Lemma~\ref{lem:lambdareduction}]
  The function $\sum_{x \in y} \lambda_x P(x) - H_L(P)$ attains its
  minimum $d$ on $\Delta_y$ at some $P$ \citep[Theorem
  27.3]{Rockafellar1970}. Let $\lambda' \in \Lambda_y$ be given by
  $\lambda'_x = \lambda_x - d$ for all $x \in y$: this defines a
  hyperplane to $H_L \restriction \Delta_y$ that is supporting at the
  minimizing $P$, proving the first part of the lemma.

  For the third part, let $\lambda$, $\lambda'$ and $P$ be as
  described. $\lambda' \leq \lambda$ implies $P\trans\lambda' \leq
  P\trans\lambda$.
  Neither can be smaller than $H_L(P)$, and since the right hand side
  must equal $H_L(P)$ because $\lambda$ is supporting at $P$, so must
  the left hand side, showing that $\lambda'$ is also supporting at
  $P$. If $P(x) = 1$ for some $x \in y$, then $P \trans \lambda'' =
  \lambda''_x$ for any $\lambda''$, so in particular $\lambda'_x =
  \lambda_x = H_L(P)$. For other $P \in \Delta_y$, we use that two
  linear functions obeying an inequality on their domain $D \isdef
  \Delta_{\setc{x \in y}{P(x) > 0}}$ and coinciding at a point in the
  relative interior of $D$ must coincide everywhere on $D$, so that
  again $\lambda'_x = \lambda_x$ for $x \in y$ with $P(x) > 0$.

  It remains to show that given such a $\lambda$, a minimal $\lambda'
  \leq \lambda$ exists in $\Lambda_y$. Consider the set $\Lambda'
  \isdef \setc{\lambda' \in \Lambda_y}{\lambda' \leq \lambda}$. The
  set of supporting hyperplanes to $H_L \restriction \Delta_y$ at $P$
  in $\Lambda_y$ is closed %
  \citep[Section 23, definition subdifferential (page
  215)]{Rockafellar1970}; $\Lambda'$ is a subset of this set (as we
  just saw), obtained by adding further non-strict linear constraints,
  so it too is closed. %
  It also has the property that if $\lambda' \in \Lambda'$ is minimal
  in that set, it is also minimal in $\Lambda_y$. Now fix any $P'$ in
  the relative interior of $\Delta_y$, and pick some $\lambda' \in
  \Lambda'$ that minimizes ${P'}\trans\lambda'$ (this minimum must be
  attained because the expression is bounded below and $\Lambda'$ is
  closed). Such a $\lambda'$ is also minimal in the partial order, so
  it is the element we are looking for.
\end{proof}

\begin{proof}[Proof of Theorem~\ref{thm:existQ}]
  Take \anoptimal{} strategy $P^*$ for the quizmaster and KT-vector
  $\lambda^*$. For each $y \in \Y$, define a vector
  \begin{equation*}
    \lambda'_x = \begin{cases}
      \lambda^*_x & \text{for } x \in y \\
      0           & \text{for } x \not\in y.
    \end{cases}
  \end{equation*}
  By the statement of Theorem~\ref{thm:existP_charP}, $\lambda' \in
  \Lambda_y$. Let $\lambda$ be a minimal element of $\Lambda_y$ with
  $\lambda \leq \lambda'$: such an element exists by parts 1 and 2 of
  Lemma~\ref{lem:lambdareduction}. (If $\lambda'$ is itself minimal,
  $\lambda = \lambda'$). By assumption, $\lambda$ is realizable on
  $y$. Let $Q^*_{\mid y}$ be given by this $Q$.

  By playing this $Q^*$, the contestant will achieve expected loss
  (against any strategy $P \in \P$ for the quizmaster, for $\lambda^*$
  any KT-vector)
  \begin{equation*}
    \sum_{x,y} P(x,y)L(x, Q^*_{\mid y})
    \leq \sum_{x,y} P(x,y) \lambda^*_x
    = \sum_x \marg{}_x \lambda^*_x.
  \end{equation*}
  The right-hand side is the maximum loss the quizmaster can achieve
  in the maximin game. By \eqref{weakduality}, the reverse inequality
  also holds, so we find that the values of the minimax and maximin
  games must be equal.
\end{proof}

\begin{proof}[Proof of Proposition~\ref{prop:existQ_01}]
  We first introduce some additional terminology in order to apply a
  corollary from \citet{Rockafellar1970}.

  A nonvertical hyperplane defined by $\lambda \in \R^\X$
  is geometrically a subset of $\R^\X \times \R$, namely $\setc{(P',
    z') \in \R^\X \times \R}{z' = {P'}\trans \lambda}$. This set is
  the boundary of the half-space $H_\lambda = \setc{(P', z')
    \in \R^\X \times \R}{z' \leq {P'}\trans \lambda}$. A hyperplane
  $\lambda$ is supporting to a concave function $f: \R^\X \to \R$ at
  the point $P \in \R^\X$ with $f(P) = P\trans \lambda$ if and only if
  the hypograph of $f$ is a subset of $H_\lambda$.

  A column vector $(-\alpha\lambda, \alpha)$ is called \emph{normal}
  to a convex set $C$ at a point $(P, z) \allowbreak\in C$ if $(P'-P,
  z'-z)\trans(-\alpha\lambda, \alpha) \leq 0$ for all $(P', z') \in C$
  \citep{Rockafellar1970}; that is, if $C \subseteq \setc{(P',
    z')}{(P'-P, z'-z)\trans(-\alpha\lambda, \alpha) \leq 0}$. This
  latter set is equal to $H_\lambda$ if $\alpha > 0$ and $z = P\trans
  \lambda$. So if $C$ is the hypograph of $f$ and $f(P) = z = P\trans
  \lambda$, then $\lambda$ is a supporting hyperplane to $f$ at $P$ if
  and only if $(-\lambda, 1)$ is normal to $C$.

  The set of all vectors normal to $C$ at $(P, z)$ is called the
  \emph{normal cone} at $(P, z)$. The normal cone to $H_\lambda$ at
  given $(P, P\trans \lambda)$ is the half-line
  $\setc{(-\alpha\lambda, \alpha)}{\alpha \in [0, \infty)}$.

  For $L$ randomized 0-1 loss, let the function $f_0: \R^\X \to \R$ be
  given by $f_0(P) = \min_{Q \in \Delta_\X} \sum_{x' \in \X} P(x')
  L(x', Q)$; note that $f_0 \restriction \Delta_\X = H_L$, and that
  for all $y \in \Y$, any minimal supporting hyperplane $\lambda \in
  \Lambda_y$ to $H_L \restriction \Delta_y$ can be extended to a
  supporting hyperplane $\lambda'$ to $f_0$ with $\lambda'_x =
  \lambda_x$ for all $x \in y$.

  The hypograph of $f_0$ is $C = \bigcap_{x \in \X}
  H_{\lambda^{(x)}}$, with $\lambda^{(x)}_{x'} = L(x', e_x)$ (where
  $e_x$ is the distribution that puts all mass on $x$). By
  \citet[Corollary 23.8.1]{Rockafellar1970}, for $C$ of this form and
  $(P, z)$ a point on the boundary of $C$, the normal cone of $C$ at
  $(P, z)$ is the sum of the individual normal cones. The normal cone
  of any set at a point in the interior of that set is just $\set{0}$,
  so we can ignore those halfspaces when determining the normal cone.
  Then the corollary says that any vector $(-\lambda, 1)$ normal to
  $f_0$ at $(P, f_0(P))$ can be written as $\sum_{x \in \X: P\trans
    \lambda^{(x)} = f_0(P)} (-\alpha_x \lambda^{(x)}, \alpha_x)$:
  $\lambda$ is a convex combination of those $\lambda^{(x)}$.

  Conclusion: any minimal supporting hyperplane $\lambda$ to $H_L
  \restriction \Delta_y$ at $P \in \Delta_y$ with randomized 0-1 loss
  is a convex combination of the hyperplanes realized by hard 0-1
  loss that are supporting at $P$. Therefore, randomizing allows the
  contestant to realize $\lambda$.
\end{proof}

\begin{proof}[Proof of Theorem~\ref{thm:charQ}]
  From Theorem~\ref{thm:existQ}, we know that a strategy exists for
  the contestant that achieves loss $\sum_x \marg{}_x \lambda^*_x$ where
  $\lambda^*$ is any KT-vector, and that this is \optimal{}. Hence $Q^*$
  is \optimal{} if and only if it achieves the same worst-case expected
  loss. The worst-case expected loss of a strategy $Q \in \Q$ is
  \begin{equation*}
    \max_{P \in \P} \sum_{x,y} P(x,y) L(x, Q_{\mid y})
    = \sum_x \marg{}_x \max_{y \ni x} L(x, Q_{\mid y}).
  \end{equation*}
  Therefore if for all $x, y$ with $x \in y$, we have $L(x, Q_{\mid
    y}) \leq \lambda^*_x$ for some KT-vector $\lambda^*$, $Q$ is
  \optimal{}.

  For the converse, pick any $Q \in \Q$ and suppose that the vector given %
  by $\lambda_x \isdef \max_y L(x, Q_{\mid y})$ is not a KT-vector.
  Then by Theorem~\ref{thm:existP_charP}, there is no $P \in \P$ such
  that $P$ and $\lambda$ satisfy the conditions of that theorem.
  Equivalently, for all $P \in \P$, there either is a message $y$ such
  that the hyperplane defined by $\lambda$ passes below $H_L$
  somewhere in $\Delta_y$, or there is a message $y$ with $P(y) > 0$
  but the hyperplane lies strictly above $H_L$ at $P(\cdot \mid y)$.
  The former contradicts the definition of $H_L$, so for $\lambda$ not
  a KT-vector, the latter must be the case. But then against any $P
  \in \P$ (in particular against \optimal{} $P$), there is a different
  strategy $Q' \in \Q$ that is equal to $Q$ except for its response to
  the message $y$: $Q'_{\mid y}$ realizes a supporting hyperplane to
  $H_L \restriction \Delta_y$ at $P(\cdot \mid y)$. This strategy $Q'$
  obtains strictly smaller expected loss than $Q$, so $Q$ is not
  \optimal{}. (In other words: in a Nash equilibrium $(P^*, Q^*)$, the
  contestant can only do worse against $P^*$ by changing strategy, but
  here he can do better.)
\end{proof}

\begin{proof}[Proof of Lemma~\ref{lem:diffb}]
  For proper loss functions, $L$ and $H_L$ are related as follows: at
  all $y \in \Y$ and $P \in \Delta_y$, if the vector $\lambda =
  L(\cdot, P)$ is finite at all $x \in y$, it describes the
  nonvertical hyperplane realized by $P$, which is a supporting
  hyperplane to $H_L \restriction \Delta_y$ at $P$.

  Now suppose that at some $P \in \Delta_y$, there exists a minimal
  supporting hyperplane $\lambda' \in \Lambda_y$ other than $\lambda
  \isdef L(\cdot \mid P)$, the supporting hyperplane realized by $P$
  (here we allow vertical hyperplanes, for which $\lambda$ may include
  infinities). Let $x \in y$ be an outcome where $\lambda'_x <
  \lambda_x$, which exists by minimality of $\lambda'$. Write $e_x$
  for the probability distribution that puts all mass on this outcome,
  and define $P_\alpha \isdef (1 - \alpha) P + \alpha e_x$ for $\alpha
  \in (0, 1]$. For each of these points $P_\alpha$, the hyperplane
  $L(\cdot, P_\alpha)$ realized by $P_\alpha$ is at most as high as
  $\lambda'$ at $P_\alpha$ (because $L(\cdot, P_\alpha)$ is supporting
  there) and at least as high as $\lambda'$ at $P$ (where $\lambda'$
  is supporting), so $L(x, P_\alpha)$ is bounded away from $\lambda_x$
  by $\lambda'_x$: $L(x, P_\alpha) \leq \lambda'_x < \lambda_x$.
  Therefore $\lim_{\alpha \downarrow 0} L(x, P_\alpha) \neq L(x, P)$,
  and $L$ is not continuous. For $L$ proper and continuous, this
  proves the `at most one' part of the lemma.

  For the `exactly one' part: by \citet[Theorem
  23.3]{Rockafellar1970}, %
  a nonvertical supporting hyperplane may only fail to exist at $P$ if
  there is a line segment through $P$ falling inside $\Delta_y$ on one
  side of $P$ and outside on the other; that is, for $P$ on the
  relative boundary of $\Delta_y$.

  Finally, suppose that $\lambda = L(\cdot, P)$ (the hyperplane
  realized by $P$) is finite at all $x \in y$ but not minimal. Then by
  Lemma~\ref{lem:lambdareduction}, a different minimal supporting
  hyperplane $\lambda'$ exists at $P$, which by the above gives a
  contradiction. This shows that if $\lambda$ is finite, it is the
  minimal supporting hyperplane.
\end{proof}

\begin{proof}[Proof of Theorem~\ref{thm:proper}]
  Theorem~\ref{thm:existP_charP} applies, showing existence of a
  KT-vector and a \optimal{} strategy for the quizmaster.

  \Anoptimal{} $Q^*$ exists and forms a Nash equilibrium with $P^*$:
  For each $y \in \Y$ and each $P \in \Delta_y$, by
  Lemma~\ref{lem:diffb} there is at most one minimal supporting
  hyperplane at $P$ which is then realized by $Q = P$. So all minimal
  supporting hyperplanes are realizable on $y$, and
  Theorem~\ref{thm:existQ} applies.

  Next we show that the characterization of $P^*$ and $\lambda^*$ in
  Theorem~\ref{thm:existP_charP} is equivalent to the one in this
  theorem. We consider $y$ with $P^*(y) > 0$ first. If $P^*$ is
  \anoptimal{} strategy for the quizmaster $\lambda^*$ defines a
  supporting hyperplane to $H_L \restriction \Delta_y$ at $P^*(\cdot
  \mid y)$, then by Lemma~\ref{lem:lambdareduction} there exists a
  minimal $\lambda' \in \Lambda_y$ which is also supporting at
  $P^*(\cdot \mid y)$ and which satisfies $\lambda'_x \leq \lambda_x$
  for $x \in y$, with equality for $P^*(x \mid y) > 0$. By
  Lemma~\ref{lem:diffb}, $\lambda'_x = L(\cdot, P^*(\cdot \mid y))$
  for all $x \in y$, showing that the conditions of this theorem hold.
  Conversely, if $\lambda^*$ satisfies the equality in this theorem,
  then $\sum_{x \in y} P^*(x \mid y) L(x, P(\cdot \mid y)) =
  H_L(P^*(\cdot \mid y))$, so $\lambda^*$ defines a supporting
  hyperplane at $P^*(\cdot \mid y)$.

  For $y$ with $P^*(y) = 0$, if the hyperplane defined by $\lambda^*$
  is nowhere below $H_L \restriction \Delta_y$ as in
  Theorem~\ref{thm:existP_charP}, then using
  Lemma~\ref{lem:lambdareduction} it can be lowered to become a
  minimal supporting hyperplane, for which a realizing $Q^*_{\mid y}$
  exists; conversely, the existence of a supporting hyperplane to $H_L
  \restriction \Delta_y$ at $Q^*_{\mid y}$ that is nowhere above
  $\lambda^*$ implies that $\lambda^*$ is itself nowhere below $H_L
  \restriction \Delta_y$.

  Uniqueness of the KT-vector: For any \optimal{} strategy $P^*$, the
  characterization in this theorem puts an equality constraint on
  $\lambda^*_x$ for each $x$, so only one vector can satisfy these
  conditions. We just saw that these conditions are equivalent to
  those in Theorem~\ref{thm:existP_charP}, so $\lambda^*$ is the
  unique KT-vector.

  Characterization of $Q^*$: By Theorem~\ref{thm:charQ}, $Q^*$ is
  \optimal{} for the contestant if and only if the (unique) KT-vector
  equals the left-hand side of \eqref{properQ}. Similarly, if a
  strategy $P^*$ is \optimal{} for the quizmaster, then the KT-vector
  equals the right-hand side of \eqref{properQ}. Therefore: if for given
  $Q^*$ \anoptimal{} $P^*$ exists for which \eqref{properQ} holds, then
  both sides equal the KT-vector and $Q^*$ is \optimal{}; if $Q^*$ is
  \optimal{}, then \eqref{properQ} holds for all \optimal{} $P^*$; and if,
  for given $Q^*$, \eqref{properQ} holds for all \optimal{} $P^*$, then
  it holds for at least one \optimal{} $P^*$ by the existence of \optimal{}
  $P^*$.
\end{proof}

\begin{proof}[Proof of Theorem~\ref{thm:charP_local}]
  For local $L$, by definition $L(x, Q) = f_x(Q(x))$ for some sequence
  of functions $f_x: [0,1] \to [0,\infty]$. Given a point $P \in
  \Delta_y$, the vector $\lambda$ given by $\lambda_x = f_x(P(x))$
  defines a supporting hyperplane to $H_L \restriction \Delta_y$ at
  $P$, because $L$ is proper. Each $f_x$ is nonincreasing. (To see
  this, consider moving the point $P$ along any line that goes through
  the vertex of the simplex $\Delta_\X$ which puts all mass on some
  $x$. Because $H_L$ is concave along this line, the farther away $P$
  is from that vertex, the higher a supporting hyperplane to $H_L$ at
  $P$ will be at that vertex.)

  Given $P^*$ and $q$ satisfying the conditions in this theorem, let
  $\lambda^*_x$ be $f_x(q_x)$ for each $x \in \X$. We show that $P^*$
  is \optimal{} by verifying that $P^*$ and $\lambda^*$ satisfy
  Theorem~\ref{thm:existP_charP}. For each $y$ with $P^*(y) > 0$,
  $\lambda^*$ defines a supporting hyperplane to $H_L \restriction
  \Delta_y$ at $P^*(\cdot \mid y)$. For each other $y$, consider a
  supporting hyperplane to $H_L \restriction \Delta_y$ at $Q_{\mid
    y}(x) = q_x / \sum_{x' \in y} q_{x'}$: because $Q_{\mid y}(x) \geq
  q_x$, $f_x(Q_{\mid y}(x)) \leq f_x(q_x) = \lambda^*_x$, the
  hyperplane defined by $\lambda^*$ is everywhere at least as high as
  this supporting hyperplane, as required.

  For the converse: For strictly concave $H_L$, $f_x$ is strictly
  decreasing. %
  Define functions $g_x$ as follows: $g_x(\lambda_x) = \inf \setc{q
    \in [0,1]}{f_x(q) \leq \lambda_x}$. %
  If $f_x$ is continuous, $g_x$ is just the ordinary inverse of $f_x$,
  but if $f_x$ has a jump discontinuity at $q$, there will be an
  interval where $g_x$ is constantly equal to $q$. In either case,
  $g_x$ satisfies $g_x(f_x(q)) = q$ for all $q \in [0,1]$.

  Take $P^*$ some \optimal{} strategy, and $\lambda^*$ a KT-vector.
  Define $q \in [0,1]^\X$ by $q_x = g_x(\lambda^*_x)$. We use
  Theorem~\ref{thm:existP_charP} to show that $q$ satisfies
  \eqref{RCARcondition}. For each $y \in \Y$, let $\lambda' \in
  \Lambda_y$ be a minimal supporting hyperplane to $H_L \restriction
  \Delta_y$ that obeys $\lambda' \leq \lambda^*$; such a $\lambda'$
  exists by Lemma~\ref{lem:lambdareduction}.
  Let $Q_{\mid y}$ be the (unique) point at which $\lambda'$ supports
  $H_L \restriction \Delta_y$. It satisfies $f_x(Q_{\mid y}(x)) =
  \lambda'_x$, from which it follows that $g_x(\lambda'_x) =
  g_x(f_x(Q_{\mid y}(x))) = Q_{\mid y}(x)$.
  Applying $g_x$ to both sides of $\lambda'_x \leq \lambda^*_x$, we
  get $Q_{\mid y}(x) \geq q_x$ for all $x \in y$, so that $\sum_{x \in
    y} q_x \leq \sum_{x \in y} Q_{\mid y}(x) = 1$. If $P^*(y) > 0$,
  then the hyperplane defined by $\lambda^*$ is itself a supporting
  hyperplane and $Q_{\mid y}$ is the point where it touches $H_L
  \restriction \Delta_y$, namely the point $P^*(\cdot \mid y)$.
  Because $Q_{\mid y}(x)$ also satisfies $Q_{\mid y}(x) =
  g_x(\lambda^*_x) = q_x$, the equality $q_x = Q_{\mid y}(x) = P^*(x
  \mid y)$ follows.
\end{proof}

\begin{proof}[Proof of Lemma~\ref{lem:RCARvector}]
  At least one vector $q$ must exist because for the game with
  logarithmic loss and $\X, \Y, \marg{}$ as in the lemma, \anoptimal{}
  strategy $P^*$ must exist by Theorem~\ref{thm:existP_charP}, and an
  associated RCAR vector $q$ must exist for it by
  Theorem~\ref{thm:charP_local} using that $H_L$ is strictly concave.

  For logarithmic loss, the RCAR vector $q$ and KT-vector $\lambda^*$
  are related by $\lambda^*_x = -\log q_x$. By
  Theorem~\ref{thm:existP_charP}, any strategy $P \in \P$ that does
  not agree with the KT-vector $\lambda^*$ is not \optimal{}, showing
  that $q$ is unique.
\end{proof}

\begin{proof}[Proof of Lemma~\ref{lem:strictlyconcave}]
  We will %
  show that any nonvertical supporting hyperplane $\lambda
  \in \Lambda_y$ is supporting at no more than one point $P \in
  \Delta_y$. By Lemma~\ref{lem:lambdareduction}, if a supporting
  hyperplane exists at $P$, then a minimum supporting hyperplane also
  exists at that point, so it suffices to restrict our attention to
  minimal $\lambda \in \Lambda_y$. We know that such a $\lambda$ is
  realizable on $y$; let $Q$ be a distribution realizing it. Then $Q$
  minimizes the expected loss against any $P$ at which $\lambda$
  supports $H_L \restriction \Delta_y$. For strictly proper $L$, there
  can be at most one such $P$, proving strict concavity.
\end{proof}

\begin{proof}[Proof of Lemma~\ref{lem:affinetransform}]
  The generalized entropy function of $L'$ is given by
  \[
  H_{L'}(P) = a H_L(P) + \sum_{\mathclap{x \in \X}} b_x P(x),
  \]
  where $a \in \R_{>0}$ and $b \in \R^\X$ are the constants in the
  affine transformation \eqref{affinetransform}. $H_{L'}$ is again
  finite and continuous. If $P^*$ is \optimal{} for the quizmaster in
  game $\G$, then by Theorem~\ref{thm:existP_charP} there exists a
  KT-vector $\lambda^*$ satisfying that theorem's conditions. Define a
  transformed vector by $\lambda' = a \lambda^* + b$. This is a
  KT-vector for $\G'$, showing that $P^*$ is also \optimal{} in that
  game.

  If the conditions of Theorem~\ref{thm:existQ} hold for $\G$, then
  they also holds for $\G'$: If $\lambda'$ is a minimal supporting
  hyperplane to $H_{L'} \restriction \Delta_y$, then $\lambda =
  (1/a)(\lambda' - b)$ is a minimal supporting hyperplane to $H_L
  \restriction \Delta_y$. By assumption, $\lambda$ is realizable on
  $y$ in game $\G$, say by $Q \in \Delta_\X$. Then the same $Q$ also
  realizes $\lambda'$ in game $\G'$.

  If $Q^*$ is \optimal{} for the contestant in $\G$, then by
  Theorem~\ref{thm:charQ}, $\lambda$ given by $\lambda_x \isdef
  \max_{y \ni x} L(x, Q^*_{\mid y})$ is a KT-vector. The transformed
  vector $\lambda' = a \lambda + b$ is then a KT-vector in $\G'$, so
  that by Theorem~\ref{thm:charQ}, $Q^*$ is \optimal{} in that game.

  Because the affine transformation from $L$ into $L'$ can be reversed
  by a second affine transformation (with $a' = 1/a$ and $b' =
  -(1/a)b$), the reverse implications follow.
\end{proof}

\begin{proof}[Proof of Lemma~\ref{lem:decomposition}]
  For each $y \in \Y$, assume without loss of generality that $y \in
  \Y_1$. Then observe that the generalized entropies for $\G$ and
  $\G_1$ are identical on $\Delta_y$; $P^*(y) > 0$ if and only if
  $P^*_1(y) > 0$; and $P^*(x \mid y) = P^*_1(x \mid y)$ for all $x \in
  y$. Now the claim follows from Theorem~\ref{thm:existP_charP}.
\end{proof}

\begin{proof}[Proof of Lemma~\ref{lem:lossexchange}]
  By Theorem~\ref{thm:existP_charP}, $\lambda^*$ is supporting to $H_L
  \restriction \Delta_{y_1}$ at $P^*(\cdot \mid y_1)$. Define $\lambda^1
  \in \Lambda_{y_1}$ equal to $\lambda^*$ on $y_1$. Then by
  Lemma~\ref{lem:lambdareduction}, any $\lambda' \in \Lambda_{y_1}$
  with $\lambda' \leq \lambda^1$ obeys $\lambda'_{x_1} =
  \lambda^1_{x_1}$.

  Again by Theorem~\ref{thm:existP_charP}, $\lambda^*$ is dominating
  to $H_L \restriction \Delta_{y_2}$. Define $\lambda^2$ by
  $\lambda^2_x = \lambda^*_x$ for $x \in y_1 \cap y_2$,
  $\lambda^2_{x_1} = \lambda^*_{x_2}$, and 0 elsewhere. Because $L$ is
  symmetric between $x_1$ and $x_2$, $\lambda^2 \in \Lambda_{y_1}$.
  For $x \in y_1 \cap y_2$, $\lambda^1_x = \lambda^2_x$. If
  $\lambda^2_{x_1} \leq \lambda^1_{x_1}$, then $\lambda^2 \leq
  \lambda^1$; then we must have $\lambda^2_{x_1} = \lambda^1_{x_1}$.
  So $\lambda^2_{x_1} < \lambda^1_{x_1}$ is impossible, and we find
  $\lambda^*_{x_1} = \lambda^1_{x_1} \leq \lambda^2_{x_1} =
  \lambda^*_{x_2}$.
\end{proof}

\begin{proof}[Proof of Theorem~\ref{thm:binary}]
  For graph games, all loss functions are essentially local. We will
  make this precise by constructing functions $f_x$, analogous to
  those in the proof of Theorem~\ref{thm:charP_local}: they have the
  property that for all $y \in \Y$, a supporting hyperplane to $H_L
  \restriction \Delta_y$ at $P \in \Delta_y$ is given by $\lambda$
  with $\lambda_x = f_x(P(x))$ for all $x \in y$. (Note that we may
  not get $f_x(Q(x)) = L(x, Q)$ as in the case of local proper loss
  functions in the proof of Theorem~\ref{thm:charP_local}, because the
  hyperplane realized by $Q$ may not be supporting at $Q$ if $L$ is
  improper.)

  For each $x \in \X$, if the only message in which $x$ occurs is
  $\set{x}$, then $f_x(q)$ is only defined for $q=1$, where it is
  $f_x(1) = H_L(e_x)$ (where $e_x$ is the unique element of
  $\Delta_y$). For other $x$, pick any message $y \in \Y$ with $y \ni
  x$ and $\lvert y \rvert = 2$. For all these $y$, the generalized
  entropies $H_L \restriction \Delta_y$ are identical copies of the
  same function, by symmetry of $L$. For each $q \in [0,1]$, pick a
  supporting hyperplane $\lambda$ to $H_L \restriction \Delta_y$ at
  the unique $P \in \Delta_y$ with $P(x) = q$, and let $f_x(q) =
  \lambda_x$. If $H_L$ is not differentiable at $P$ (including when $q
  \in \set{0,1}$), we can choose a supporting hyperplane arbitrarily
  as long as the same one is used to define $f_x(q)$ and $f_{x'}(1-q)$
  wherever $\set{x, x'} \in \Y$. (In particular this means that if a
  connected component of $\Y$ viewed as a graph contains an odd cycle,
  $f_x(1/2)$ must take the same value for all $x$ in that component.)

  As for local $L$, each $f_x$ is nonincreasing because $H_L$ is
  concave, and $f_x$ is strictly decreasing if $H_L$ is strictly
  concave. The rest of the proof is the same as for
  Theorem~\ref{thm:charP_local}.
\end{proof}

\begin{proof}[Proof of Theorem~\ref{thm:matroid}]
  We know from Theorem~\ref{thm:charP_local} that a quizmaster %
  strategy $P^*$ is
  \optimal{} for logarithmic loss if and only if it is RCAR, and from
  Theorem~\ref{thm:existP_charP} that such a $P^*$ exists. Take any
  such $P^*$. Let $\lambda$ be the KT-vector with respect to
  logarithmic loss, and $\Y_{P^*} = \setc{y \in \Y}{P^*(y) > 0}$. For
  any pair $y \in \Y_{P^*}, y' \in \Y$, we will show that there exists
  a bi\/jection $\pi$ from $y \setminus y'$ to $y' \setminus y$ such
  that $\lambda_x \leq \lambda_{\pi(x)}$ for all $x \in y \setminus
  y'$. This follows from \citet[Corollary 39.12a]{SchrijverB}, but
  here we give a direct proof by induction on $\lvert y' \setminus y
  \rvert$:
  \begin{itemize}
  \item $\lvert y' \setminus y \rvert = 1$: Apply
    Lemma~\ref{lem:lossexchange} to $y_1 = y$ and $y_2 = y'$ (using
    that for an RCAR strategy, $P^*(y_1) > 0$ implies $P^*(x_1, y_1) >
    0$ for all $x_1 \in y_1$) to find the required inequality.
  \item $\lvert y' \setminus y \rvert > 1$:
    Let $y'_1 = y'$ and pick any $x_1 \in y
    \setminus y'$. Starting with $i = 1$, apply the basis exchange
    property on $y \setminus \set{x_i}$ and $y'_i$ to find $x'_i$ (it
    will be in $y'_i \setminus y \subseteq y'$); then apply it again
    on $y'_i \setminus \set{x'_i}$ and $y$ to find $x_{i+1} \in y
    \setminus y'_i$, defining the message $y'_{i+1} = y'_i \setminus
    \set{x'_i} \cup \set{x_{i+1}}$ (which may not be in $\Y_{P^*}$).
    Continue until $x_{i+1} = x_1$. Now $\pi$ defined by $\pi(x_1) =
    x'_1, \ldots, \pi(x_i) = x'_i$ is a bi\/jection from $\set{x_1,
      \ldots, x_i} = (y \cap y'_{i+1}) \setminus y' \subseteq y
    \setminus y'$ to $\set{x'_1, \ldots, x'_i} = y' \setminus y'_{i+1}
    \subseteq y' \setminus y$ (to see this, note that an element
    $x'_j$ found in the basis exchange from $y$ is then removed from
    $y'_{j+1}$ so that it will not be found again; an element
    $x_{j+1}$ found in the other basis exchange is added to $y'_{j+1}$
    with the same result), and for each $1 \leq j \leq i$, applying
    Lemma~\ref{lem:lossexchange} to $y$ and $y \cup \set{x'_j}
    \setminus \set{x_j}$ tells us that $\lambda_{x_j} \leq
    \lambda_{x'_j}$ as required. If $y'_{i+1} = y$, then this is the
    bi\/jection we are looking for; otherwise, it can be completed by
    combining it with a bi\/jection from $y \setminus y'_{i+1}$ to
    $y'_{i+1} \setminus y$, which exists by the induction hypothesis.
  \end{itemize}
  If also $y' \in \Y_{P^*}$, a bi\/jection $\pi'$ from $y' \setminus
  y$ to $y \setminus y'$ such that $\lambda_{x'} \leq
  \lambda_{\pi'(x')}$ is found by the same argument. Together, $\pi$
  and $\pi'$ divide the outcomes in the two sets into disjoint cycles
  that must all have the same value for $\lambda$, defining an
  equivalence relation $\sim$ on $y \oplus y'$. For logarithmic loss,
  the RCAR vector $q$ obeys $q_x = e^{-\lambda_x}$, so it must also be
  constant within each equivalence class of $\sim$. Because the
  entropy of logarithmic loss is strictly concave, the conditionals of
  $P^*$ must agree with $q$ by Theorem~\ref{thm:charP_local}.

  Now take an arbitrary loss function $L$ satisfying the conditions in
  the theorem, and the same strategy $P^*$. At an arbitrary message
  $y$ with $P^*(y) > 0$, choose a supporting hyperplane $\lambda' \in
  \Lambda_y$ to $H_L \restriction \Delta_y$ at $P^*(\cdot \mid y)$
  with the property that $\lambda'_x = \lambda'_{x'}$ for all $x, x'
  \in y$ with $\lambda_x = \lambda_{x'}$ and between which $L$ is
  symmetric: there $P^*(x \mid y) = P^*(x' \mid y)$, so such a
  supporting hyperplane exists. For all $x, x' \in y$ with $q_x >
  q_{x'}$ (equivalently, $\lambda_x < \lambda_{x'}$) between which $L$
  is symmetric, this $\lambda'$ satisfies $\lambda'_x \leq
  \lambda'_{x'}$. (A supporting hyperplane to $H_L \restriction
  \Delta_y$ at $P^*(\cdot \mid y)$ with $\lambda'_x > \lambda'_{x'}$
  would be lower at $(P^*)^{x \leftrightarrow x'}(\cdot \mid y)$ than
  at $P^*(\cdot \mid y)$, while by symmetry $H_L$ is the same at those
  points: a contradiction.)

  For each $y' \in \Y_{P^*}$ other than $y$, we saw that each $x' \in
  y' \setminus y$ is in the same equivalence class of $\sim$ as some
  $x \in y \setminus y'$. In addition to $\lambda_x = \lambda_{x'}$,
  $x \sim x'$ implies that $L$ is symmetric between $x$ and $x'$
  (because $L$ is symmetric with respect to exchanges in $\Y$, and
  $\pi$ and $\pi'$ were constructed using such exchanges). We extend
  the definition of $\lambda'$ to $y'$ by setting $\lambda'_{x'} =
  \lambda'_x$. This way, $\lambda'$ defines a supporting hyperplane to
  $H_L \restriction \Delta_{y'}$ at $P^*(\cdot \mid y')$. We repeat
  this for all $y' \in \Y_{P^*}$, defining $\lambda'$ on all of $\X$.
  
  Also, for each $y' \in \Y \setminus \Y_{P^*}$ and $y \in \Y_{P^*}$,
  we have that a bi\/jection $\pi$ exists from $y \setminus y'$ to $y'
  \setminus y$ such that for all $x \in y \setminus y'$, $L$ is
  symmetric between $x$ and $\pi(x)$, and $\lambda_x \leq
  \lambda_{\pi(x)}$; then also $\lambda'_x \leq \lambda'_{\pi(x)}$, so
  $\lambda'$ defines a dominating hyperplane to $H_L \restriction
  \Delta_{y'}$. Thus $\lambda'$ is a KT-vector certifying that $P^*$
  is also \optimal{} for $L$.
  
  For the converse: If $H_L$ is strictly concave, the supporting
  hyperplanes de\-fined by a KT-vector $\lambda'$ each touch $H_L
  \restriction \Delta_y$ at only one point, so that any \optimal{}
  strategy $P'$ for the quizmaster must have $P'(x \mid y) = q_x$ for
  all $x \in y$ with $P'(y) > 0$. Therefore any \optimal{} $P'$ must
  be RCAR.
\end{proof}

\begin{proof}[Proof of Theorem~\ref{thm:nonmatroid_nongraph}]
  We will first show how to construct a vector $q \in \R_{>0}^\X$ that
  satisfies $\sum_{x \in y} q_x \leq 1$ for all $y \in \Y$, and for
  all $x \in \X$, there is a message $x \in y \in \Y$ with $\sum_{x
    \in y} q_x = 1$. Then we will determine a marginal so that this
  vector $q$ is the RCAR vector of the game with that marginal. We
  will additionally find two intersecting messages, both having sum 1,
  such that $q$ represents the uniform distribution on one, but not on
  the other.

  Two different constructions are given: one for nonuniform and one
  for uniform games.

  If the game is not uniform, let $k_2$ be the size of the largest
  message in $\Y$. By connectedness, there exists a message of size
  less than $k_2$ that has nonempty intersection with a message of
  size $k_2$. From among such messages, let $y_1$ be one of maximum
  size, and let $k_1 < k_2$ be that size. Finally, let $y_2$ be a
  messages of size $k_2$ that maximizes $\lvert y_2 \cap y_1 \rvert$.
  Set initial values for $q$ as follows:
  \begin{equation*}
    q_x = \begin{cases}
      \frac{1}{k_1} & \text{for } x \in y_1;\\
      \frac{\lvert y_1 \setminus y_2 \rvert}{\lvert y_2 \setminus
        y_1 \rvert} \cdot \frac{1}{k_1}
          & \text{for } x \in y_2 \setminus y_1;\\
      \frac{1}{\lvert y_2 \setminus y_1 \rvert} \cdot \frac{1}{k_1}
          & \text{otherwise.} %
    \end{cases}
  \end{equation*}
  Note that the three cases of $q_x$ are listed in nonincreasing
  order. Now $\sum_{x \in y_1} q_x \allowbreak= \sum_{x \in y_2} q_x = 1$, while
  $\sum_{x \in y} q_x \leq 1$ for general $y \in \Y$: $\max_x q_x =
  1/k_1$, so a message $y \in \Y$ with $\lvert y \rvert \leq k_1$ will
  have sum at most 1; a message with $\lvert y \rvert = k_2$ will
  share no more outcomes with $y_1$ than $y_2$ does and thus cannot
  have a larger sum; and because a message with $k_1 < \lvert y \rvert
  < k_2$ has empty intersection with $y_2$,
  the $k_1 - 1$ largest elements of $(q_x)_{x \in y}$ sum to at most
  $(k_1-1)/k_1$, while the fewer than $\lvert y_2 \setminus y_1
  \rvert$ remaining elements all equal $1/(\lvert y_2 \setminus y_1
  \rvert \cdot k_1)$ and hence sum to less than $1/k_1$.

  A greedy algorithm that repeatedly increments some $q_x$ until none
  can be increased further, while maintaining the inequality $\sum_{x
    \in y} q_x \leq 1$ on each $y$, will terminate with a $q$
  satisfying the conditions stated at the beginning of the proof. This
  $q$ will be unchanged and thus still be uniform on $y_1$, while on
  the intersecting message $y_2$, $q$ also still sums to 1 but is not
  uniform.

  For the case of uniform games, the construction is similar. Let $k$
  be the size of the game's messages. By \citet[Corollary
  2.1.5]{Oxley_matroid}, a nonempty family of sets $\Y$ is the
  collection of bases of a matroid if and only if for all $y_1, y_2
  \in \Y$ and $x_2 \in y_2 \setminus y_1$,
  \begin{equation}\label{eq:dualexchange}
    y_1 \cup \set{x_2} \setminus \set{x_1} \in \Y
    \text{ for some } x_1 \in y_1 \setminus y_2.
  \end{equation}
  Because our $\Y$ is not a matroid, it follows that there exist $y_1,
  y_2 \in \Y$ and $x_2 \in y_2 \setminus y_1$ for which no
  corresponding $x_1$ exists. For $k \geq 3$ (which holds because the
  game we consider is not a graph game), we claim something stronger:
  that there exist $y_1, y_2, x_2$ as above with the additional
  property that $y_1$ and $y_2$ intersect. The proof of this claim is
  below.

  Using such $y_1$ and $x_2$ and some $0 < \epsilon < 1/k$, initialize
  $q$ as follows:
  \begin{equation*}
    q_x = \begin{cases}
      \frac{1}{k} & \text{for } x \in y_1;\\
      \frac{1}{k} + \epsilon & \text{for } x = x_2;\\
      \frac{1}{k} - \epsilon & \text{otherwise.}
    \end{cases}
  \end{equation*}
  Because any message containing $x_2$ also contains at least one
  other outcome not in $y_1$, we again have $\sum_{x \in y} q_x \leq
  1$ for all $y \in \Y$.

  For $k \geq 3$, the initial $q$ has the property that the set of
  outcomes $x$ for which $q_x$ cannot be increased further (we call
  these outcomes \emph{maximized}) is connected by messages $y$ with
  $\sum_{x \in y} x = 1$ (that is, the maximized outcomes cannot be
  partitioned into two nonempty sets such that each sum-1 message is
  contained in one of these sets); this is because $y_1$ has sum 1,
  and any other message with sum 1 must intersect $y_1$. (For $k = 2$,
  this would not be the case: the only messages having sum 1 would be
  $y_1$ and all messages that contain $x_2$, but $y_1$ would not
  intersect any of these.) We can have the greedy algorithm maintain
  this as an invariant: Because the game is connected, there is always
  a message partially in the set of maximized outcomes and partially
  outside. We call such a message a \emph{crossing} message. Each
  round, we pick an outcome $x$ that is not maximized yet and is
  contained in a crossing message; if $x_2$ is not maximized, we
  always pick $x = x_2$ (using that it is contained in the crossing
  message $y_2$). The tightest constraint on increasing $q_x$ will
  come from a crossing message, because for any non-crossing message
  $y \ni x$, we have $\sum_{x' \in y \setminus \set{x}} q_{x'} =
  (k-1)(1/k - \epsilon)$, which is the smallest possible value of this
  sum. So increasing $q_x$ as much as possible will cause a crossing
  message to get sum equal to 1. This message connects $x$ to the set
  of previously maximized outcomes, and any other outcomes that were
  maximized by this increment must be contained in some message that
  also contains $x$.

  When the greedy algorithm terminates, $q$ will still be uniform on
  $y_1$, while there will be another message on which $q$ sums to one
  but is not uniform. (This may not be $y_2$, which may not have sum
  1.) Because all outcomes are connected by sum-1 messages, we can
  also find a pair of intersecting messages, one of which is uniform
  and one of which is not. Use these two messages as $y_1$ and $y_2$
  in the sequel.

  Having found, for both nonuniform and uniform games, a vector $q$
  and messages $y_1$ and $y_2$ as described above,
  we let strategy $P$ be RCAR with vector $q$ and $P(y)$
  uniform on $\setc{y}{\sum_{x \in y} q_x = 1}$. This $P$ is
  \anoptimal{} strategy for the game with logarithmic loss and
  marginal $\marg{}_x = \sum_{y \ni x} P(x, y)$, and $q$ is its unique
  RCAR vector.

  We will show that $P$ is not \optimal{} for the game with the same
  marginal and Brier loss. Brier loss is proper and continuous, so by
  Theorem~\ref{thm:proper}, $L(x, P(\cdot \mid y_1)) = L(x, P(\cdot
  \mid y_2))$ for \optimal{} $P$. These are squared Euclidean
  distances from a vertex of the simplex to the predicted
  distribution. However, the equality will not hold for $P$:

  Among all predictions in $\Delta_{y_i}$ with $Q(x) = q_x$ for each
  $x \in y_1 \cap y_2$ (this set of predictions is the intersection of
  $\Delta_{y_i}$ with an affine subspace), the squared Euclidean
  distance $L(x, Q)$ between such $Q$ and given vertex $x \in y_1 \cap
  y_2$ is uniquely minimized by $Q$ uniform on the outcomes not in
  $y_1 \cap y_2$ (this is the orthogonal projection of the vertex onto
  that subspace).
  For a uniform game, $P(\cdot \mid y_1)$ is uniform and thus $L(x,
  P(\cdot \mid y_1))$ equals this minimum; $P(\cdot \mid y_1)$ differs
  from the uniform distribution at some outcomes not in $y_1 \cap y_2$
  and thus $L(x, P(\cdot \mid y_2))$ is larger than the minimum.

  For a nonuniform game, $P(\cdot \mid y_1)$ is uniform on $y_1
  \setminus y_2$ and $P(\cdot \mid y_2)$ is uniform on $y_2 \setminus
  y_1$, so both minimize the distance to the vertex in their
  respective subspaces. However, the subspace for $y_2$ is isomorphic
  to a subspace contained in the subspace for $y_1$ and not containing
  $P(\cdot \mid y_1)$. Therefore $L(x, P(\cdot \mid y_1)) < L(x,
  P(\cdot \mid y_2))$.
\end{proof}

\begin{proof}[Proof of claim]
  Suppose for a contradiction that any pair of intersecting messages
  $y, y'$ obeys the above exchange property \eqref{dualexchange} for
  all $x' \in y' \setminus y$. Let $y_1, y_2$ be two messages that
  fail \eqref{dualexchange} for some outcome $x_2 \in y_2 \setminus
  y_1$; it follows from our assumption that they are disjoint. Because
  $\Y$ is connected, there exists a sequence of messages starting with
  $y_1$ and ending with $y_2$ in which adjacent messages intersect.
  Using \eqref{dualexchange}, we can extend this sequence to one where
  adjacent messages differ by the exchange of one outcome: given
  intersecting $y, y'' \in \Y$ with $d \isdef \lvert y'' \setminus y
  \rvert > 1$, we find $y' \in \Y$ with $\lvert y' \setminus y \rvert
  = 1$ and $\lvert y'' \setminus y' \rvert = d-1$. Write the entire
  sequence as $y^0 = y_1, y^1, \ldots, y^n = y_2$.

  We have $n \geq k$, because $n < k$ would imply that $y_1 \cap y_2
  \neq \myempty$. If $n > k$, we can find a shorter sequence as follows:
  pick $0 \leq i < j \leq n - k$ for which $y^i \cap y^{j+1} \neq
  \myempty$; this holds if $j+1 - i < k$, so such $i, j$ can always be
  found if $k \geq 3$. Let $x'$ be the unique outcome in $y^{j+1}
  \setminus y^j$.
  \begin{itemize}
  \item If $x' \not\in y^{j+k}$ (intuitively, adding $x'$ leads us on
    a detour that can be avoided when going to $y^{j+k}$): In each of
    the $k$ exchange steps from $y^j$ to $y^{j+k}$, one outcome was
    removed. One of those outcomes was $x'$, which is not in $y^j$, so
    at most $k-1$ outcomes from $y^j$ were removed. Thus $y^j$ and
    $y^{j+k}$ intersect, and a shorter path between them can be found
    using \eqref{dualexchange}.
  \item If $x' \in y^{j+k}$ and $x' \in y^i$ (removing $x'$ is the
    start of a detour): We can use \eqref{dualexchange} to find a
    shorter path between $y^i$ and $y^{j+k}$.
  \item If $x' \in y^{j+k}$ but $x' \not\in y^i$ (adding $x'$ is
    apparently useful, but can be done sooner): Apply
    \eqref{dualexchange} to messages $y^i$ and $y^{j+1}$ (which
    intersect) and outcome $x'$ (which is in $y^{j+1}$ but not in $y^i$)
    to find a message $y'$ that is one step away from $y^i$ and
    contains $x'$. From $y'$, we can find a path to $y^{j+k}$ by
    \eqref{dualexchange} taking fewer than $k$ steps. Thus we can get
    from $y^i$ to $y^{j+k}$ in at most $k$ steps.
  \end{itemize}
  Thus we can always find a sequence with $n = k$.

  Given such a sequence $y^0, y^1, \ldots, y^n$, we will now show a
  contradiction with the assumption that $y_1 = y^0$ and $y_2 = y^n$
  fail \eqref{dualexchange} by showing that for any $x_2 \in y_2$, a
  message exists that differs from $y_1$ by adding $x_2$ and removing
  one other outcome. If $x_2 \in y^1$, then $y^1$ is such a message
  and we are done. Otherwise, we can apply \eqref{dualexchange} to
  $y^1$ and $x_2 \in y_2$ to find a message $y'$ containing $x_2$;
  because $k \geq 3$, this message still intersects $y_1$, so applying
  \eqref{dualexchange} to $y_1$ and $x_2 \in y'$ gives the message we
  are looking for. This shows by contradiction that if a connected
  uniform game with $k \geq 3$ is not a matroid game, there exists a
  pair of intersecting messages $y_1, y_2$ and an outcome $x_2 \in y_2
  \setminus y_1$ that do not satisfy \eqref{dualexchange}.
\end{proof}